\documentclass[a4paper,11pt]{article}
\pdfoutput=1 

\usepackage{jheppub} 

\usepackage[T1]{fontenc} 
\usepackage[numbers]{natbib}
\usepackage{filecontents}
\usepackage{dynkin-diagrams}
\usepackage{tikz}
\usepackage{float}
\usepackage{placeins}
\usepackage{stackengine}
\tikzset{Rightarrow/.style={double equal sign distance,>={Implies},->},
triple/.style={-,preaction={draw,Rightarrow}},
quadruple/.style={preaction={draw,shorten >=0pt},shorten >=1pt,-,double,double
distance=0.2pt}}
\usetikzlibrary{positioning}
\usepackage{graphicx}
\usepackage{pdflscape}
\usepackage{caption}
\usepackage{subcaption}
\usepackage{comment}
\usepackage{subfiles}
\usepackage{todonotes}
\usepackage{booktabs}

\usetikzlibrary{hobby,intersections} 
\usetikzlibrary{shapes.geometric}
\usetikzlibrary{shapes.misc}

\def\ns#1#2{
	\node[circle, draw, fill=white] (#2) at (#1){};
	\node[cross out, draw] at (#1){};
}
\tikzset{flavour/.style={draw=none,minimum size=0.3mm,fill=white, regular polygon,regular polygon sides=4,draw}}
\tikzset{flavourr/.style={draw=none,minimum size=0.3mm,fill=red, regular polygon,regular polygon sides=4,draw}}
\tikzset{gaugeBig/.style={inner sep=1mm,draw=none,fill=white,minimum size=2mm,circle, draw}}
\tikzset{bd/.style={circle, draw=black, inner sep=0pt, fill=black, minimum size=2mm}}
\tikzset{wd/.style={circle, draw=black, inner sep=0pt, fill=white, minimum size=2mm}}
\tikzset{Dynkin/.style={circle, draw=black, inner sep=0pt, fill=white, minimum size=2mm}}
\tikzstyle{ligne}=[draw, very thick] 
\tikzstyle{gridline}=[draw, gray] 
\tikzset{gauge/.style={circle, draw,inner sep=2.5pt}}
\tikzset{gaugeo/.style={circle, draw,inner sep=2.5pt,fill=orange}}
\tikzset{gauger/.style={circle, draw,inner sep=2.5pt,fill=red}}
\tikzset{gaugeb/.style={circle, draw,inner sep=2.5pt,fill=blue}}
\tikzset{gaugeg/.style={circle, draw,inner sep=2.5pt,fill=green}}
\tikzset{gaugegoodgreen/.style={circle, draw,inner sep=2.5pt,fill=goodgreen}}
\tikzset{gaugem/.style={circle, draw,inner sep=2.5pt,fill=magenta}}
\tikzset{hasse/.style={circle, fill,inner sep=2pt}}
\tikzset{d2/.style={circle, fill,inner sep=1.3pt}}
\tikzset{shrinky/.style={circle, fill,inner sep=1pt}}
\tikzset{sized/.style={circle, draw, inner sep=1.5pt}}
\tikzset{seven/.style={circle, draw,inner sep=3pt}}

\makeatletter
\DeclareRobustCommand{\rvdots}{%
  \vbox{
    \baselineskip4\p@\lineskiplimit\z@
    \kern-\p@
    \hbox{.}\hbox{.}\hbox{.}
  }}
\makeatother

\newcommand{\Figref}[1]{Figure~\ref{#1}}
\newcommand{\Quiver}[1]{$\mathcal Q_{\ref{#1}}$}
\newcommand{\surm}{\mathrm{SU}}

\newcommand{\urm}{\mathrm{U}}
\newcommand{\sorm}{\mathrm{SO}}
\newcommand{\orm}{\mathrm{O}}
\newcommand{\sprm}{\mathrm{Sp}}
\newcommand{\hs}{\mathrm{HS}}

\newcommand{\hwg}{\mathrm{HWG}}
\newcommand{\pe}{\mathrm{PE}}
\newcommand{\pl}{\mathrm{PL}}

\title{\boldmath Orthosymplectic Quotient Quiver Subtraction}

\author{Sam Bennett,}
\author{Amihay Hanany,}
\author{and Guhesh Kumaran}

\affiliation{Theoretical Physics Group, The Blackett Laboratory, Imperial College London, Prince Consort Road, SW7 2AZ, UK}

\emailAdd{a.hanany@imperial.ac.uk}
\emailAdd{samuel.bennett18@imperial.ac.uk}
\emailAdd{guhesh.kumaran18@imperial.ac.uk}

\abstract{The technique of \textit{orthosymplectic quotient quiver subtraction} is introduced. This involves subtraction of an \textit{orthosymplectic quotient quiver} from a $3d\;\mathcal N=4$ orthosymplectic quiver gauge theory which has the effect of gauging subgroups of the IR Coulomb branch global symmetry. Orthosymplectic quotient quivers for $\mathrm{SU}(2),\;\mathrm{SU}(3),\;G_2,$ and $\mathrm{SO}(7)$ are found and derived from Type IIA brane systems involving negatively charged branes for certain $6d\;\mathcal N=(1,0)$ gauge theories. Orthosymplectic quotient quiver subtraction is applied to magnetic quivers for nilpotent orbit closures providing new orthosymplectic counterparts to known unitary quivers. New Coulomb branch constructions are found such as for two height four nilpotent orbit closures of $F_4$ and one of height three. A novel application is to find magnetic quivers and Type IIA brane systems for the $6d\;\mathcal N=(1,0)$ worldvolume theory of two $\frac{1}{2}$M5 branes on $E_6$ Klein singularity and for $6d\;\mathcal N=(1,0)\;(E_6,E_6)$ conformal matter. These give a perturbative Lagrangian realisation to the dynamics of strongly interacting M5 branes. The magnetic quiver for $6d\;\mathcal N=(1,0)\;(E_6,E_6)$ conformal matter is star-shaped and can also be interpreted as a magnetic quiver for a class $\mathcal S$ theory specified by $\mathrm{SO}(26)$ algebra on a three-punctured sphere.}
\begin{document}
\preprint{Imperial/TP/24/AH/02}
\maketitle
\flushbottom
\section{Introduction}
Gauging a subgroup of the flavour symmetry of a $3d\;\mathcal N=4$ quiver gauge theory is a classical problem \cite{Argyres:2007cn} which is well understood. The Higgs branch is a symplectic singularity and the action of gauging flavour symmetry subgroups corresponds to the action of a hyper-K\"ahler quotient on the Higgs branch. This procedure typically proceeds the same in the UV as in the IR. Performing an analogous gauging of a subgroup of the Coulomb branch global symmetry is much more difficult. The Coulomb branch in the UV is a smooth variety with the global symmetry being the Abelian topological symmetry given by the center of the gauge group. In the IR, due to perturbative and non-perturbative effects, the Coulomb branch is deformed to also be a symplectic singularity. Additionally, the Abelian global symmetry in the UV can be enhanced to a non-Abelian symmetry in the IR \cite{Intriligator:1996ex}. The focus of this paper will lie on the IR Coulomb branch which is a symplectic singularity and may enjoy a richer set of non-Abelian global symmetry subgroups that may be gauged. The reason that the Coulomb branch global symmetry is difficult to gauge is because there is strong coupling in the IR and perturbative methods fail.

Recent efforts have tackled the problem by introducing new combinatoric operations on certain $3d\;\mathcal N=4$ quiver gauge theories with unitary gauge groups. These operations are called \textit{quotient quiver subtraction} \cite{Hanany:2023tvn}  and \textit{quiver polymerisation} \cite{Hanany:2024fqf}. The gauging of the Coulomb branch global symmetry corresponds to an analogous hyper-Kähler quotient on the Coulomb branch.
Both of these techniques are inspired from known constructions in supersymmetric gauge theory and in string theory. Quiver polymerisation takes inspiration from the (partial) gluing of Riemann surfaces from $4d\;\mathcal N=2$ class $\mathcal S$ theories and the corresponding action on their \textit{magnetic quivers} \cite{Gaiotto:2009we,Benini:2010uu}. The authors of \cite{Benini:2010uu} also gave an alternative construction using 5-brane webs \cite{Aharony:1997bh, Benini:2009gi}. More pertinent to this work is the inspiration for quotient quiver subtraction which comes from studying the gauging of an $\surm(3)$ flavour symmetry subgroup of the anomaly-free $6d\;\mathcal N=(1,0)\;\sprm(0)$ gauge theory at infinite coupling \cite{Hanany:2022itc}. Specifically, through the description of the Higgs branches of these $6d$ theories with magnetic quivers \cite{Cabrera:2019izd,Cabrera:2019dob}. These magnetic quivers have origin from Type IIA brane systems with negatively charged branes \cite{Mekareeya:2016yal,Hanany:2022itc}.


A magnetic quiver provides a construction of a moduli space of vacua (typically the Higgs branch) of a $d=3,4,5,6$ gauge theory with eight supercharges as a moduli space of dressed monopole operators \cite{Cabrera:2019izd,Cabrera:2019dob}. This is particularly useful in cases where perturbative methods do not apply, such as for strongly coupled theories or theories with no Lagrangian description. Using intuition gained from Type IIB D3-/D5-/NS5-brane systems, magnetic quivers can be constructed from a variety of D$p$-/D$(p+2)$-brane configurations in both Type IIA and Type IIB.

The magnetic quiver programme has been particularly successful in developing simple and consistent combinatorial techniques on unitary magnetic quivers which have straightforward realisations as actions on moduli spaces of vacua \cite{Cabrera:2018ann,Hanany:2023tvn,Bourget:2022ehw,Bourget:2022tmw,Bourget:2020bxh,Bourget:2021xex,Hanany:2018vph,Hanany:2018dvd,Hanany:2023uzn,Hanany:2023uzn,Bourget:2023dkj,Bourget:2024mgn,Dancer:2024lra,Hanany:2024fqf}. These techniques have been applied to understand the structure and actions on moduli spaces of vacua of theories with eight supercharges in $d=3,4,5,6$ \cite{Cabrera:2019izd,Bourget:2019aer,Bourget:2019rtl,Cabrera:2019dob,Grimminger:2020dmg,Bourget:2020gzi,Bourget:2020asf,Bourget:2020xdz,Beratto:2020wmn,Closset:2020scj,Akhond:2020vhc,vanBeest:2020kou,Bourget:2020mez,VanBeest:2020kxw,Giacomelli:2020ryy,Akhond:2021knl,Carta:2021whq,Arias-Tamargo:2021ppf,Bourget:2021xex,Gledhill:2021cbe,vanBeest:2021xyt,Carta:2021dyx,Sperling:2021fcf,Nawata:2021nse,Akhond:2022jts,Giacomelli:2022drw,Kang:2022zsl,Hanany:2022itc,Gu:2022dac,Fazzi:2022hal,Bourget:2022tmw,Gledhill:2022hrz,Fazzi:2022yca,Bhardwaj:2023zix,Bourget:2023uhe,Bourget:2023cgs,DelZotto:2023myd,DelZotto:2023nrb,Hanany:2023tvn,Hanany:2023uzn,Lawrie:2023uiu,Bourget:2023dkj,Benvenuti:2023qtv,Mansi:2023faa,Fazzi:2023ulb,Bourget:2024mgn,Lawrie:2024zon}. Many of these theories arise as worldvolume gauge theories of brane systems in Type IIA, Type IIB, M-theory, and F-theory. 

Despite the plethora of techniques on unitary magnetic quivers, there is a marked gap in the development of analogous techniques for orthosymplectic theories. So far, only the folding of orthosymplectic magnetic quivers \cite{Bourget:2021xex} has been shown systematically and only in very specific cases is the action of discrete gauging understood. Although it is known how to realise the small $E_8$ instanton transition \cite{Hanany:2018uhm}, this is only within the context of magnetic quivers for six dimensional theories.

In this work, inspired once again by $6d\;\mathcal N=(1,0)$ gauge theories, the technique of \textit{orthosymplectic quotient quiver subtraction} is developed for gauging $\surm(2),\;\surm(3),\;G_2,$ and $\sorm(7)$ subgroups of the Coulomb branch global symmetry of orthosymplectic $3d\;\mathcal N=4$ quivers (\textit{i.e.} containing only orthogonal and symplectic nodes). These are found by comparing the magnetic quiver for $\sprm(k)$ gauge theory at infinite coupling with $\sorm(4k+16)$ flavour symmetry\footnote{Except for the case $k=0$ where there is enhancement $\sorm(16)\rightarrow E_8$} before and after gauging $\surm(3),\;G_2,\;\sorm(7)$ subgroups of the flavour symmetry, for $k=0,1,2$ respectively. This is exactly the same methodology used to derive the quotient quiver subtraction in the unitary case for $\surm(3)$. The orthosymplectic quotient quiver subtraction for $\surm(2)$ is derived slightly differently but follows from the Higgsing pattern $\sorm(7)\rightarrow G_2\rightarrow \surm(3)\rightarrow \surm(2)$. The quivers studied here are restricted to unframed orthosymplectic quivers containing only $\sorm(2n)$ and $\sprm(n)$ gauge nodes.

Orthosymplectic quotient quiver subtraction shares much in common with its unitary counterpart. In particular, since the orthosymplectic quotient quivers for $\surm(2),\;\surm(3),$ $\;G_2,$ and $\sorm(7)$ contain gauge nodes with negative imbalance \cite{Gaiotto:2008ak}, their Coulomb branches are potentially smooth. Another similarity is the possibility of unions of moduli spaces arising from an `overshoot' at a `junction' of the starting quiver. There are also notable differences, which will be expanded upon in Section \ref{sec:Method}. Briefly, the prescription for rebalancing gauge nodes after subtraction differs between the orthosymplectic and unitary cases. This prescription may explicitly break the alternating $\sorm-\sprm$ gauge node pattern in some cases.

One successful outcome of unitary quotient quiver subtraction was in its use to derive new mathematical relationships between nilpotent orbit closures under hyper-Kähler quotients. Following this, orthosymplectic quotient quiver subtraction is also applied to $3d\;\mathcal N=4$ quivers whose Coulomb branches are nilpotent orbit closures. In doing so, previously established relationships found through unitary quotient quiver subtraction and through polymerisation are realised again using orthosymplectic quivers, providing a check of orthosymplectic quotient quiver subtraction. Additionally, new orthosymplectic counterparts to known unitary quivers are found. These quivers are counterparts in the sense that their Coulomb branches and Higgs branches are the same. New relationships between nilpotent orbit closures can be found using the orthosymplectic quotient quiver subtraction for which a unitary construction is not yet known.


Benefiting from the physical implications of this algorithm, orthosymplectic quotient quiver subtraction is also applied to understand certain $6d\;\mathcal N=(1,0)$ Higgs branches. Magnetic quivers were originally conceived as auxiliary descriptions of $6d\;\mathcal N=(1,0)$ Higgs branches of the worldvolume gauge theories arising from M5-branes on $AD$ Klein singularities \cite{Cabrera:2019izd,Cabrera:2019dob}, for which the dual Type IIA configuration is crucial. Extension to the case of $E_{6,7,8}$ Klein singularities has been a longstanding challenge, not least for the fact that there is no dual Type IIA description.
However, there do exist descriptions in F-theory \cite{Aspinwall:1998xj,DelZotto:2014hpa} from which an electric quiver may be found in both the finite and infinite coupling limits, although there is currently no systematic approach. There also exist descriptions of the $6d\;\mathcal N=(1,0)$ worldvolume gauge theory of one and two M5 branes on Klein $E_6$ singularity as a class $\mathcal S$ theory, which was derived from compactification on $T^2$ both without \cite{Ohmori:2015pua,Ohmori:2015pia} and with magnetic fluxes \cite{Ohmori:2018ona}.

Focusing on the case of a single M5 brane on an $E_6$ Klein singularity, orthosymplectic quotient quiver subtraction is used to provide an orthosymplectic magnetic quiver for two phases of the $6d\;\mathcal N=(1,0)$ gauge theory. The first phase is the $6d\;\mathcal N=(1,0)$ worldvolume theory of two $\frac{1}{2}$M5 branes on a Klein $E_6$ singularity. This theory is not conformal as there is a scale from the non-trivial vev of the tensor multiplet which is also the (finite) gauge coupling in this case. The orthosymplectic magnetic quiver is consistent with its unitary counterpart \cite{Hanany:2024fqf}, however only the orthosymplectic magnetic quiver admits a suitable description from a Type IIA brane system. The second phase is the infinite coupling limit of this $6d\;\mathcal N=(1,0)$ theory also known as minimal $(E_6,E_6)$ conformal matter\footnote{Hereafter dropping ``minimal".}. This is believed to be a strongly coupled SCFT. The $(E_6,E_6)$ conformal matter phase is related to the finite coupling phase by a small $E_8$ instanton transition \cite{Ganor:1996mu,Hanany:2018uhm}. The orthosymplectic magnetic quiver conjectured for $6d\;\mathcal N=(1,0)\;(E_6,E_6)$ conformal matter is the first of its kind and also admits a description from a Type IIA brane system.

\paragraph{Notation and Conventions}
\begin{itemize}
    \item All red nodes are $D_n$ nodes which are taken to be $\sorm(2n)$ rather than $\orm(2n)$
    \item All blue nodes are $C_n$ nodes which are taken to be $\sprm(n)$ with the convention $\sprm(1)\simeq \surm(2)$
    \item All Coulomb branch Hilbert series are computed as the sum of contributions from the integer and half-integer magnetic lattice \cite{Bourget:2020xdz}
    \item All perturbative Hilbert series and their $\pl$ are given to the order at which the first syzygy appears
    \item For some electric quivers, the self-intersection number of the F-theory curve corresponding to a particular gauge node are also given
    \item Although the analysis in this work is largely restricted to operations on quivers themselves, the few Type IIA brane systems that are used follow the conventions of \cite{Cabrera:2019dob, Cabrera:2019izd, Hanany:2022itc}
    \item The unrefined Hilbert series presented in this work display palindromic numerators; for particularly cumbersome examples, ellipses $\cdots$ will be used after the last unique coefficient, with the understanding that for a numerator of degree $4n$ the coefficient of the term at $t^{2n+2k}$ will be the same as that for $t^{2n-2k}$ for $1 \leq k \leq n$
    \item Nilpotent orbits of exceptional algebras will be labelled by their Characteristic \cite{Collingwood1993NilpotentAlgebras, Hanany:2017ooe} and Bala-Carter label
\end{itemize}
\section{Derivation of Orthosymplectic Quotient Quivers}
\label{sec:Derivation}
The starting point for the analysis is the following electric $6d\;\mathcal N=(1,0)$ gauge theory \begin{equation}
    \begin{tikzpicture}
        \node[gaugeb, label=below:$\sprm(k)$] (Ck)[]{};
        \node[flavourr, label=above:$\sorm(4k+16)$] (Dkp4) [above =of Ck]{};
        \draw[-] (Ck)--(Dkp4);
    \end{tikzpicture}
    \label{eq:CkDkp4E}
\end{equation}
at infinite coupling, whose corresponding magnetic quiver,
\begin{equation}
    \begin{tikzpicture}
        \node[gauger, label=below:$D_1$] (D1l)[]{};
        \node[gaugeb, label=below:$C_1$] (C1l) [right=of D1l]{};
        \node[] (cdotsl) [right=of C1l]{$\cdots$};
        \node[gaugeb, label=below:$C_{k+3}$] (Ckp3l) [right=of cdotsl]{};
        \node[gauger, label=below:$D_{k+4}$] (Dkp4) [right=of Ckp3l]{};
        \node[gaugeb, label=below:$C_{k+3}$] (Ckp3r) [right=of Dkp4]{};
        \node[] (cdotsr) [right=of Ckp3r]{$\cdots$};
        \node[gaugeb, label=below:$C_1$] (C1r) [right=of cdotsr]{};
        \node[gauger, label=below:$D_1$] (D1r)[right=of C1r]{};
        \node[gaugeb, label=above:$C_1$] (C1t) [above=of Dkp4]{};
        \draw[-] (D1l)--(C1l)--(cdotsl)--(Ckp3l)--(Dkp4)--(Ckp3r)--(cdotsr)--(C1r)--(D1r) (Dkp4)--(C1t);
    \end{tikzpicture}\label{eq:CkDkp4M}
\end{equation} gives a construction of the Higgs branch $\mathcal H_\infty(\eqref{eq:CkDkp4E})$ of \eqref{eq:CkDkp4E} as a moduli space of dressed monopole operators \cite{Cabrera:2019dob}.\footnote{Note that the $\sorm(4k+16)$ global symmetry of $\mathcal H_\infty(\ref{eq:CkDkp4E})$ undergoes enhancement to $E_8$ at $k=0$.} The Highest Weight Generating function (HWG) for the moduli space takes the following form \begin{equation}
    \hwg\left[\mathcal H_\infty(\eqref{eq:CkDkp4E})\right]=\pe\left[\sum_{i=1}^{k+2}\mu_{2i}t^{2i}+t^4+\mu_{2k+6}\left(t^{k+1}+t^{k+3}\right)\right],\label{eq:CkDkp4HWG}
\end{equation}where each of the $\mu_i$ are highest weight fugacities for $\sorm(2k+16)$.  


The objective is now to gauge a subgroup of the flavour symmetry of the electric theory \eqref{eq:CkDkp4E} and find its associated magnetic quiver. The magnetic quiver for the gauged theory can be compared to that of the pregauged theory \eqref{eq:CkDkp4M} in order to isolate the particular operation on the magnetic quiver corresponding to the gauging. As $6d\;\mathcal N=(1,0)$ theories are heavily constrained by anomalies there are typically very few permissible gaugings which in turn determines the action on the magnetic quiver precisely. 

In general, gauging a flavour symmetry subgroup $G \subset G_F$ geometrically corresponds to a $G$ hyper-Kähler quotient on the Higgs branch. If there is complete Higgsing of this symmetry, which is always the case in this work, this hyper-K\"ahler quotient is straightforwardly identified using Hilbert series methods. More generally, given a moduli space $\mathcal M$ with global symmetry $G_{\mathcal M}$ and refined Hilbert series $\hs\left[\mathcal M\right](t;u_1,\cdots,u_{\textrm{rank}(G_{\mathcal M})})$, where $t$ and $u_i$ are are fugacities for conformal dimension and global symmetries respectively, gauging a subgroup $G$, where $G\times G'\subset G_{\mathcal M}$ (with $G'$ being the commutant of $G$ inside $G_{\mathcal M}$),  corresponds to a Weyl integration on $\hs\left[\mathcal M\right]$ as shown in \eqref{eq:weyl}.

\begin{equation}
    \hs\left[\mathcal M///G\right](t;x_1,\cdots,x_{r})=\oint_{G}d\mu_G\frac{ \hs\left[\mathcal M\right](t;x_1,\cdots,x_r,y_1,\cdots,y_{\mathrm{rank}(G)})}{\pe[\chi_{\mathrm{Adj}}^G(y_1,\cdots,y_{\mathrm{rank}(G)})t^2]},
\label{eq:weyl}
\end{equation}
Note that $x_i$ are fugacities for the ungauged symmetry subgroup $G'$ of rank $r$, and the $y_i$ are fugacities for the gauged symmetry subgroup $G$.

The equation \eqref{eq:weyl} has a simple physical interpretation. The plethystic exponential in the denominator gives symmetrisations of the adjoint representation of $G$ graded by $t^2$, which have the interpretation of imposing additional F-terms that arise when $G$ is gauged. If $G$ is gauged with complete Higgsing (which is assumed and is the case throughout), the $F^{\flat}$-space is a complete intersection and so the (quaternionic) dimension of $\mathcal{M}$ reduces by $\mathrm{dim}\,G$. This change in dimension hence provides a simple check of complete Higgsing.

Anomaly cancellation conditions in six-dimensions leave only a small set of $\sorm(4k+16)$ flavour subgroups that admit a consistent gauging, each corresponding to a different value of $k$. For $k=0,1,2$ the possible gaugings are of $\surm(3)$, $G_2$, and $\sorm(7)$ respectively, with $\surm(2)$ a special case discussed in Section \ref{subsec:SU2}. 

Although gauging a flavour subgroup is relatively straightforward in the electric theory, the derivation of a corresponding magnetic quiver requires the introduction of a Type IIA brane system. $6d\;\mathcal N=(1,0)$ theories arise as worldvolume theories of D6-branes suspended between NS5-branes on an orientifold plane with flavour D8-branes. In fact, all such gauge theories which admit a formal description in Type IIA string theory -- including systems with negatively charged branes -- were classified in \cite{Mekareeya:2016yal}, and explicit brane constructions following this were given in \cite{Hanany:2022itc} with exploration of finite and infinite coupling phases. The magnetic phase of such brane systems involves suspending the D6-branes between D8-branes through a series of brane moves \cite{Hanany:1996ie,Feng:2000eq}. 
Then to find the magnetic quiver of the electric theory at infinite coupling, one simply merges $\frac{1}{2}$NS5-branes pairwise and moves them off the orientifold plane.

In \cite{Hanany:2022itc}, a set of Type IIA brane configurations involving negatively-charged D6-branes were studied alongside their corresponding electric and magnetic theories \cite{Mekareeya:2016yal,Hanany:1996ie,Feng:2000eq} -- the following section will derive from these the quotient quivers associated to the gauging of flavour subgroups. 

These six dimensional theories also have a construction in F-theory. However, it is not known to date how to extract the magnetic quiver from this. It is useful to make connection to the F-theory construction due to the specific effects on the Higgs branch that are seen from collapsing various curves of self-intersection.



\subsection{Gauging $\surm(3)$}
\label{subsec:SU3}
Consider first the electric theory \eqref{eq:CkDkp4E} with $k=0$. Although in other dimensions an $\sprm(0)$ gauge theory is trivial, this is different in six dimensions due to the presence of a tensor multiplet which has dynamics. The case $k=0$ is special since the expected $\sorm(16)$ flavour symmetry of \eqref{eq:CkDkp4E} is enhanced $\sorm(16)\rightarrow E_8$ because $\sprm(0)$ is at infinite coupling, however the HWG \eqref{eq:CkDkp4HWG} still holds with the appropriate embedding. Gauging an $\surm(3)$ subgroup of the $E_8$ flavour symmetry gives \eqref{eq:SU3Sp0E6E}, where the $\sprm(0)$ remains at infinite coupling and the $\surm(3)$ is at finite coupling.
\begin{equation}
\begin{tikzpicture}
    \node[gauge] (SU3) []{};
    \node[align=center,anchor=north] (SU3lab) at (SU3.south) {$\surm(3)$};
    \node[gauge] (Sp0) [right=of SU3]{};
    \node[align=center,anchor=north] (Sp0lab) at (Sp0.south) {$\sprm(0)$};
    \node[flavour, label=right:$E_6$] (O10) [above=of Sp0]{};
    \draw[-] (SU3)--(Sp0)--(O10);
\end{tikzpicture}\label{eq:SU3Sp0E6E}
\end{equation}
To derive the magnetic quiver, first consider an alternative presentation of \eqref{eq:SU3Sp0E6E} from F-theory which is amenable to a description in Type IIA. Start with the electric $\surm(3)\times\sprm(0)$ anomaly-free theory with both gauge groups at finite coupling, given in \eqref{eq:SU3Sp0O10}. 
\begin{equation}
    \begin{tikzpicture}
        \node[gauge] (SU3) []{};
        \node[align=center,anchor=north] (SU3lab) at (SU3.south) {$\surm(3)$\\$-3$};
        \node[gauge] (Sp0) [right=of SU3]{};
        \node[align=center,anchor=north] (Sp0lab) at (Sp0.south) {$\sprm(0)$\\$-1$};
        \node[flavour, label=right:$\sorm(10)\times\urm(1)$] (O10) [above=of Sp0]{};
        \draw[-] (SU3)--(Sp0)--(O10);
    \end{tikzpicture}\label{eq:SU3Sp0O10}
\end{equation}
The $\surm(3)$ gauge group is supported on a $(-3)$-curve and the $\sprm(0)$ on a $(-1)$-curve. Collapsing the $(-1)$-curve corresponds to a small $E_8$-instanton transition and tunes the $\sprm(0)$ gauge coupling to infinity, leaving the rank-1 E-string coupled to $\surm(3)$. 

This change in the Higgs branch can also be seen through brane systems in Type IIA and from magnetic quivers. The quiver \eqref{eq:SU3Sp0O10} at finite coupling is the D6 brane worldvolume theory of the following Type IIA brane system \begin{equation}
    \begin{tikzpicture}
        \def\x{1cm};
        \draw (1,-\x)--(1,\x);
        \node[label=above:$2$] at (1,\x) {};
        \draw (5,-\x)--(5,\x);
        \node[label=above:$10$] at (5,\x) {};
        \ns{0,0};
        \ns{2,0};
        \ns{4,0};
        \ns{6,0};
        \draw[-,red] (0,0.2)--(2,0.2);
        \node[label=above:$-2$] at (1.5,0.2){};
        \draw[-,red] (0,-0.2)--(2,-0.2);
        \draw[dotted] (0.2,0)--(1.8,0);

        \draw[-] (2,0.2)--(4,0.2);
        \node[label=above:$3$] at (3,0.2){};
        \draw[-] (2,-0.2)--(4,-0.2);

        \draw[dotted] (4,0)--(6,0);
    \end{tikzpicture}
\end{equation}
The collapse of the $(-1)$-curve, which tunes the gauge coupling of $\sprm(0)$ to infinity is realised in the magnetic phase of the brane system by firstly suspending D6 between D8 branes and then bringing all four $\frac{1}{2}$-NS5 branes together pairwise and lifting them off the orientifold plane. Doing this results in the following brane system 
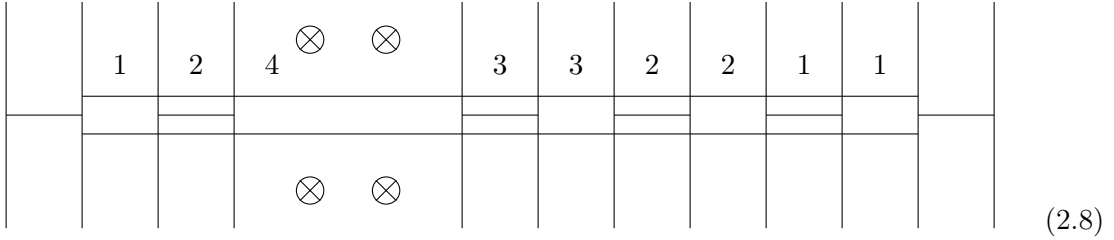
\begin{equation}
    \begin{tikzpicture}
         \def\x{1.5cm};
         \draw[-] (0,-\x)--(0,\x);
         \draw[-] (1,-\x)--(1,\x);
         \draw[-] (2,-\x)--(2,\x);
         \draw[-] (3,-\x)--(3,\x);
         \draw[-] (6,-\x)--(6,\x);
         \draw[-] (7,-\x)--(7,\x);
         \draw[-] (8,-\x)--(8,\x);
         \draw[-] (9,-\x)--(9,\x);
         \draw[-] (10,-\x)--(10,\x);
         \draw[-] (11,-\x)--(11,\x);
         \draw[-] (12,-\x)--(12,\x);
         \draw[-] (13,-\x)--(13,\x);
         \ns{4,1};
         \ns{5,1};
         \ns{4,-1};
         \ns{5,-1};
         \draw[-] (0,0)--(1,0) (2,0)--(3,0) (6,0)--(7,0) (8,0)--(9,0) (10,0)--(11,0) (12,0)--(13,0);

         \draw[-] (1,0.25)--(12,0.25);
         \draw[-] (1,-0.25)--(12,-0.25);

         \node[label=above:$1$] at (1.5,0.2){};
         \node[label=above:$2$] at (2.5,0.2){};
         \node[label=above:$4$] at (3.5,0.2){};
         \node[label=above:$3$] at (6.5,0.2){};
         \node[label=above:$3$] at (7.5,0.2){};
         \node[label=above:$2$] at (8.5,0.2){};
         \node[label=above:$2$] at (9.5,0.2){};
         \node[label=above:$1$] at (10.5,0.2){};
         \node[label=above:$1$] at (11.5,0.2){};

    \end{tikzpicture}
\end{equation}from which one is able to read off the following magnetic quiver for the Higgs branch of \eqref{eq:SU3Sp0E6E} 
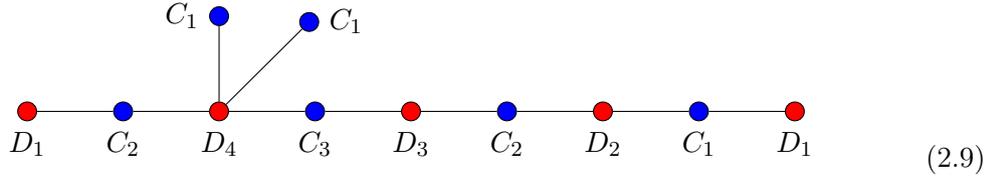
\begin{equation}
    \begin{tikzpicture}
             \node[gaugeb, label=left:$C_1$] (C1t) []{};
             \node[gauger, label=below:$D_4$] (D4) [below=of C1t]{};
            \node[gaugeb, label=below:$C_2$] (C2l) [left=of D4]{};
            \node[gauger, label=below:$D_1$] (D1l) [left=of C2l]{};
            \node[gaugeb, label=below:$C_3$] (C3r) [right=of D4]{};
           \node[gauger, label=below:$D_3$] (D3r) [right=of C3r]{};
           \node[gaugeb, label=below:$C_2$] (C2r) [right=of D3r]{};
           \node[gauger, label=below:$D_2$] (D2r) [right=of C2r]{};
           \node[gaugeb, label=below:$C_1$] (C1r) [right=of D2r]{};
           \node[gauger, label=below:$D_1$] (D1r) [right=of C1r]{};
            \node[gaugeb, label=right:$C_1$] (C1tr) [above right=of D4]{};

    \draw[-] (C1tr)--(D4)--(C1t);
    \draw[-] (D1r)--(C1r)--(D2r)--(C2r)--(D3r)--(C3r)--(D4)--(C2l)--(D1l);
        \end{tikzpicture}\label{eq:SU3Sp0E6M}
\end{equation} as was done in \cite{Hanany:2022itc}. The bouquet of two $C_1$ gauge nodes attached to the $D_4$ gauge node is indicative of $\surm(3)$ being finitely coupled.

As the $6d$ theory \eqref{eq:SU3Sp0E6E} comes from gauging an $\surm(3)$ subgroup of the flavour symmetry of \eqref{eq:CkDkp4E} (for $k=0$), comparing their magnetic quivers \eqref{eq:SU3Sp0E6M} and \eqref{eq:CkDkp4M} will realise gauging $\surm(3)$ subgroups of Coulomb branch global symmetries.

The key observation that translates the notion of gauging flavour symmetries in six-dimensions to the magnetic quiver is a simple and combinatorial one. The two magnetic quivers are related by the following \textit{quiver subtraction}
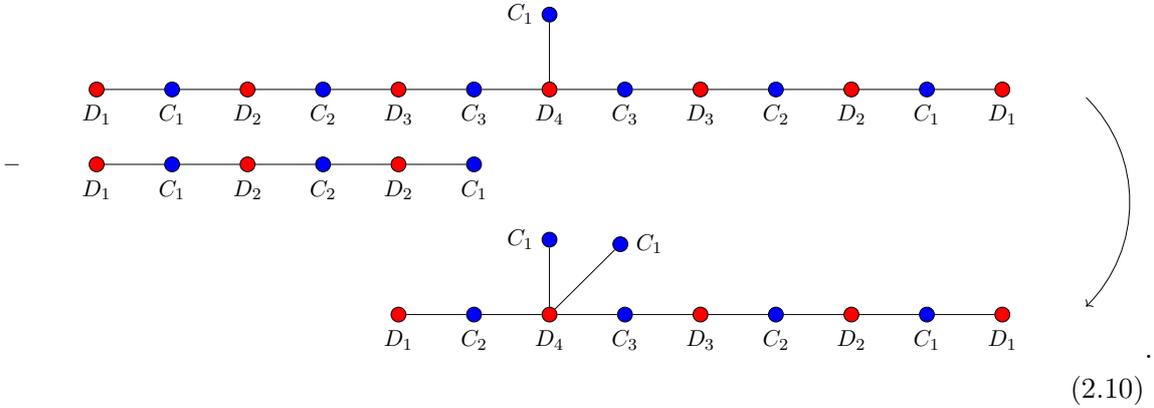
\begin{equation}
    \resizebox{\linewidth}{!}{\begin{tikzpicture}
        \node[gauger, label=below:$D_1$] (D1l)[]{};
        \node[gaugeb, label=below:$C_1$] (C1l) [right=of D1l]{};
        \node[gauger, label=below:$D_2$] (D2l) [right=of C1l]{};
        \node[gaugeb, label=below:$C_2$] (C2l) [right=of D2l]{};
        \node[gauger, label=below:$D_3$] (D3l) [right=of C2l]{};
        \node[gaugeb, label=below:$C_3$] (C3l) [right=of D3l]{};
        \node[gauger, label=below:$D_4$] (D4) [right=of C3l]{};
        \node[gaugeb, label=below:$C_3$] (C3r) [right=of D4]{};
        \node[gauger, label=below:$D_3$] (D3r) [right=of C3r]{};
        \node[gaugeb, label=below:$C_2$] (C2r) [right=of D3r]{};
        \node[gauger, label=below:$D_2$] (D2r) [right=of C2r]{};
        \node[gaugeb, label=below:$C_1$] (C1r) [right=of D2r]{};
        \node[gauger, label=below:$D_1$] (D1r)[right=of C1r]{};
        \node[gaugeb, label=left:$C_1$] (C1t) [above=of D4]{};

        \draw[-] (D1l)--(C1l)--(D2l)--(C2l)--(D3l)--(C3l)--(D4)--(C3r)--(D3r)--(C2r)--(D2r)--(C1r)--(D1r) (D4)--(C1t);

        \node[gaugeb, label=below:$C_1$] (c1rs) [below=of C3l]{};
        \node[gauger, label=below:$D_2$] (d2rs) [left=of c1rs]{};
        \node[gaugeb, label=below:$C_2$] (c2s) [left=of d2rs]{};
        \node[gauger, label=below:$D_2$] (d2s) [left=of c2s]{};
        \node[gaugeb, label=below:$C_1$] (c1s) [left=of d2s]{};
        \node[gauger, label=below:$D_1$] (d1s) [left=of c1s]{};
        
        \draw[-] (c1rs)--(d2rs)--(c2s)--(d2s)--(c1s)--(d1s);
        
        \node[] (minus) [left=of d1s]{$-$};

         \node[] (ghost) [below=of D4]{};
         \node[gaugeb, label=left:$C_1$] (c1t) [below=of ghost]{};
         \node[gauger, label=below:$D_4$] (d4) [below=of c1t]{};
        \node[gaugeb, label=below:$C_2$] (c2l) [left=of d4]{};
        \node[gauger, label=below:$D_1$] (d1l) [left=of c2l]{};
        \node[gaugeb, label=below:$C_3$] (c3r) [right=of d4]{};
       \node[gauger, label=below:$D_3$] (d3r) [right=of c3r]{};
       \node[gaugeb, label=below:$C_2$] (c2r) [right=of d3r]{};
       \node[gauger, label=below:$D_2$] (d2r) [right=of c2r]{};
       \node[gaugeb, label=below:$C_1$] (c1r) [right=of d2r]{};
       \node[gauger, label=below:$D_1$] (d1r) [right=of c1r]{};
        \node[gaugeb, label=right:$C_1$] (c1tr) [above right=of d4]{};

        \draw[-] (c1tr)--(d4)--(c1t);
        \draw[-] (d1r)--(c1r)--(d2r)--(c2r)--(d3r)--(c3r)--(d4)--(c2l)--(d1l);

        \node[] (topghost) [right=of D1r] {};
        \node[] (bottomghost) [right=of d1r]{};

        \draw[->] (topghost)to [out=-45,in=45,looseness=1](bottomghost);
    
    \end{tikzpicture}}
    \label{eq:minE8SU3Quot}.
\end{equation}

The following quiver 
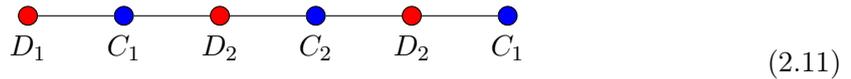
\begin{equation}
    \begin{tikzpicture}
        \node[gaugeb, label=below:$C_1$] (c1rs) []{};
        \node[gauger, label=below:$D_2$] (d2rs) [left=of c1rs]{};
        \node[gaugeb, label=below:$C_2$] (c2s) [left=of d2rs]{};
        \node[gauger, label=below:$D_2$] (d2s) [left=of c2s]{};
        \node[gaugeb, label=below:$C_1$] (c1s) [left=of d2s]{};
        \node[gauger, label=below:$D_1$] (d1s) [left=of c1s]{};
        
        \draw[-] (c1rs)--(d2rs)--(c2s)--(d2s)--(c1s)--(d1s);
    \end{tikzpicture}
\end{equation}is hence proposed as the $\surm(3)$ orthosymplectic quotient quiver. In particular one should note that the $C_2$ gauge node and the $C_1$ gauge node on the right have negative balance \cite{Gaiotto:2008ak} and hence the Coulomb branch is smooth and its Hilbert series is not computable with the monopole formula \cite{Cremonesi:2013lqa}. However, the unitary $\surm(3)$ quotient quiver has a Coulomb branch which is $T^*\left(\mathrm{SL}(3)\right)$. It is reasonable to conjecture that the Coulomb branch of the orthosymplectic $\surm(3)$ quotient quiver is $T^*\left(\mathrm{GL}(3)\right)$ which is of dimension 9.

A full explanation of the orthosymplectic quotient quiver subtraction algorithm will be done in Section \ref{sec:Method}.
\subsection{Gauging $G_2$}
\label{subsec:G2}
 A similar set of considerations apply to the $G_2$ case. In order to gauge a $G_2$ subgroup of the flavour symmetry of \eqref{eq:CkDkp4E}, the choice $k=1$ must be made to cancel anomalies. The resulting gauge theory is \begin{equation}
     \begin{tikzpicture}
         \node[gauge] (SU3) []{};
        \node[align=center,anchor=north] (SU3lab) at (SU3.south) {$G_2$};
        \node[gauge] (Sp0) [right=of SU3]{};
        \node[align=center,anchor=north] (Sp0lab) at (Sp0.south) {$\sprm(1)$};
        \node[flavour, label=right:$\sorm(13)$] (O10) [above=of Sp0]{};
        \draw[-] (SU3)--(Sp0)--(O10);
    \end{tikzpicture}\label{eq:G2Sp1SO13E}
\end{equation}where again the $\sprm(1)$ is tuned to infinite coupling and the $G_2$ is at finite coupling.

As before, the magnetic quiver may be found from a slightly different derivation of \eqref{eq:G2Sp1SO13E}. Take the non-anomalous $G_2\times\sprm(1)$ gauge theory with both gauge groups at finite coupling. The quiver is \begin{equation}
    \begin{tikzpicture}
         \node[gauge] (SU3) []{};
        \node[align=center,anchor=north] (SU3lab) at (SU3.south) {$G_2$\\$-3$};
        \node[gauge] (Sp0) [right=of SU3]{};
        \node[align=center,anchor=north] (Sp0lab) at (Sp0.south) {$\sprm(1)$\\$-1$};
        \node[flavour, label=right:$\sorm(13)$] (O10) [above=of Sp0]{};
        \draw[-] (SU3)--(Sp0)--(O10);
    \end{tikzpicture}\label{eq:G2Sp1SO13EFinite}
\end{equation}which is similar to \eqref{eq:G2Sp1SO13E} except that in F-theory the $\sprm(1)$ is supported on a $(-1)$-curve and the $G_2$ is supported on a $(-3)$-curve. There are two tensor multiplets associated to each gauge group. The $(-1)$-curve may be collapsed which tunes the gauge coupling of $\sprm(1)$ to infinity. There is also a small $E_8$ instanton transition as the tensor multiplet associated to the $\sprm(1)$ is lost. The electric theory is again \eqref{eq:G2Sp1SO13E}. Compared to the case of gauging $\surm(3)$, which involved an $\sprm(0)$ gauge node, there is non-trivial matter which is coupled. 

The Type IIA brane system which gives \eqref{eq:G2Sp1SO13EFinite} as the D6 brane worldvolume theory is \begin{equation}
    \begin{tikzpicture}
        \def\x{1cm};
        \draw (1,-\x)--(1,\x);
        \node[label=above:$1$] at (1,\x) {};
        \draw (5,-\x)--(5,\x);
        \node[label=above:$13$] at (5,\x) {};
        \ns{0,0};
        \ns{2,0};
        \ns{4,0};
        \ns{6,0};
        \draw[-,red] (0,0.2)--(2,0.2);
        \node[label=above:$2$] at (1.5,0.2){};
        \draw[-,red] (0,-0.2)--(2,-0.2);
        \draw[dotted] (0.2,0)--(1,0);
        \draw[dashed] (1,0)--(1.8,0);
        \draw[-] (2.2,0)--(3.8,0);
        \draw[dashed] (4.2,0)--(5,0);
        \draw[dotted] (5,0)--(5.8,0);
        \node[label=above:$1$] at (4.5,0.2){};

        \draw[-] (2,0.2)--(6,0.2);
        \node[label=above:$3$] at (3,0.2){};
        \draw[-] (2,-0.2)--(6,-0.2);
        
    \end{tikzpicture}
\end{equation}
Then the collapse of the $(-1)$-curve, which tunes the coupling of $\sprm(1)$ to infinity, is seen by moving to the magnetic phase by first suspending D6 branes between D8 and then merging all four $\frac{1}{2}$-NS5 branes pairwise and lifting them off the orientifold plane. The resulting brane system is
\begin{equation}
    \begin{tikzpicture}
        \def\x{1.5cm};
         \draw[-] (0,-\x)--(0,\x);
         \draw[-] (1,-\x)--(1,\x);
         \draw[-] (2,-\x)--(2,\x);
         \draw[-] (3,-\x)--(3,\x);
         \draw[-] (6,-\x)--(6,\x);
         \draw[-] (7,-\x)--(7,\x);
         \draw[-] (8,-\x)--(8,\x);
         \draw[-] (9,-\x)--(9,\x);
         \draw[-] (10,-\x)--(10,\x);
         \draw[-] (11,-\x)--(11,\x);
         \draw[-] (12,-\x)--(12,\x);
         \draw[-] (13,-\x)--(13,\x);
         \draw[-] (14,-\x)--(14,\x);
         \draw[-] (15,-\x)--(15,\x);
         
         \ns{4,1};
         \ns{5,1};
         \ns{4,-1};
         \ns{5,-1};
         \draw[-] (0,0)--(1,0) (2,0)--(3,0) (6,0)--(7,0) (8,0)--(9,0) (10,0)--(11,0) (12,0)--(13,0) (14,0)--(15,0);

         \draw[-] (1,0.25)--(14,0.25);
         \draw[-] (1,-0.25)--(14,-0.25);

        \node[label=above:$2$] at (1.5,0.2){};
         \node[label=above:$3$] at (2.5,0.2){};
         \node[label=above:$5$] at (3.5,0.2){};
         \node[label=above:$4$] at (6.5,0.2){};
         \node[label=above:$4$] at (7.5,0.2){};
         \node[label=above:$3$] at (8.5,0.2){};
         \node[label=above:$3$] at (9.5,0.2){};
         \node[label=above:$2$] at (10.5,0.2){};
         \node[label=above:$2$] at (11.5,0.2){};
         \node[label=above:$1$] at (12.5,0.2){};
         \node[label=above:$1$] at (13.5,0.2){};   
    \end{tikzpicture}
\end{equation}from which one is able to read off the following magnetic quiver \begin{equation}
    \begin{tikzpicture}
        \node[gauger, label=below:$D_2$] (D2l) []{};
        \node[gaugeb, label=below:$C_3$] (C3l) [right=of D2l]{};
        \node[gauger, label=below:$D_5$] (D5) [right=of C3l]{};
        \node[gaugeb, label=below:$C_4$] (C4r) [right=of D5]{};
        \node[gauger, label=below:$D_4$] (D4r) [right=of C4r]{};
        \node[gaugeb, label=below:$C_3$] (C3r) [right=of D4r]{};
        \node[gauger, label=below:$D_3$] (D3r) [right=of C3r]{};
        \node[gaugeb, label=below:$C_2$] (C2r) [right=of D3r]{};
        \node[gauger, label=below:$D_2$] (D2r) [right=of C2r]{};
        \node[gaugeb, label=below:$C_1$] (C1r) [right=of D2r]{};
        \node[gauger, label=below:$D_1$] (D1r) [right=of C1r]{};
        \node[gaugeb, label=left:$C_1$] (C1t) [above=of D5]{};
        \node[gaugeb, label=right:$C_1$] (C1tr) [above right=of D5]{};

        \draw[-] (D2l)--(C3l)--(D5)--(C4r)--(D4r)--(C3r)--(D3r)--(C2r)--(D2r)--(C1r)--(D1r) (C1t)--(D5)--(C1tr);
    \end{tikzpicture}\label{eq:G2Sp1SO13M}
\end{equation}for the Higgs branch of \eqref{eq:G2Sp1SO13E}. The bouquet of $C_1$ gauge nodes attached to the $D_5$ gauge node is indicative that the $G_2$ gauge node in \eqref{eq:G2Sp1SO13E} is at finite coupling.

As the $6d$ theory \eqref{eq:G2Sp1SO13E} comes from gauging a $G_2$ subgroup of the flavour symmetry of \eqref{eq:CkDkp4E} (for $k=1$), comparing their magnetic quivers \eqref{eq:G2Sp1SO13M} and \eqref{eq:CkDkp4M} will realise gauging $G_2$ subgroups of Coulomb branch global symmetries.

The two magnetic quivers are related by the following quiver subtraction \begin{equation}
    \resizebox{\linewidth}{!}{\begin{tikzpicture}
         \node[gauger, label=below:$D_1$] (d1l) []{};
    \node[gaugeb, label=below:$C_1$] (c1l) [right=of d1l]{};
    \node[gauger, label=below:$D_2$] (d2l) [right=of c1l]{};
    \node[gaugeb, label=below:$C_2$] (c2l) [right=of d2l]{};    \node[gauger, label=below:$D_3$] (d3l) [right=of c2l]{};
    \node[gaugeb, label=below:$C_3$] (c3l) [right=of d3l]{};
    \node[gauger, label=below:$D_4$] (d4l) [right=of c3l]{};
    \node[gaugeb, label=below:$C_4$] (c4l) [right=of d4l]{};
    \node[gauger, label=below:$D_5$] (d5) [right=of c4l]{};
    \node[gaugeb, label=below:$C_4$] (c4r) [right=of d5]{};
    \node[gauger, label=below:$D_4$] (d4r) [right=of c4r]{};
    \node[gaugeb, label=below:$C_3$] (c3r) [right=of d4r]{};
    \node[gauger, label=below:$D_3$] (d3r) [right=of c3r]{};
    \node[gaugeb, label=below:$C_2$] (c2r) [right=of d3r]{};
    \node[gauger, label=below:$D_2$] (d2r) [right=of c2r]{};
    \node[gaugeb, label=below:$C_1$] (c1r) [right=of d2r]{};
    \node[gauger, label=below:$D_1$] (d1r) [right=of c1r]{};
     \node[gaugeb, label=left:$C_1$] (c1t) [above=of d5]{};
     \draw[-] (d1l)--(c1l)--(d2l)--(c2l)--(d3l)--(c3l)--(d4l)--(c4l)--(d5)--(c4r)--(d4r)--(c3r)--(d3r)--(c2r)--(d2r)--(c1r)--(d1r);
     \draw[-] (c1t)--(d5);

        \node[gauger, label=below:$D_1$] (d1s) [below=of d1l]{};
        \node[gaugeb, label=below:$C_1$] (c1ls) [right=of d1s]{};
        \node[gauger, label=below:$D_2$] (d2ls) [right=of c1ls]{};
        \node[gaugeb, label=below:$C_2$] (c2s) [right=of d2ls]{};
        \node[gauger, label=below:$D_3$] (d3s) [right=of c2s]{};
        \node[gaugeb, label=below:$C_3$] (c3s) [right=of d3s]{};
        \node[gauger, label=below:$D_2$] (d2rs) [right=of c3s]{};
        \node[gaugeb, label=below:$C_1$] (c1rs) [right=of d2rs]{};

        \draw[-] (d1s)--(c1ls)--(d2ls)--(c2s)--(d3s)--(c3s)--(d2rs)--(c1rs);
        
        \node[] (minus) [left=of d1s]{$-$};

         \node[] (ghost) [below=of d2rs]{};
          \node[gauger, label=below:$D_2$] (D2l) [below=of ghost]{};
        \node[gaugeb, label=below:$C_3$] (C3l) [right=of D2l]{};
        \node[gauger, label=below:$D_5$] (D5) [right=of C3l]{};
        \node[gaugeb, label=below:$C_4$] (C4r) [right=of D5]{};
        \node[gauger, label=below:$D_4$] (D4r) [right=of C4r]{};
        \node[gaugeb, label=below:$C_3$] (C3r) [right=of D4r]{};
        \node[gauger, label=below:$D_3$] (D3r) [right=of C3r]{};
        \node[gaugeb, label=below:$C_2$] (C2r) [right=of D3r]{};
        \node[gauger, label=below:$D_2$] (D2r) [right=of C2r]{};
        \node[gaugeb, label=below:$C_1$] (C1r) [right=of D2r]{};
        \node[gauger, label=below:$D_1$] (D1r) [right=of C1r]{};
        \node[gaugeb, label=left:$C_1$] (C1t) [above=of D5]{};
        \node[gaugeb, label=right:$C_1$] (C1tr) [above right=of D5]{};

        \draw[-] (D2l)--(C3l)--(D5)--(C4r)--(D4r)--(C3r)--(D3r)--(C2r)--(D2r)--(C1r)--(D1r) (C1t)--(D5)--(C1tr);

        \node (topghost) [right=of d1r]{};
        \node (bottomghost) [right=of D1r]{};

        \draw[->] (topghost)to [out=-45,in=45,looseness=1](bottomghost);
    \end{tikzpicture}}
    \label{fig:G2Sp1SO13QuivSubtraction}
\end{equation}

The following quiver \begin{equation}
    \begin{tikzpicture}
        \node[gauger, label=below:$D_1$] (d1) []{};
        \node[gaugeb, label=below:$C_1$] (c1l) [right=of d1]{};
        \node[gauger, label=below:$D_2$] (d2l) [right=of c1l]{};
        \node[gaugeb, label=below:$C_2$] (c2) [right=of d2l]{};
        \node[gauger, label=below:$D_3$] (d3) [right=of c2]{};
        \node[gaugeb, label=below:$C_3$] (c3) [right=of d3]{};
        \node[gauger, label=below:$D_2$] (d2r) [right=of c3]{};
        \node[gaugeb, label=below:$C_1$] (c1r) [right=of d2r]{};

        \draw[-] (d1)--(c1l)--(d2l)--(c2)--(d3)--(c3)--(d2r)--(c1r);
    \end{tikzpicture}
\end{equation}is hence proposed as the $G_2$ orthosymplectic quotient quiver. This quiver also contains gauge nodes with negative balance \cite{Gaiotto:2008ak}, the $C_3$ gauge node and the $C_1$ gauge node on the right. The Coulomb branch is smooth and its Hilbert series is not computable with the monopole formula \cite{Cremonesi:2013lqa} and remains a challenge.

A full explanation of the orthosymplectic quotient quiver subtraction algorithm will be done in Section \ref{sec:Method}.

\subsection{Gauging $\sorm(7)$}
\label{subsec:SO7}
 A similar set of considerations apply once again to the $\sorm(7)$ case. In order to gauge a $\sorm(7)$ subgroup of the flavour symmetry of \eqref{eq:CkDkp4E}, the choice $k=2$ must be made to cancel anomalies. The resulting gauge theory is \begin{equation}
     \begin{tikzpicture}
         \node[gauge] (SU3) []{};
        \node[align=center,anchor=north] (SU3lab) at (SU3.south) {$\sorm(7)$};
        \node[gauge] (Sp0) [right=of SU3]{};
        \node[align=center,anchor=north] (Sp0lab) at (Sp0.south) {$\sprm(2)$};
        \node[flavour, label=right:$\sorm(16)$] (O10) [above=of Sp0]{};
        \draw[-] (Sp0)--(O10);
        \draw[-,red] (Sp0)--(SU3);
    \end{tikzpicture}\label{eq:SO7Sp2SO16E},
\end{equation}where the $\sprm(2)$ is tuned to infinite coupling and the $\sorm(7)$ is at finite coupling. The red edge indicates that the hypermultiplet transforms in the bispinor representation when $\sprm(2)$ is treated as $\mathrm{Spin}(5)$.

As before, the magnetic quiver may be found from a slightly different derivation of \eqref{eq:SO7Sp2SO16E}. Take the non-anomalous $\sorm(7)\times\sprm(2)$ gauge theory with both gauge groups at finite coupling. The quiver is \begin{equation}
    \begin{tikzpicture}
         \node[gauge] (SU3) []{};
        \node[align=center,anchor=north] (SU3lab) at (SU3.south) {$\sorm(7)$\\$-3$};
        \node[gauge] (Sp0) [right=of SU3]{};
        \node[align=center,anchor=north] (Sp0lab) at (Sp0.south) {$\sprm(2)$\\$-1$};
        \node[flavour, label=right:$\sorm(16)$] (O10) [above=of Sp0]{};
        \draw[-] (Sp0)--(O10);
        \draw[-,red] (Sp0)--(SU3);
    \end{tikzpicture}\label{eq:SO7Sp2SO16EFinite}
\end{equation}which is similar to \eqref{eq:SO7Sp2SO16E} except that in F-theory the $\sprm(2)$ is supported on a $(-1)$-curve and the $\sorm(7)$ is supported on a $(-3)$-curve. There are two tensor multiplets associated to each gauge group. The $(-1)$-curve may be collapsed which tunes the gauge coupling of $\sprm(2)$ to infinity. There is also a small $E_8$ instanton transition as the tensor multiplet associated to the $\sprm(2)$ is lost. The electric theory is again \eqref{eq:SO7Sp2SO16E}.

The Type IIA brane system which gives \eqref{eq:SO7Sp2SO16EFinite} as the D6 brane worldvolume theory is
\begin{equation}
    \begin{tikzpicture}
         \def\x{1cm};
        \draw (5,-\x)--(5,\x);
        \node[label=above:$16$] at (5,\x) {};
        \ns{0,0};
        \ns{2,0};
        \ns{4,0};
        \ns{6,0};
        \draw[-,red] (0,0.2)--(2,0.2);
        \node[label=above:$2$] at (1,0.2){};
        \draw[-,red] (0,-0.2)--(2,-0.2);
        \draw[dotted] (0.2,0)--(1.8,0);
        \draw[dotted] (4.2,0)--(5.8,0);
        \node[label=above:$4$] at (3,0.2){};
        \node[label=above:$2$] at (5.5,0.2){};

        \draw[-] (2,0.2)--(6,0.2);
        \draw[-] (2,-0.2)--(6,-0.2);
        
    \end{tikzpicture}
\end{equation}
The collapse of the $(-1)$-curve which tunes the coupling of $\sprm(2)$ to infinity is seen in the brane system by moving to the magnetic phase. This is done by first suspending D6 branes between D8 and then merging all four $\frac{1}{2}$-NS5 branes pairwise and lifting them off the orientifold plane. The resulting brane system is
\begin{equation}
    \resizebox{\linewidth}{!}{\begin{tikzpicture}
        \def\x{1.5cm};
         \draw[-] (0,-\x)--(0,\x);
         \draw[-] (1,-\x)--(1,\x);
         \draw[-] (2,-\x)--(2,\x);
         \draw[-] (3,-\x)--(3,\x);
         \draw[-] (6,-\x)--(6,\x);
         \draw[-] (7,-\x)--(7,\x);
         \draw[-] (8,-\x)--(8,\x);
         \draw[-] (9,-\x)--(9,\x);
         \draw[-] (10,-\x)--(10,\x);
         \draw[-] (11,-\x)--(11,\x);
         \draw[-] (12,-\x)--(12,\x);
         \draw[-] (13,-\x)--(13,\x);
         \draw[-] (14,-\x)--(14,\x);
         \draw[-] (15,-\x)--(15,\x);
         \draw[-] (16,-\x)--(16,\x);
         \draw[-] (17,-\x)--(17,\x);
         
         \ns{4,1};
         \ns{5,1};
         \ns{4,-1};
         \ns{5,-1};
         \draw[-] (0,0)--(1,0) (2,0)--(3,0) (6,0)--(7,0) (8,0)--(9,0) (10,0)--(11,0) (12,0)--(13,0) (14,0)--(15,0) (16,0)--(17,0);

         \draw[-] (1,0.25)--(16,0.25);
         \draw[-] (1,-0.25)--(16,-0.25);

        \node[label=above:$1$] at (0.5,0.2){};
        \node[label=above:$3$] at (1.5,0.2){};
         \node[label=above:$4$] at (2.5,0.2){};
         \node[label=above:$6$] at (3.5,0.2){};
         \node[label=above:$5$] at (6.5,0.2){};
         \node[label=above:$5$] at (7.5,0.2){};
         \node[label=above:$4$] at (8.5,0.2){};
         \node[label=above:$4$] at (9.5,0.2){};
         \node[label=above:$3$] at (10.5,0.2){};
         \node[label=above:$3$] at (11.5,0.2){};
         \node[label=above:$2$] at (12.5,0.2){};
         \node[label=above:$2$] at (13.5,0.2){};   
         \node[label=above:$1$] at (14.5,0.2){};
         \node[label=above:$1$] at (15.5,0.2){};   
    \end{tikzpicture}}
\end{equation}from which one is able to read off the following magnetic quiver \begin{equation}
    \resizebox{\linewidth}{!}{\begin{tikzpicture}
        \node[gaugeb, label=below:$C_1$] (C1l) []{};
        \node[gauger, label=below:$D_3$] (D3l) [right=of C1l]{};
        \node[gaugeb, label=below:$C_4$] (C4l) [right=of D3l]{};
        \node[gauger, label=below:$D_6$] (D6) [right=of C4l]{};
        \node[gaugeb, label=below:$C_5$] (C5r) [right=of D6]{};
        \node[gauger, label=below:$D_5$] (D5r) [right=of C5r]{};
        \node[gaugeb, label=below:$C_4$] (C4r) [right=of D5r]{};
        \node[gauger, label=below:$D_4$] (D4r) [right=of C4r]{};
        \node[gaugeb, label=below:$C_3$] (C3r) [right=of D4r]{};
        \node[gauger, label=below:$D_3$] (D3r) [right=of C3r]{};
        \node[gaugeb, label=below:$C_2$] (C2r) [right=of D3r]{};
        \node[gauger, label=below:$D_2$] (D2r) [right=of C2r]{};
        \node[gaugeb, label=below:$C_1$] (C1r) [right=of D2r]{};
        \node[gauger, label=below:$D_1$] (D1r) [right=of C1r]{};
        \node[gaugeb, label=left:$C_1$] (C1t) [above=of D6]{};
        \node[gaugeb, label=right:$C_1$] (C1tr) [above right=of D6]{};

        \draw[-] (C1l)--(D3l)--(C4l)--(D6)--(C5r)--(D5r)--(C4r)--(D4r)--(C3r)--(D3r)--(C2r)--(D2r)--(C1r)--(D1r) (C1t)--(D6)--(C1tr);
    \end{tikzpicture}}\label{eq:SO7Sp2SO16M}
\end{equation} for the Higgs branch of the $6d$ theory \eqref{eq:SO7Sp2SO16E}. The bouquet of $C_1$ gauge nodes attached to the $D_6$ gauge node is indicative of the $\sorm(7)$ gauge node in \eqref{eq:SO7Sp2SO16E} being at finite coupling.

As the $6d$ theory \eqref{eq:SO7Sp2SO16E} comes from gauging an $\sorm(7)$ subgroup of the flavour symmetry of \eqref{eq:CkDkp4E} (for $k=2$), comparing their magnetic quivers \eqref{eq:SO7Sp2SO16M} and \eqref{eq:CkDkp4M} will realise gauging $\sorm(7)$ subgroups of Coulomb branch global symmetries.

The two magnetic quivers are related by the quiver subtraction shown below \begin{equation}
    \resizebox{\linewidth}{!}{\begin{tikzpicture}
       \node[gauger, label=below:$D_1$] (d1l) []{};
       \node[gaugeb, label=below:$C_1$] (c1l) [right=of d1l]{};
       \node[gauger, label=below:$D_2$] (d2l) [right=of c1l]{};
       \node[gaugeb, label=below:$C_2$] (c2l) [right=of d2l]{};
       \node[gauger, label=below:$D_3$] (d3l) [right=of c2l]{};
       \node[gaugeb, label=below:$C_3$] (c3l) [right=of d3l]{};
       \node[gauger, label=below:$D_4$] (d4l) [right=of c3l]{};
       \node[gaugeb, label=below:$C_4$] (c4l) [right=of d4l]{};
       \node[gauger, label=below:$D_5$] (d5l) [right=of c4l]{};
       \node[gaugeb, label=below:$C_5$] (c5l) [right=of d5l]{};
       \node[gauger, label=below:$D_6$] (d6) [right=of c5l]{};
       \node[gaugeb, label=below:$C_5$] (c5r) [right=of d6]{};
       \node[gauger, label=below:$D_5$] (d5r) [right=of c5r]{};
       \node[gaugeb, label=below:$C_4$] (c4r) [right=of d5r]{};
       \node[gauger, label=below:$D_4$] (d4r) [right=of c4r]{};
       \node[gaugeb, label=below:$C_3$] (c3r) [right=of d4r]{};
       \node[gauger, label=below:$D_3$] (d3r) [right=of c3r]{};
       \node[gaugeb, label=below:$C_2$] (c2r) [right=of d3r]{};
       \node[gauger, label=below:$D_2$] (d2r) [right=of c2r]{};
       \node[gaugeb, label=below:$C_1$] (c1r) [right=of d2r]{};
       \node[gauger, label=below:$D_1$] (d1r) [right=of c1r]{};
        \node[gaugeb, label=left:$C_1$] (c1t) [above=of d6]{};
        \draw[-] (d1l)--(c1l)--(d2l)--(c2l)--(d3l)--(c3l)--(d4l)--(c4l)--(d5l)--(c5l)--(d6)--(c5r)--(d5r)--(c4r)--(d4r)--(c3r)--(d3r)--(c2r)--(d2r)--(c1r)--(d1r);
        \draw[-] (c1t)--(d6);

         \node[gauger, label=below:$D_1$] (d1s) [below=of d1l]{};
        \node[gaugeb, label=below:$C_1$] (c1ls) [right=of d1s]{};
        \node[gauger, label=below:$D_2$] (d2ls) [right=of c1ls]{};
        \node[gaugeb, label=below:$C_2$] (c2s) [right=of d2ls]{};
        \node[gauger, label=below:$D_3$] (d3ls) [right=of c2s]{};
        \node[gaugeb, label=below:$C_3$] (c3ls) [right=of d3ls]{};
        \node[gauger, label=below:$D_4$] (d4s) [right=of c3ls]{};
        \node[gaugeb, label=below:$C_3$] (c3rs) [right=of d4s]{};
        \node[gauger, label=below:$D_2$] (d2rs) [right=of c3rs]{};
        \node[gaugeb, label=below:$C_1$] (c1rs) [right=of d2rs]{};

        \draw[-] (d1s)--(c1ls)--(d2ls)--(c2s)--(d3ls)--(c3ls)--(d4s)--(c3rs)--(d2rs)--(c1rs);
        
        \node[] (minus) [left=of d1s]{$-$};

         \node[] (ghost) [below=of c3rs]{};
          \node[gaugeb, label=below:$C_1$] (C1l) [below=of ghost]{};
        \node[gauger, label=below:$D_3$] (D3l) [right=of C1l]{};
        \node[gaugeb, label=below:$C_4$] (C4l) [right=of D3l]{};
        \node[gauger, label=below:$D_6$] (D6) [right=of C4l]{};
        \node[gaugeb, label=below:$C_5$] (C5r) [right=of D6]{};
        \node[gauger, label=below:$D_5$] (D5r) [right=of C5r]{};
        \node[gaugeb, label=below:$C_4$] (C4r) [right=of D5r]{};
        \node[gauger, label=below:$D_4$] (D4r) [right=of C4r]{};
        \node[gaugeb, label=below:$C_3$] (C3r) [right=of D4r]{};
        \node[gauger, label=below:$D_3$] (D3r) [right=of C3r]{};
        \node[gaugeb, label=below:$C_2$] (C2r) [right=of D3r]{};
        \node[gauger, label=below:$D_2$] (D2r) [right=of C2r]{};
        \node[gaugeb, label=below:$C_1$] (C1r) [right=of D2r]{};
        \node[gauger, label=below:$D_1$] (D1r) [right=of C1r]{};
        \node[gaugeb, label=left:$C_1$] (C1t) [above=of D6]{};
        \node[gaugeb, label=right:$C_1$] (C1tr) [above right=of D6]{};

        \draw[-] (C1l)--(D3l)--(C4l)--(D6)--(C5r)--(D5r)--(C4r)--(D4r)--(C3r)--(D3r)--(C2r)--(D2r)--(C1r)--(D1r) (C1t)--(D6)--(C1tr);
    \end{tikzpicture}}
    \label{fig:SO7Sp2SO16QuivSubtraction}
\end{equation}

The following quiver \begin{equation}
    \begin{tikzpicture}
        \node[gauger, label=below:$D_1$] (d1) []{};
        \node[gaugeb, label=below:$C_1$] (c1l) [right=of d1]{};
        \node[gauger, label=below:$D_2$] (d2l) [right=of c1l]{};
        \node[gaugeb, label=below:$C_2$] (c2) [right=of d2l]{};
        \node[gauger, label=below:$D_3$] (d3l) [right=of c2]{};
        \node[gaugeb, label=below:$C_3$] (c3l) [right=of d3l]{};
        \node[gauger, label=below:$D_4$] (d4) [right=of c3l]{};
        \node[gaugeb, label=below:$C_3$] (c3r) [right=of d4]{};
        \node[gauger, label=below:$D_2$] (d2r) [right=of c3r]{};
        \node[gaugeb, label=below:$C_1$] (c1r) [right=of d2r]{};

        \draw[-] (d1)--(c1l)--(d2l)--(c2)--(d3l)--(c3l)--(d4)--(c3r)--(d2r)--(c1r);
    \end{tikzpicture}
\end{equation}is hence proposed as the $\sorm(7)$ orthosymplectic quotient quiver. This quiver also contains gauge nodes with negative balance \cite{Gaiotto:2008ak}, the $D_4$ gauge node and the $C_1$ gauge node on the right. The Coulomb branch is smooth and its Hilbert series is not computable with the monopole formula \cite{Cremonesi:2013lqa} and remains a challenge.

A full explanation of the orthosymplectic quotient quiver subtraction algorithm will be done in Section \ref{sec:Method}.

\subsection{Gauging $\surm(2)$}
\label{subsec:SU2}
 The $\surm(2)$ case differs from the other three in that it is not motivated directly from a six dimensional theory, but rather by the Higgsing pattern observed between the gauge groups introduced thus far. After Higgsing $\sorm(7) \rightarrow G_2 \rightarrow \surm(3)$, the next gauge group to be broken to is $\surm(2)$, and applying the intuition gathered from the previous set of examples leads to the following candidate $\surm(2)$ orthosymplectic quotient quiver
 \begin{equation}
     \begin{tikzpicture}
         \node[gauger, label=below:$D_1$] (d1l)[]{};
         \node[gaugeb, label=below:$C_1$] (c1l)[right=of d1l]{};
         \node[gauger, label=below:$D_1$] (d1r) [right=of c1l]{};
         \node[gaugeb, label=below:$C_1$] (c1r) [right=of d1r]{};

         \draw[-] (d1l)--(c1l)--(d1r)--(c1r);
     \end{tikzpicture}
 \end{equation}
 Both $C_1$ gauge nodes in this quiver have negative balance \cite{Gaiotto:2008ak}. The Coulomb branch is smooth and its Hilbert series is not computable with the monopole formula \cite{Cremonesi:2013lqa}. However, the Coulomb branch of the unitary $\surm(2)$ quotient quiver is $T^*\left(\mathrm{SL}(2)\right)$ and so it is reasonable to conjecture that the Coulomb branch of the orthosymplectic $\surm(2)$ quotient quiver is $T^*\left(\mathrm{GL}(2)\right)$.
\subsection{A comment on other groups}
\label{subsec:Others}
Despite its successes for $\surm(2)$, $\surm(3)$, $G_2$, and $\sorm(7)$, the above analysis does not determine orthosymplectic quotient quivers for any other group. As aforementioned, this is principally a consequence of the anomaly cancellation conditions in six dimensions, which do not ensure linearity of the $6d$ quivers. Non-linear quivers also cannot be described by Type IIA brane systems. The most problematic issue with non-linear quivers for this analysis, is that one needs to couple additional matter, this is non-trivial to do on magnetic quivers and no prescription currently exists.


Let this point be illustrated with a specific example of a $6d\;\mathcal N=(1,0)\;\sorm(8)$ gauge theory on a $(-3)$-curve. This non-anomalous $\sorm(8)$ gauge theory enjoys a triality so the gauge node is coupled to hypermultiplets in the vector, spinor, and co-spinor representations, given in the following quiver \eqref{quiv:SO8WithTriality} \begin{equation}
    \begin{tikzpicture}
        \node[gauger](d4)[]{};
        \node[align=center,anchor=north] (d4lab) at (d4.south) {$\sorm(8)$\\$-3$};
        \node[flavour, label=below:$\sprm(1)$,fill=blue] (fl) [left=of d4]{};
        \node[flavour, label=below:$\sprm(1)$,fill=blue] (fr) [right=of d4]{};
        \node[flavour, label=above:$\sprm(1)$,fill=blue] (ft) [above=of d4]{};
        \draw[-] (fl)--node[midway, above]{$V$} (d4)--node[midway, above]{$S$} (fr) (d4)--node[midway, right]{$C$}(ft);
    \end{tikzpicture}
\label{quiv:SO8WithTriality}
\end{equation}

Suppose one starts with the non-anomalous $\sprm(1)$ gauge theory \eqref{eq:CkDkp4E} and attempts to gauge an $\sorm(8)$ subgroup of the $\sorm(20)$ flavour symmetry. In order to ensure that the $\sorm(8)$ gauge group is anomaly free, additional matter must be coupled before/after the action of gauging $\sorm(8)$.

The obstruction of the analysis used in the previous cases of $\surm(3),\;G_2,$ and $\sorm(7)$ to the above case of $\sorm(8)$ is that it is not clear how to realise the coupling of additional matter to magnetic quivers in general. This is essential in the treatment of non-linear $6d$ quivers.

Non-linearity similarly afflicts all $\sorm(n)$ gauge theories for $n\geq 8$, and as such the techniques used in this work cannot derive the associated $\sorm(n)$ quotient quivers.

Quotient quivers for the groups $\surm(n)$ for $n\geq 4$, $\sprm(n)$ for $n\geq 2$, and the other exceptional groups remain a challenge due to a lack of constructions of the $6d$ theories from branes or other means and hence there are a lack of magnetic quivers.

\subsection{A comment on the Higgs branch}
When a subgroup $G$ of the flavour symmetry of a magnetic quiver is gauged, the Higgs branch dimension decreases by $\textrm{dim}\;G$. The action on the Higgs branch is a hyper-Kähler quotient by $G$ with complete Higgsing. The Coulomb branch dimension increases by $\textrm{rank}\;G$ due to the additional vector multiplet. The action on the Coulomb branch is a hyper-Kähler quotient in reverse, which typically involves incomplete Higgsing since $\textrm{dim}\;G$ is typically larger than $\textrm{rank}\;G$ (except for $G=\urm(1)$ when there is equality). Since there is incomplete Higgsing, Hilbert series methods are difficult to apply.

Analogously when a subgroup $G$ of the Coulomb branch global symmetry of a magnetic quiver is gauged, the action on the Coulomb branch is a hyper-Kähler quotient by $G$. There is complete Higgsing if the decrease in the Coulomb branch dimension is $\textrm{dim}\;G$. The corresponding action on the Higgs branch is a reverse hyper-Kähler quotient by $G$. For all of the examples in this work, since we gauge only $G=\surm(2),\;\surm(3),\;G_2,\;\sorm(7)$, the Higgs branch should increase in dimension by $\textrm{rank}\;G$.
\section{Method of Orthosymplectic Quotient Quiver Subtraction}
\label{sec:Method}
In this Section, the method for orthosymplectic quotient quiver subtraction is explicitly described. The orthosymplectic quotient quiver subtraction is similar in spirit to the unitary quotient quiver subtraction \cite{Hanany:2023tvn}. In fact some quirks of the unitary quotient quiver subtraction are shared with the orthosymplectic quotient quiver subtraction.

Consider the subtraction of a quotient quiver from some target quiver. The method proceeds as:\begin{itemize}
    \item Align the quotient quiver against a maximal leg of the target quiver, permitting the quotient quiver to go one node beyond a junction. Ensure that D-type nodes are aligned and C-type nodes are aligned.
    \item Subtract the ranks of the nodes of the quotient quiver from the target quiver, ensuring that all nodes have positive rank and have positive (or zero) imbalance.
    \item Rebalance only the nodes which do not participate in the subtraction with a gauge node of $C_1$.
    \item If the quotient quiver has gone one node past a junction, the result is the union of all of the possible alignments.
\end{itemize}

Although there is much similarity with the unitary and orthosymplectic quotient quiver subtraction, there are some differences. In the unitary case, all gauge nodes retain their balance after subtraction, however in the orthosymplectic case, only the gauge nodes not participating in the subtraction retain their balance. In particular, the balance of each of the remaining gauge nodes in the maximal leg of the target quiver generally changes after subtraction.

Another difference is that the orthosymplectic quotient quivers can only be subtracted from maximal legs on the target quiver. In the unitary case, only a leg of $(1)-\cdots-(n)-$ is required to subtract a unitary $\surm(n)$ quotient quiver.
 
In the unitary case, rebalancing can be done with a $\urm(1)$ gauge node. In the orthosymplectic case, even though there are D-type and C-type nodes, rebalancing is always done with a $C_1$ gauge node. This may explicitly break the alternating pattern of D-type and C-type gauge nodes in the quiver. 

When the result of the quotient quiver subtraction gives a union, the intersection may easily be found through the Kraft-Procesi transition quiver subtraction \cite{Cabrera:2016vvv,Cabrera:2017njm,Cabrera:2018ann}. To date there is no systematic method of realising the Kraft-Procesi transition on orthosymplectic magnetic quivers. In the cases where a union arises, the typical slice to the intersection is an $A_1$ transition which corresponds to Kraft-Procesi subtraction of the following quiver 

\begin{equation}
\begin{tikzpicture}
    \node[gaugeb, label=below:$C_1$] (c1l) []{};
    \node[gaugeb, label=below:$C_1$] (c1r) [right=of c1l]{};

    \draw[transform canvas={yshift=+1.5pt}] (c1l)--(c1r);
    \draw[transform canvas={yshift=-1.5pt}] (c1l)--(c1r);
\end{tikzpicture}\label{eq:A1Slice}    
\end{equation}The Coulomb branch Hilbert series of this quiver is not computable with the monopole formula.

The gauge nodes not participating in the Kraft-Procesi subtraction are again rebalanced with a $C_1$ gauge node.

\section{Examples of $\surm(2)$ Orthosymplectic Quotient Quiver Subtraction}
\label{sec:SU2Examples}
\subsection{$\overline{min. E_7}///\surm(2)$}
The orthosymplectic quiver with Coulomb branch $\overline{min. E_7}$, given at the top of \Figref{fig:minE7A1Sub}, presents one of the simplest examples on which to test $\surm(2)$ orthosymplectic quotient quiver subtraction. This example is chosen first as the moduli space of $\overline{min. E_7}$ is well understood and additionally, the orthosymplectic magnetic quiver is star-shaped and constructed using fixtures of $D_3$. There is a unitary counter-part due to the Lie algebra isomorphism $D_3\simeq A_3$ and hence there can be comparison to the unitary case.

Performing the $\surm(2)$ quotient quiver subtraction as in \Figref{fig:minE7A1Sub} results in $\mathcal{Q}_{\ref{fig:minE7A1Sub}}$, whose Coulomb branch Hilbert series is
\begin{equation}
\hs\left[\mathcal C\left(\text{\Quiver{fig:minE7A1Sub}}\right)\right]=\frac{(1+t^2)^2\left(\begin{aligned}1&+36 t^2+590 t^4+4853 t^6+21516 t^8+52933 t^{10}+71802 t^{12}\\&+52933 t^{14}+21516 t^{16}+4853 t^{18}+590 t^{20}+36 t^{22}+t^{24}\end{aligned}\right)}{(1-t^2)^{28}}.
    \label{HS:minE7SU(2)}
\end{equation}
This is the same Hilbert series as for $\overline{n. n. min. D_6}$. Hence the conclusion is that \begin{equation}
    \overline{min. E_7}///\surm(2)=\overline{n. n.min. D_6}.
\end{equation}
This result is in fact already known from the unitary quotient quiver subtraction \cite{Hanany:2023tvn}, which gives \eqref{eq:nnminD6Unitary} as the resulting unitary magnetic quiver for $\overline{n. n. min. D_6}$.
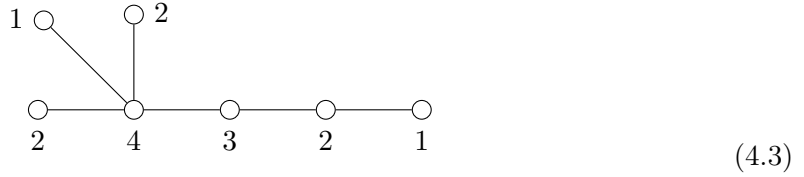
\begin{equation}
    \begin{tikzpicture}
        \node[gauge, label=right:$2$] (2) []{};
    \node[gauge, label=below:$4$] (4) [below=of 2]{};
    \node[gauge,label=below:$2$] (2L) [left=of 4]{};
    \node[gauge,label=below:$3$] (3) [right=of 4]{};
    \node[gauge, label=below:$2$] (2R) [right=of 3]{};
    \node[gauge, label=below:$1$] (1) [right=of 2R]{};
    \node[gauge,label=left:$1$] (1F) [above left=of 4]{};
    
    \draw[-] (2L)--(4)--(3)--(2R)--(1);
    \draw[-] (1F)--(4)--(2);
    \end{tikzpicture}
    \label{eq:nnminD6Unitary}
\end{equation}
Both the unitary and orthosymplectic magnetic quiver for $\overline{n. n. min. D_6}$ are star-shaped quivers, and consist of slices in $\surm(4)$, or equivalently, $\sorm(6)$ glued together. The isomorphism $A_3\simeq D_3$ maps the partition data as in Table \ref{tab:FixtureTable}, which gives a further check between the orthosymplectic and unitary quotient quiver subtraction.
\begin{figure}[H]
    \centering
    \resizebox{\textwidth}{!}{\begin{tikzpicture}[main/.style={draw,circle}]
    \node[gauger, label=below:$D_1$,fill=red] (d1l) []{};
    \node[gaugeb, label=below:$C_1$,fill=blue] (c1l) [right=of d1l]{};
    \node[gauger, label=below:$D_2$,fill=red] (d2l) [right=of c1l]{};
    \node[gaugeb, label=below:$C_2$,fill=blue] (c2l) [right=of d2l]{};
    \node[gauger, label=below:$D_3$,fill=red] (d3l) [right=of c2l]{};
    \node[gaugeb, label=below:$C_2$,fill=blue] (c2r) [right=of d3l]{};
    \node[gauger, label=below:$D_2$,fill=red] (d2r) [right=of c2r]{};
    \node[gaugeb, label=below:$C_1$,fill=blue] (c1r) [right=of d2r]{};
    \node[gauger, label=below:$D_1$,fill=red] (d1r) [right=of c1r]{};
    
    \node[gaugeb, label=left:$C_1$,fill=blue] (c1t) [above=of d3l]{};
    \node[gauger, label=left:$D_1$,fill=red] (d1t) [above=of c1t]{};

    \draw[-] (d1l)--(c1l)--(d2l)--(c2l)--(d3l)--(c2r)--(d2r)--(c1r)--(d1r);
    \draw[-] (d1t)--(c1t)--(d3l);

    \node[gauger, label=below:$D_1$,fill=red] (d1ls) [below=of d1l]{};
    \node[gaugeb, label=below:$C_1$,fill=blue] (c1ls) [right=of d1ls]{};
    \node[gauger, label=below:$D_1$,fill=red] (d2ls) [right=of c1ls]{};
    \node[gaugeb, label=below:$C_1$,fill=blue] (c2ls) [right=of d2ls]{};
    
    \draw[-] (d1ls)--(c1ls)--(d2ls)--(c2ls);

    \node[] (minus) [left=of d1ls]{$-$};

    \node[gauger, label=right:$D_1$,fill=red] (D1t) [below =of d3l]{};
    \node[gaugeb, label=right:$C_1$,fill=blue] (C1t) [below=of D1t]{};
    \node[gauger, label=below:$D_3$,fill=red] (D3) [below =of C1t]{};
    \node[gaugeb, label=below:$C_1$,fill=blue] (C1l) [left=of D3]{};
    \node[gauger, label=below:$D_1$,fill=red] (D1l) [left=of C1l]{};
    \node[gaugeb, label=left:$C_1$,fill=blue] (C1diag) [above left=of D3]{};

    \node[gaugeb, label=below:$C_2$,fill=blue] (C2r) [right=of D3]{};
    \node[gauger, label=below:$D_2$,fill=red] (D2r) [right=of C2r]{};
    \node[gaugeb, label=below:$C_1$,fill=blue] (C1r) [right=of D2r]{};
    \node[gauger, label=below:$D_1$,fill=red] (D1r) [right=of C1r]{};

    \draw[-] (D1t)--(C1t)--(D3)--(C1l)--(D1l);
    \draw[-] (C1diag)--(D3)--(C2r)--(D2r)--(C1r)--(D1r);
    
    \node[] (topghost) [right=of d1r]{};
    \node[] (bottomghost) [right=of D1r]{};

    \draw[->] (topghost)to [out=-45,in=45,looseness=1](bottomghost);
        
    \end{tikzpicture}}
    \caption{Subtraction of the $\surm(2)$ orthosymplectic quotient quiver from the orthosymplectic magnetic quiver for $\overline{min. E_7}$ to produce \Quiver{fig:minE7A1Sub}.}
    \label{fig:minE7A1Sub}
\end{figure}

\begin{table}[H]
    \centering
    \begin{tabular}{|c|c||c|c|}
    \hline
        Orthosymplectic Fixture & Partition of 6&Unitary Fixture & Partition of 4  \\\hline
         $C_1-[D_3]$& $(2^2,1^2)$& $(1)-[4]$&$(2,1^2)$\\\hline
         $D_1-C_1-[D_3]$&$(3,1^3)$& $(2)-[4]$&$(2^2)$ \\\hline
         $D_1-C_2-[D_3]$&$(3^2)$& $(1)-(2)-[4]$&$(3,1)$ \\\hline
         $D_1-C_1-D_2-C_2-[D_3]$ &$(5,1)$& $(1)-(2)-(3)-[4]$&$(4)$ \\\hline
    \end{tabular}
    \caption{The orthosymplectic fixtures of $\sorm(6)$ alongside the corresponding fixture for $\surm(4)$.}
    \label{tab:FixtureTable}
\end{table}

The $\surm(2)$ orthosymplectic quotient quiver subtraction shown in \Figref{fig:minE7A1Sub} provides another realisation of the Higgs branch side of the $4d\;\mathcal N=2$ duality proposed in \cite{Argyres:2007cn}. This duality says that $\sprm(2)$ with $6$ flavours at infinite coupling is dual to the $\surm(2)$ flavour symmetry gauging of the rank 1 $E_7$ SCFT.

This is the second time this result has been realised using quotient quiver subtraction; the first with unitary quivers \cite{Hanany:2023tvn} and the second with orthosymplectic quivers in \Figref{fig:minE7A1Sub}.

Due to the isomorphism $A_3\simeq D_3$, the Higgs branch of \Quiver{fig:minE7A1Sub} is $\mathcal S^{D_6}_{\mathcal N,(7,5)}$ which is the Spaltenstein dual to $\overline{n. n. min. D_6}$. This is seen with the computation of the Higgs branch Hilbert series \begin{equation}
    \hs\left[\mathcal H(\text{\Quiver{fig:minE7A1Sub}})\right]=\pe\left[2t^4+t^6+2t^8+t^{10}-t^{16}-t^{20}\right],
\end{equation}which matches the Hilbert series of $\mathcal S^{D_6}_{\mathcal N,(7,5)}$.
\subsection{$\overline{min. E_6}///\surm(2)$}
The next simplest example is the $\surm(2)$ quotient quiver subtraction on the magnetic quiver for $\overline{min. E_6}$. This is shown in \Figref{fig:minE6A1Sub} to produce \Quiver{fig:minE6A1Sub}, where rebalancing the gauge nodes introduces a loop into \Quiver{fig:minE6A1Sub}. The Higgs branch dimension increases by $\textrm{rank}\;\surm(2)=1$ after $\surm(2)$ quotient quiver subtraction as expected.

\begin{figure}[H]
    \centering
    \begin{tikzpicture}[main/.style={draw,circle}]
    \node[gauger, label=below:$D_1$] (D1l) []{};
    \node[gaugeb, label=below:$C_1$] (C1l) [right=of D1l]{};
    \node[gauger, label=below:$D_2$] (D2l) [right=of C1l]{};
    \node[gaugeb, label=below:$C_2$] (C2) [right=of D2l]{};
    \node[gauger, label=below:$D_2$] (D2r) [right=of C2]{};
    \node[gaugeb, label=below:$C_1$] (C1r) [right=of D2r]{};
    \node[gauger, label=below:$D_1$] (D1r) [right=of C1r]{};
    \node[gauger, label=left:$D_1$] (D1t) [above=of C2]{};
    \node (1) [left=of d1l]{};
    \draw[-] (D1l)--(C1l)--(D2l)--(C2)--(D2r)--(C1r)--(D1r);
    \draw[-] (D1t)--(C2);

   \node[gauger, label=below:$D_1$] (D1ls) [below=of D1l]{};
   \node[gaugeb, label=below:$C_1$] (C1ls) [right=of D1ls]{};
   \node[gauger, label=below:$D_1$] (D2ls) [right=of C1ls]{};
   \node[gaugeb, label=below:$C_1$] (C2s) [right=of D2ls]{};
    
    \draw[-] (D1ls)--(C1ls)--(D2ls)--(C2s);

    \node[] (minus) [left=of D1ls]{$-$};

    \node[gauger, label=left:$D_1$] (d1t) [below=of C2s]{};
   \node[gaugeb, label=below:$C_1$] (c2) [below =of d1t]{};
   \node[gauger, label=below:$D_1$] (d1l) [left=of c2]{};
   \node[gauger, label=below:$D_2$] (d2r) [right=of c2]{};
   \node[gaugeb, label=below:$C_1$] (c1r) [right=of d2r]{};
   \node[gauger, label=below:$D_1$] (d1r) [right=of c1r]{};
    \node[gaugeb, label=right:$C_1$] (c1t) [above=of d2r]{};

    \draw[-] (d1r)--(c1r)--(d2r)--(c1t)--(d1t)--(c2)--(d1l);
    \draw[-] (c2)--(d2r);

    \node[] (topghost) [right=of D1r]{};
    \node[] (bottomghost) [right=of d1r]{};

    \draw[->] (topghost)to [out=-45,in=45,looseness=1](bottomghost);
        
    \end{tikzpicture}
    \caption{Subtraction of the $\surm(2)$ orthosymplectic quotient quiver from the orthosymplectic magnetic quiver for $\overline{min. E_6}$ to produce \Quiver{fig:minE6A1Sub}.}
    \label{fig:minE6A1Sub}
\end{figure}

The Coulomb branch Hilbert series is
\begin{equation}
    \hs\left[\mathcal C\left(\mathcal Q_{\ref{fig:minE6A1Sub}}\right)\right]=\frac{(1 + t^2)^2 (1 + 17 t^2 + 119 t^4 + 251 t^6 + 119 t^8 + 17 t^{10} + 
   t^{12})}{(1 - t^2)^{16}}
   \label{HS:E6SU2}
\end{equation}
which exactly matches that of $\overline{n. min. A_5}$, and verifies that 
\begin{equation}
    \overline{min. E_6}///\surm(2)=\overline{n.min. A_5}.
\end{equation}

The above result is known from the unitary quotient quiver subtraction which gives \eqref{eq:nminA5Unitary} as the resulting unitary magnetic quiver for $\overline{n.min. A_5}$ \begin{equation}
    \begin{tikzpicture}
        \node[gauge, label=below:$1$] (1l) []{};
        \node[gauge, label=below:$2$] (2l) [right=of 1l]{};
        \node[gauge, label=below:$2$] (2m) [right=of 2l]{};
        \node[gauge, label=below:$2$] (2r) [right=of 2m]{};
        \node[gauge, label=below:$1$] (1r) [right=of 2r]{};
        \node[gauge, label=above:$1$] (1t) [above=of 2m]{};

        \draw[-] (1l)--(2l)--(2m)--(2r)--(1r) (2l)--(1t)--(2r);
    \end{tikzpicture}
    \label{eq:nminA5Unitary}
\end{equation}

In fact, the Higgs branch Hilbert series of \Quiver{fig:minE6A1Sub} is computed as \begin{equation}
    \hs\left[\mathcal H\left(\mathcal Q_{\ref{fig:minE6A1Sub}}\right)\right]=\frac{1 + t^4 + t^6 + t^8 + t^{12}}{(1 - t^2) (1 - t^4)^2 (1 - t^6)}
   \label{HS:E6SU2Higgs}
\end{equation}which is the same Hilbert series as for $\mathcal S^{A_5}_{\mathcal N,(4,2)}$ which is the Spaltenstein dual to $\overline{n. min. A_5}$. The Higgs branch dimension increases by $\textrm{rank}\;\surm(2)=1$ after $\surm(2)$ quotient quiver subtraction as expected.

The orthosymplectic $\surm(2)$ quotient quiver subtraction in \Figref{fig:minE6A1Sub} thus gives a novel orthosymplectic quiver with a gauge node loop which is a counterpart to the unitary quiver \eqref{eq:nminA5Unitary}. This is because they have the same Coulomb branch and the same Higgs branch.
\subsection{$\overline{n.min. A_5}///\surm(2)$}
The quiver \Quiver{fig:minE6A1Sub}, the magnetic quiver for $\overline{n. min. A_5}$, resulting from an orthosymplectic $\surm(2)$ quotient quiver subtraction on $\overline{min.E_6}$ may itself be the starting point for a further orthosymplectic $\surm(2)$ quotient quiver subtraction. In this case, there are two alignments of the $\surm(2)$ quotient quiver along the long leg of \Quiver{fig:minE6A1Sub} shown in \Figref{fig:nminA5A1Sub1} and \Figref{fig:nminA5A1Sub2} resulting in \Quiver{fig:nminA5A1Sub1} and \Quiver{fig:nminA5A1Sub2} respectively. Their intersection is \Quiver{fig:nminA5A1SubInt}, where the rule of subtracting \eqref{eq:A1Slice} (not shown) discussed in Section \ref{sec:Method} is used.

The Coulomb branch Hilbert series of \Quiver{fig:nminA5A1Sub1} is
 \begin{equation}
    \hs[\mathcal C(\text{\Quiver{fig:nminA5A1Sub1}})]=\frac{1 + 5 t^2 + 14 t^4 + 14 t^6 + 5 t^8 + t^{10}}{(1 - t^2)^{10}},
\end{equation}from which the Coulomb branch is identified as $\mathcal C(\text{\Quiver{fig:nminA5A1Sub1})}=\overline{\mathcal O}^{A_3}_{(3,1)}$. For \Quiver{fig:nminA5A1Sub2}, the Coulomb branch Hilbert series is 
\begin{equation}
    \hs[\mathcal C(\text{\Quiver{fig:nminA5A1Sub2}})]=\frac{1 + 11 t^2 + 57 t^4 + 170 t^6 + 324 t^8 + 398 t^{10} +\cdots+ t^{20}}{(1 - t^2)^5 (1 - t^4)^5},
\end{equation}from which the Coulomb branch can be identified as $\mathcal C(\text{\Quiver{fig:nminA5A1Sub2}})=\overline{[\mathcal W_{D_4}]}^{[0,1,0,2]}_{[0,0,0,2]}$.
\begin{figure}
    \centering
    \begin{subfigure}{\textwidth}
    \centering
    \begin{tikzpicture}
    \node[gauger, label=left:$D_1$] (d1t) []{};
    \node[gaugeb, label=below:$C_1$] (c2) [below =of d1t]{};
    \node[gauger, label=below:$D_1$] (d1l) [left=of c2]{};
    \node[gauger, label=below:$D_2$] (d2r) [right=of c2]{};
    \node[gaugeb, label=below:$C_1$] (c1r) [right=of d2r]{};
    \node[gauger, label=below:$D_1$] (d1r) [right=of c1r]{};
    \node[gaugeb, label=right:$C_1$] (c1t) [above=of d2r]{};
    \node (1) [left=of d1l]{};
    \draw[-] (d1r)--(c1r)--(d2r)--(c1t)--(d1t)--(c2)--(d1l);
    \draw[-] (c2)--(d2r);

    \node[gauger, label=below:$D_1$] (d1rs) [below=of d1r]{};
    \node[gaugeb, label=below:$C_1$] (c1rs) [left=of d1rs]{};
    \node[gauger, label=below:$D_1$] (d1ls) [left=of c1rs]{};
    \node[gaugeb, label=below:$C_1$] (c1ls) [left=of d1ls]{};

    \draw[-] (d1rs)--(c1rs)--(d1ls)--(c1ls);

    \node[gauger, label=above:$D_1$] (D1t) [below=of d1ls]{};
    \node[gaugeb, label=below:$C_1$] (C1l) [below left={1/sqrt(2)} and {1/sqrt(2)} of D1t]{};
    \node[gaugeb, label=below:$C_1$] (C1r) [below right={1/sqrt(2)} and {1/sqrt(2)} of D1t]{};
    \node[gauger, label=below:$D_1$] (D1l) [left=of C1l]{};
    \node[gauger, label=below:$D_1$] (D1r) [right=of C1r]{};

    \draw[-] (D1l)--(C1l)--(D1t)--(C1r)--(D1r);

    \draw[transform canvas={yshift=1.5pt}] (C1l)--(C1r);
    \draw[transform canvas={yshift=-1.5pt}] (C1l)--(C1r);

    \node[] (minus) [right=of d1rs]{$-$};

    \node[] (topghost) [left=of d1l]{};
    \node[] (bottomghost) [left=of D1l]{};

    \draw[->] (topghost)to [out=-135,in=135,looseness=1](bottomghost);
    \end{tikzpicture}
     \caption{}
    \label{fig:nminA5A1Sub1}
    \end{subfigure}
    \centering
    \begin{subfigure}{\textwidth}
    \centering
    \begin{tikzpicture}

    \node[gauger, label=left:$D_1$] (d1t) []{};
    \node[gaugeb, label=below:$C_1$] (c2) [below =of d1t]{};
    \node[gauger, label=below:$D_1$] (d1l) [left=of c2]{};
    \node[gauger, label=below:$D_2$] (d2r) [right=of c2]{};
    \node[gaugeb, label=below:$C_1$] (c1r) [right=of d2r]{};
    \node[gauger, label=below:$D_1$] (d1r) [right=of c1r]{};
    \node[gaugeb, label=right:$C_1$] (c1t) [above=of d2r]{};
    \node (1) [left=of d1l]{};
    \draw[-] (d1r)--(c1r)--(d2r)--(c1t)--(d1t)--(c2)--(d1l);
    \draw[-] (c2)--(d2r);

   \node[gaugeb, label=right:$C_1$] (c1s) [below=of d2r]{};
   \node[gauger, label=below:$D_1$] (d1ls) [below=of c1s]{};
   \node[gaugeb, label=below:$C_1$] (c1rs) [right=of d1ls]{};
   \node[gauger, label=below:$D_1$] (d1rs) [right=of c1rs]{};

    \draw[-] (d1rs)--(c1rs)--(d1ls)--(c1s);

    \node[] (ghost) [below=of d1ls]{};

    \node[gauger, label=below:$D_1$] (D1r) [below=of ghost]{};
    \node[gaugeb, label=below:$C_1$] (C1l) [left=of D1r]{};
    \node[gauger, label=below:$D_1$] (D1l) [left=of C1l]{};
    \node[gauger, label=left:$D_1$] (D1t) [above left={1/sqrt(2)} and {1/sqrt(2)} of C1l]{};
    \node[gaugeb, label=right:$C_1$] (C1t) [above right={1/sqrt(2)} and {1/sqrt(2)} of C1l]{};

    \draw[-] (D1l)--(C1l)--(D1r) (D1t)--(C1l) (D1t)--(C1t);
     \draw[transform canvas={xshift=+1.3pt,yshift=-1.3pt}] (C1l)--(C1t);
    \draw[transform canvas={xshift=-1.3pt,yshift=+1.3pt}] (C1l)--(C1t);

    \node[] (minus) [right=of d1rs]{$-$};

    \node[] (topghost) [left=of d1l]{};
    \node[] (bottomghost) [left=of D1l]{};

    \draw[->] (topghost)to [out=-135,in=135,looseness=1](bottomghost);

    \end{tikzpicture}
    \caption{}
    \label{fig:nminA5A1Sub2}
    \end{subfigure}
    \centering
    \begin{subfigure}{\textwidth}\centering
        \begin{tikzpicture}
            \node[gaugeb, label=below:$C_1$] (C1)[]{};
            \node[gauger, label=below:$D_1$] (D1l) [left=of C1]{};
            \node[gauger, label=below:$D_1$] (D1r) [right=of C1]{};
            \node[gauger, label=above:$D_1$] (D1t) [above=of C1]{};
            \node (1) [left=of D1l]{};
            \draw[-] (D1l)--(C1)--(D1r);
            \draw[transform canvas={xshift=+1.5pt}] (C1)--(D1t);
            \draw[transform canvas={xshift=-1.5pt}] (C1)--(D1t);
        \end{tikzpicture}
        \caption{}
        \label{fig:nminA5A1SubInt}
    \end{subfigure}
    \caption{Both alignments of the $\surm(2)$ quotient quiver against \Quiver{fig:minE6A1Sub} to produce \Quiver{fig:nminA5A1Sub1} and \Quiver{fig:nminA5A1Sub1}. Their intersection is \Quiver{fig:nminA5A1SubInt}.}
    \label{fig:nminA5A1Sub}
\end{figure}
The intersection of \Quiver{fig:nminA5A1Sub1} and \Quiver{fig:nminA5A1Sub2} is conjectured to be \Quiver{fig:nminA5A1SubInt}. There is no established method of quiver subtraction for Kraft-Procesi transitions therefore \Quiver{fig:nminA5A1SubInt} was found by subtracting \eqref{eq:A1Slice}. The Coulomb branch Hilbert series for \Quiver{fig:nminA5A1SubInt} is 
\begin{equation}
    \hs[\mathcal C(\text{\Quiver{fig:nminA5A1SubInt}})]=\frac{(1 + t^2)^2 (1 + 5 t^2 + t^4)}{(1 - t^2)^8},
\end{equation}which identifies the Coulomb branch as $\mathcal C(\text{\Quiver{fig:nminA5A1SubInt}})=\overline{\mathcal O}^{A_2}_{(2^2)}$.

Therefore the conclusion is that \begin{equation}
    \overline{n. min. A_5}///\surm(2)=\overline{\mathcal O}^{A_3}_{(3,1)}\cup \overline{[\mathcal W_{D_4}]}^{[0,1,0,2]}_{[0,0,0,2]},
\end{equation}which is consistent with the unitary quotient quiver subtraction \cite{Hanany:2023tvn}. In particular the unitary counterparts to \Quiver{fig:nminA5A1Sub1}, \Quiver{fig:nminA5A1Sub2}, and \Quiver{fig:nminA5A1SubInt} are \eqref{eq:A331Unitary}, \eqref{eq:D4AGUnitary}, and \eqref{eq:A322Unitary} respectively. The orthosymplectic counterpart \Quiver{fig:nminA5A1SubInt} to \eqref{eq:A322Unitary} follows from the Lie algebra isomorphism $\urm(1)\simeq D_1$ and $\surm(2)\simeq\sprm(1)$, however the orthosymplectic counterparts \Quiver{fig:nminA5A1Sub1} and \Quiver{fig:nminA5A1Sub2} to \eqref{eq:A331Unitary} and \eqref{eq:D4AGUnitary} are non-trivial. 

\begin{equation}
    \begin{tikzpicture}
        \node[gauge, label=below:$1$] (1l) at (0,0) {};
        \node[gauge, label=below:$2$] (2l) at (1,0) {};
        \node[gauge, label=below:$2$] (2r) at (2,0){};
        \node[gauge, label=above:$1$] (1T) at ({1+cos(60)},{sin(60)}){};
        \draw[-] (1l)--(2l)--(2r) (2l)--(1T);
        \draw[transform canvas={xshift=-1.29903810568 pt, yshift=-0.75 pt}] (2r)--(1T);
        \draw[transform canvas={xshift=+1.29903810568 pt, yshift=+0.75 pt}] (2r)--(1T);
    \end{tikzpicture}
    \label{eq:A331Unitary}
\end{equation}
\begin{equation}
    \begin{tikzpicture}
        \node[gauge, label=below:$1$] (1l) at (0,0) {};
        \node[gauge, label=below:$2$] (2l) at (1,0) {};
        \node[gauge, label=below:$1$] (1r) at (2,0){};
        \node[gauge, label=right:$1$] (1tr) at ({1+cos(60)},{sin(60)}){};
        \node[gauge, label=left:$1$] (1tl) at ({1-cos(60)},{sin(60)}){};
        \draw[-] (1l)--(2l)--(2r) (1tl)--(2l)--(1tr);
        \draw[transform canvas={xshift=0 pt, yshift=1.5 pt}] (1tl)--(1tr);
        \draw[transform canvas={xshift=0 pt, yshift=-1.5 pt}] (1tl)--(1tr);
    \end{tikzpicture}
    \label{eq:D4AGUnitary}
\end{equation}
\begin{equation}
    \begin{tikzpicture}
        \node[gauge, label=below:$1$] (1l) at (0,0){};
        \node[gauge, label=below:$2$] (2) at (1,0){};
        \node[gauge, label=below:$1$] (1r) at (2,0){};
        \node[gauge, label=above:$1$] (1t) at (1,1){};
        \draw[-] (1l)--(2)--(1r);
        \draw[transform canvas={xshift=1.5 pt, yshift=0 pt}] (2)--(1t);
        \draw[transform canvas={xshift=-1.5 pt, yshift=0 pt}] (2)--(1t);
    \end{tikzpicture}
    \label{eq:A322Unitary}
\end{equation}

This can be verified with computation of the Higgs branch Hilbert series for each quiver. Unlike the Coulomb branch Hilbert series, the Higgs branch Hilbert series admits partial refinement since each quiver has a pair of bifundamental hypers between two gauge nodes, contributing a factor of $\sprm(1)$ to the global symmetry. Additionally, the quivers \Quiver{fig:nminA5A1Sub1} and \Quiver{fig:nminA5A1Sub2} contain a larger loop of gauge nodes and therefore there is an invariant associated to them which contributes an additional factor of $\urm(1)$ to the global symmetry. It is unknown how to give a fugacity to this $\urm(1)$. Firstly for \Quiver{fig:nminA5A1Sub1}, \begin{equation}
    \hs[\mathcal H(\text{\Quiver{fig:nminA5A1Sub1}})]=\pe\left[\left([2]_{\sprm(1)}+1\right)t^2+2[1]_{\sprm(1)}t^3-t^6-t^8\right],
\end{equation}where the Dynkin label is a shorthand for the character of the given $\sprm(1)$ representation. With guidance from Spaltenstein duality, the Hilbert series for $\mathcal S^{A_3}_{\mathcal N,(2,1^2)}$ - the Spaltenstein dual of $\overline{\mathcal O}^{A_3}_{(3,1)}$ - is \begin{equation}
    \hs\left[\mathcal S^{A_3}_{\mathcal N,(2,1^2)}\right]=\pe\left[\left([2]_{\sprm(1)}+1\right)t^2+(q+1/q)[1]_{\sprm(1)}t^3-t^6-t^8\right],
\end{equation}where $q$ is a $\urm(1)$ fugacity. This produces the same Hilbert series as $\mathcal H(\text{\Quiver{fig:nminA5A1Sub1}})$ in the limit $q\rightarrow 1$.

The Higgs branch Hilbert series of \Quiver{fig:nminA5A1Sub2} may also be partially refined although the resulting Hilbert series is not a complete intersection. The unrefined Hilbert series is given as \begin{equation}
     \hs[\mathcal H(\text{\Quiver{fig:nminA5A1Sub2}})]=\frac{\left(\begin{aligned}1 &+ 3 t^2 + 2 t^3 + 6 t^4 + 4 t^5 + 10 t^6 + 6 t^7 + 9 t^8 \\&+ 6 t^9 + 
 10 t^{10} + 4 t^{11} + 6 t^{12} + 2 t^{13} + 
 3 t^{14} + t^{16}\end{aligned}\right)}{(1 - t^2)  (1 - t^3)^2 (1 - t^4) (1 - t^5)^2}
\end{equation}There is no simple analogue of Spaltenstein duality for slices in the affine Grassmannian, however it is simple to check that the Higgs branch Hilbert series of \eqref{eq:D4AGUnitary} matches.

Finally, for \Quiver{fig:nminA5A1SubInt} the Higgs branch Hilbert series is computed as \begin{equation}
     \hs\left[\mathcal S^{A_3}_{\mathcal N,(2,1^2)}\right]=\pe\left[[2]_{\sprm(1)}t^2+[2]_{\sprm(1)}t^4-t^6-t^8\right],
\end{equation}which is the same Hilbert series as for $\mathcal S^{A_3}_{(2^2)}$, the Spaltenstein dual of $\overline{\mathcal O}^{A_3}_{(2^2)}$. The Higgs branch dimension increases by $\textrm{rank}\;\surm(2)=1$ after $\surm(2)$ quotient quiver subtraction as expected.

The novelty of the $\surm(2)$ orthosymplectic quotient quiver subtraction in \Figref{fig:nminA5A1Sub} is in the discovery of orthosymplectic counterparts to \eqref{eq:A331Unitary}, \eqref{eq:D4AGUnitary}, and \eqref{eq:A322Unitary} which are \Quiver{fig:nminA5A1Sub1}, \Quiver{fig:nminA5A1Sub2}, and \Quiver{fig:nminA5A1SubInt} respectively. It is also interesting to note that the alternating $\sorm/\sprm$ gauge node structure typical of orthosymplectic quivers is not present in \Quiver{fig:nminA5A1Sub1} and \Quiver{fig:nminA5A1Sub2}.

\subsection{$\overline{min. E_8}///\surm(2)$}
The $\surm(2)$ orthosymplectic quotient quiver may be subtracted from the orthosymplectic magnetic quiver for $\overline{min. E_8}$, as shown in \Figref{fig:minE8SU2}, to produce \Quiver{fig:minE8SU2}.

\begin{figure}[H]
    \centering
    \resizebox{\textwidth}{!}{\begin{tikzpicture}[main/.style={draw,circle}]

        \node[gauger, label=below:$D_1$] (d1l) []{};
        \node[gaugeb, label=below:$C_1$] (c1l) [right=of d1l]{};
        \node[gauger, label=below:$D_2$] (d2l) [right=of c1l]{};
        \node[gaugeb, label=below:$C_2$] (c2l) [right=of d2l]{};
        \node[gauger, label=below:$D_3$] (d3l) [right=of c2l]{};
        \node[gaugeb, label=below:$C_3$] (c3l) [right=of d3l]{};
        \node[gauger, label=below:$D_4$] (d4) [right=of c3l]{};
        \node[gaugeb, label=below:$C_3$] (c3r) [right=of d4]{};
        \node[gauger, label=below:$D_3$] (d3r) [right=of c3r]{};
        \node[gaugeb, label=below:$C_2$] (c2r) [right=of d3r]{};
        \node[gauger, label=below:$D_2$] (d2r) [right=of c2r]{};
        \node[gaugeb, label=below:$C_1$] (c1r) [right=of d2r]{};
        \node[gauger, label=below:$D_1$] (d1r) [right=of c1r]{};
        \node[gaugeb, label=right:$C_1$] (c1t) [above=of d4]{};

        \draw[-] (d1l)--(c1l)--(d2l)--(c2l)--(d3l)--(c3l)--(d4)--(c3r)--(d3r)--(c2r)--(d2r)--(c1r)--(d1r) (d4)--(c1t);
    
    \node[gauger, label=below:$D_1$] (d1lk) [below=of d1l]{};
    \node[gaugeb, label=below:$C_1$] (c1lk) [right=of d1lk]{};
    \node[gauger, label=below:$D_1$] (d1rk) [right=of c1lk]{};
    \node[gaugeb, label=below:$C_1$] (c1rk) [right=of d1rk]{};

    \draw[-] (c1rk)--(d1rk)--(c1lk)--(d1lk);
    
    \node[gauger, label=below:$D_1$,fill=red] (d1ls) [below={3} of d1rk]{};
    \node[gaugeb, label=below:$C_1$,fill=blue] (c1ls) [right=of d1ls]{};
    \node[gauger, label=below:$D_3$,fill=red] (d3ls) [right=of c1ls]{};
    \node[gaugeb, label=below:$C_3$,fill=blue] (c3ls) [right=of d3ls]{};
    \node[gauger, label=below:$D_4$,fill=red] (d4s) [right=of c3ls]{};
    \node[gaugeb, label=below:$C_3$,fill=blue] (c3rs) [right=of d4s]{};
    \node[gauger, label=below:$D_3$,fill=red] (d3rs) [right=of c3rs]{};
    \node[gaugeb, label=below:$C_2$] (c2rs) [right=of d3rs]{};
    \node[gauger, label=below:$D_2$] (d2rs) [right=of c2rs]{};
    \node[gaugeb, label=below:$C_1$,fill=blue] (c1rs) [right=of d2rs]{};
    \node[gauger, label=below:$D_1$,fill=red] (d1rs) [right=of c1rs]{};
    
    \node[gaugeb, label=left:$C_1$,fill=blue] (c1t1s) [above=of d3ls]{};
    \node[gaugeb, label=right:$C_1$,fill=blue] (c1t2s) [above=of d4s]{};

    \draw[-] (d1ls)--(c1ls)--(d3ls)--(c3ls)--(d4s)--(c3rs)--(d3rs)--(c2rs)--(d2rs)--(c1rs)--(d1rs);
    \draw[-] (c1t1s)--(d3ls);
    \draw[-] (c1t2s)--(d4s);
    \node[] (minus) [left=of d1lk]{$-$};
    \node[] (topghost) [right=of d1r]{};
    \node[] (bottomghost) [right=of d1rs]{};
    \draw[->] (topghost)to [out=-45,in=+45,looseness=1](bottomghost);
    \end{tikzpicture}}
    \caption{Subtraction of the $\surm(2)$ orthosymplectic quotient quiver from the orthosymplectic magnetic quiver for $\overline{min. E_8}$ to produce \Quiver{fig:minE8SU2}.}
    \label{fig:minE8SU2}
\end{figure}

The Coulomb branch Hilbert series is \begin{equation}
    \hs\left[\mathcal C\left(\text{\Quiver{fig:minE8SU2}}\right)\right]=\frac{(1 + t^2)^2 \left(\begin{aligned}1 &+ 79 t^2 + 3161 t^4 + 75291 t^6 + 
   1158376 t^8 + 12099785 t^{10} \\&+ 88650725 t^{12} + 465895118 t^{14} + 
   1783653576 t^{16} + 5026645901 t^{18} \\&+ 10497603729 t^{20} + 
   16309233956 t^{22} + 18885794304 t^{24} + \cdots + t^{48}\end{aligned}\right)}{(1 - 
  t^2)^{52}}
\end{equation}which confirms the Coulomb branch as $\overline{n. min. E_7}$.

This is again consistent with the unitary quotient quiver subtraction \cite{Hanany:2023tvn} which gives the following unitary magnetic quiver for $\overline{n. min. E_7}$\begin{equation}
    \begin{tikzpicture}
        \node[gauge,label=right:$1$] (oneF) []{};
    \node[gauge, label=below:$4$] (four) [below =of oneF]{};
    \node[gauge, label=below:$2$] (twoL) [left=of four]{};
    \node[gauge, label=below:$5$] (five) [right=of four]{};
    \node[gauge, label=below:$6$] (six) [right=of five]{};
    \node[gauge, label=below:$4$] (fourR) [right=of six]{};
    \node[gauge,label=below:$2$] (twoR) [right=of fourR]{};
    \node[gauge,label=right:$3$] (three) [above=of six]{};
    
    \draw[-] (twoL)--(four)--(five)--(six)--(fourR)--(twoR);
    \draw[-] (three)--(six);
    \draw[-] (oneF)--(four);
    \end{tikzpicture}\label{eq:nminE7Unitary}
\end{equation}

The Higgs branch Hilbert series of \Quiver{fig:minE8SU2} is computed as \begin{equation}
     \hs\left[\mathcal H\left(\text{\Quiver{fig:minE8SU2}}\right)\right]=\pe\left[t^4 + t^8 + t^{10} + t^{12} + t^{16} + t^{18} - t^{28} - t^{36}\right]
\end{equation}which is the same Hilbert series as $\mathcal S^{E_7}_{\mathcal N,[2, 2, 0, 2, 0, 2, 2]}$ (Bala-Carter label $E_7(a_2)$) which is Spaltenstein dual to $\overline{n. min. E_7}$. 

The orthosymplectic quotient quiver subtraction hence gives a new orthosymplectic counterpart to \eqref{eq:nminE7Unitary}.
\subsection{$\overline{min.E_8}///\left(\surm(2)\times\surm(2)\right)$}
\label{sec:E8///SU2SU2}
All of the examples so far have recovered known results from either Type IIA string theory, class $\mathcal S$ theories, and in particular from unitary quotient quiver subtraction \cite{Hanany:2023tvn}. It is now time to explore the differences between unitary and orthosymplectic magnetic quivers. In doing so, results which can be obtained from orthosymplectic quivers, but not from unitary quivers (at present), are discussed.

The unitary and orthosymplectic minimal $E$-type quivers share the same Coulomb and Higgs branches but they are far from identical. Take for example the affine $\hat{E}_8^{(1)}$ quiver \eqref{eq:E8Unitary}, which is the unitary magnetic quiver for 
$\overline{min. E_8}$.
\begin{equation}
    \begin{tikzpicture}
        \node[gauge, label=below:$1$] (1l)[]{};
        \node[gauge, label=below:$2$] (2l)[right=of 1l]{};
        \node[gauge, label=below:$3$] (3l) [right=of 2l]{};
        \node[gauge, label=below:$4$] (4l) [right=of 3l]{};
        \node[gauge, label=below:$5$] (5l)[right=of 4l]{};
        \node[gauge, label=below:$6$] (6l) [right=of 5l]{};
        \node[gauge, label=below:$4$] (4r) [right=of 6l]{};
        \node[gauge, label=below:$2$] (2r) [right=of 4r]{};
        \node[gauge, label=right:$3$] (3t) [above=of 6l]{};
        \draw[-] (1l)--(2l)--(3l)--(4l)--(5l)--(6l)--(4r)--(2r) (6l)--(3t);
    \end{tikzpicture}
    \label{eq:E8Unitary}
\end{equation}
This has one long leg of gauge nodes of the form $(1)-(2)-\cdots-(6)-$, making it amenable to one subtraction of either an $\surm(2),\;\surm(3),$ or $\surm(4)$ unitary quotient quiver.  In contrast, the orthosymplectic magnetic quiver for $\overline{min.E_8}$ has two long legs exchanged under an $S_2$ outer automorphism symmetry, and is hence amenable to the subtraction of an $\surm(2)$ or $\surm(3)$ orthosymplectic quotient quiver on either long leg. The action on the Coulomb branch is the hyper-Kähler quotient by $G\times G'$ where $G$ and $G'$ are either $\surm(2)$ or $\surm(3)$.

Since the case $\overline{min.E_8}///\left(\surm(3) \times \surm(3)\right)$ was previously studied in \cite{Hanany:2022itc}, the discussion below will focus on $\overline{min.E_8}///\left(\surm(2) \times \surm(2)\right)$, here, and $\overline{min.E_8}///\left(\surm(2) \times \surm(3)\right)$ in Section \ref{sec:minE8A1A2Quot}.
\begin{figure}[]
    \centering
    \resizebox{\textwidth}{!}{\begin{tikzpicture}[main/.style={draw,circle}]

        \node[gauger, label=below:$D_1$] (d1l) []{};
        \node[gaugeb, label=below:$C_1$] (c1l) [right=of d1l]{};
        \node[gauger, label=below:$D_2$] (d2l) [right=of c1l]{};
        \node[gaugeb, label=below:$C_2$] (c2l) [right=of d2l]{};
        \node[gauger, label=below:$D_3$] (d3l) [right=of c2l]{};
        \node[gaugeb, label=below:$C_3$] (c3l) [right=of d3l]{};
        \node[gauger, label=below:$D_4$] (d4) [right=of c3l]{};
        \node[gaugeb, label=below:$C_3$] (c3r) [right=of d4]{};
        \node[gauger, label=below:$D_3$] (d3r) [right=of c3r]{};
        \node[gaugeb, label=below:$C_2$] (c2r) [right=of d3r]{};
        \node[gauger, label=below:$D_2$] (d2r) [right=of c2r]{};
        \node[gaugeb, label=below:$C_1$] (c1r) [right=of d2r]{};
        \node[gauger, label=below:$D_1$] (d1r) [right=of c1r]{};
        \node[gaugeb, label=right:$C_1$] (c1t) [above=of d4]{};

        \draw[-] (d1l)--(c1l)--(d2l)--(c2l)--(d3l)--(c3l)--(d4)--(c3r)--(d3r)--(c2r)--(d2r)--(c1r)--(d1r) (d4)--(c1t);
        
    \node[gauger, label=below:$D_1$] (d1rj) [below=of d1r]{};
    \node[gaugeb, label=below:$C_1$] (c1rj) [left=of d1rj]{};
    \node[gauger, label=below:$D_1$] (d1lj) [left=of c1rj]{};
    \node[gaugeb, label=below:$C_1$] (c1lj) [left=of d1lj]{};

    \draw[-] (d1rj)--(c1rj)--(d1lj)--(c1lj);
    
    \node[gauger, label=below:$D_1$] (d1lk) [below=of d1l]{};
    \node[gaugeb, label=below:$C_1$] (c1lk) [right=of d1lk]{};
    \node[gauger, label=below:$D_1$] (d1rk) [right=of c1lk]{};
    \node[gaugeb, label=below:$C_1$] (c1rk) [right=of d1rk]{};

    \draw[-] (c1rk)--(d1rk)--(c1lk)--(d1lk);
    
    \node[gauger, label=below:$D_1$,fill=red] (d1ls) [below={3} of d1rk]{};
    \node[gaugeb, label=below:$C_1$,fill=blue] (c1ls) [right=of d1ls]{};
    \node[gauger, label=below:$D_3$,fill=red] (d3ls) [right=of c1ls]{};
    \node[gaugeb, label=below:$C_3$,fill=blue] (c3ls) [right=of d3ls]{};
    \node[gauger, label=below:$D_4$,fill=red] (d4s) [right=of c3ls]{};
    \node[gaugeb, label=below:$C_3$,fill=blue] (c3rs) [right=of d4s]{};
    \node[gauger, label=below:$D_3$,fill=red] (d3rs) [right=of c3rs]{};
    \node[gaugeb, label=below:$C_1$,fill=blue] (c1rs) [right=of d3rs]{};
    \node[gauger, label=below:$D_1$,fill=red] (d1rs) [right=of c1rs]{};
    
    \node[gaugeb, label=left:$C_1$,fill=blue] (c1t1s) [above=of d3ls]{};
    \node[gaugeb, label=above:$C_1$,fill=blue] (c1t2s) [above=of d4s]{};
    \node[gaugeb, label=right:$C_1$,fill=blue] (c1t3s) [above=of d3rs]{};

    \draw[-] (d1ls)--(c1ls)--(d3ls)--(c3ls)--(d4s)--(c3rs)--(d3rs)--(c1rs)--(d1rs);
    \draw[-] (c1t1s)--(d3ls);
    \draw[-] (c1t2s)--(d4s);
    \draw[-] (c1t3s)--(d3rs);
    \node[] (minus) [right=of d1rj]{$-$};
    \node[] (minus) [left=of d1lk]{$-$};
    \node[] (topghost) [below=of d1lk]{};
    \node[] (bottomghost) [left=of d1ls]{};
    \draw[->] (topghost)to [out=-135,in=+135,looseness=1](bottomghost);
    \end{tikzpicture}}
    \caption{Subtraction of two $\surm(2)$ orthosymplectic quotient quivers from each long leg of the orthosymplectic magnetic quiver for $\overline{min. E_8}$. The result is quiver \Quiver{fig:minE8SU2SU2}.}
    \label{fig:minE8SU2SU2}
\end{figure}
The subtraction of two $\surm(2)$ orthosymplectic quotient quivers from each leg of the orthosymplectic magnetic quiver for $\overline{min. E_8}$ is shown in \Figref{fig:minE8SU2SU2} to produce quiver \Quiver{fig:minE8SU2SU2}. From counting the ranks of the gauge nodes, the Coulomb branch of \Quiver{fig:minE8SU2SU2} has dimension $23$ which precludes an exact computation of its Coulomb branch Hilbert series. Nevertheless, the unrefined perturbative Hilbert series may be computed and is presented here up to order $t^{10}$ together with its $\pl$.
\begin{align}
\hs\left[\mathcal C(\text{\Quiver{fig:minE8SU2SU2}})\right]&=1+66 t^2+2706 t^4+79651 t^6+1810160 t^8+33147464 t^{10}+O\left(t^{11}\right)\\
 \pl\left[\hs\left[\mathcal C(\text{\Quiver{fig:minE8SU2SU2}})\right]\right]&=66 t^2+495 t^4-3135 t^6-64636 t^8+884156 t^{10}+O\left(t^{11}\right)
\label{HS:minE8SU2SU2}
\end{align}
The $t^2$ term is $66$ is indicative of an $\sorm(12)$ global symmetry. The only other generator, at order $t^4$, appears to transform in the $\bigwedge^4\left([1,0,0,0,0,0]_{D_6}\right)$ indicating that the moduli space is not a slice in the affine Grassmannian. But otherwise it is difficult to identify the moduli space exactly.

To compare, the unrefined Hilbert series for $\overline{min. E_8}///\left(\surm(2)\times\surm(2)\right)$ can be computed exactly and is evaluated as \begin{align}
    &\hs\left[\overline{min. E_8}///\left(\surm(2)\times\surm(2)\right)\right]\nonumber\\&=\frac{\left(\begin{aligned}1 &+ 43 t^2 + 1418 t^4 + 31351 t^6 + 521884 t^8 + 6691001 t^{10} + 
  68405360 t^{12} + 569379869 t^{14} \\&+ 3933117266 t^{16} + 
  22876102070 t^{18} + 113436359067 t^{20} + 484431480383 t^{22} \\&+ 
  1797063507391 t^{24} + 5832469898724 t^{26} + 16662719682327 t^{28} + 
  42117041766495 t^{30} \\&+ 94593877569560 t^{32} + 189460950668991 t^{34} + 
  339409608683291 t^{36} + 545161969037398 t^{38} \\&+ 
  786610379435045 t^{40} + 1021071962986483 t^{42} + 
  1193603944481102 t^{44} \\&+ 1257280044576400 t^{46} +\cdots+t^{92}\end{aligned}\right)}{(1-t^2)^{23}(1-t^4)^{23}}\label{eq:minE8SU2SU2HS}\\&=1+66 t^2+2706 t^4+79651 t^6+1810160 t^8+33147464 t^{10}+O\left(t^{11}\right),
\end{align} which matches the Coulomb branch Hilbert series of \Quiver{fig:minE8SU2SU2} up to order $t^{10}$.

The Hilbert series \eqref{eq:minE8SU2SU2HS} was evaluated using the embedding of $E_8\hookleftarrow\sorm(12)\times\surm(2)\times\surm(2)$ which decomposes the fundamental as \begin{equation}
    \left(\mu_7\right)_{E_8}\rightarrow \mu_2 + \nu^2+\rho^2+ \mu_5 \nu  + \mu_6 \rho + \mu_1\nu\rho, 
\end{equation}where on the right-hand side the $\mu_i,\nu,$ and $\rho$ are highest weight fugacities for $\sorm(12), \surm(2),$ and $\surm(2)$ respectively.

This verifies that \begin{equation}
    \mathcal C\left(\text{\Quiver{fig:minE8SU2SU2}}\right)=\overline{min. E_8}///\left(\surm(2)\times\surm(2)\right),
\end{equation}where it is worth reiterating that such a result has not been seen using unitary quivers.

\section{Examples of $\surm(3)$ Orthosymplectic Quotient Quiver Subtraction}
\label{sec:SU3Examples}
\subsection{$\overline{min. E_7}///\surm(3)$}
Similar to the $\surm(2)$ quotient quiver subtraction, the orthosymplectic magnetic quiver for $\overline{min.E_7}$ also admits an $\surm(3)$ quotient quiver subtraction with two alignments, as shown in \Figref{fig:minE7SU3}. The resulting quivers are \Quiver{fig:minE7SU31} and \Quiver{fig:minE7SU32}, whose intersection is \Quiver{fig:minE7SU3Int}, which is the same as \Quiver{fig:minE6A1Sub}, whose Coulomb branch is $\overline{n. min. A_5}$.

\begin{figure}
    \centering
    \begin{subfigure}{\textwidth}
    \centering
        \begin{tikzpicture}
        \node[gauger, label=below:$D_1$,fill=red] (d1l) []{};
    \node[gaugeb, label=below:$C_1$,fill=blue] (c1l) [right=of d1l]{};
    \node[gauger, label=below:$D_2$,fill=red] (d2l) [right=of c1l]{};
    \node[gaugeb, label=below:$C_2$,fill=blue] (c2l) [right=of d2l]{};
    \node[gauger, label=below:$D_3$,fill=red] (d3l) [right=of c2l]{};
    \node[gaugeb, label=below:$C_2$,fill=blue] (c2r) [right=of d3l]{};
    \node[gauger, label=below:$D_2$,fill=red] (d2r) [right=of c2r]{};
    \node[gaugeb, label=below:$C_1$,fill=blue] (c1r) [right=of d2r]{};
    \node[gauger, label=below:$D_1$,fill=red] (d1r) [right=of c1r]{};
    
    \node[gaugeb, label=left:$C_1$,fill=blue] (c1t) [above=of d3l]{};
    \node[gauger, label=left:$D_1$,fill=red] (d1t) [above=of c1t]{};

    \draw[-] (d1l)--(c1l)--(d2l)--(c2l)--(d3l)--(c2r)--(d2r)--(c1r)--(d1r);
    \draw[-] (d1t)--(c1t)--(d3l);

    \node[gaugeb, label=right:$C_1$] (c1rs) [below=of d3l]{};
    \node[gauger, label=below:$D_2$] (d2rs) [below=of c1rs]{};
    \node[gaugeb, label=below:$C_2$] (c2s) [left=of d2rs]{};
    \node[gauger, label=below:$D_2$] (d2s) [left=of c2s]{};
    \node[gaugeb, label=below:$C_1$] (c1s) [left=of d2s]{};
    \node[gauger, label=below:$D_1$] (d1s) [left=of c1s]{};

    \draw[-] (c1rs)--(d2rs)--(c2s)--(d2s)--(c1s)--(d1s);

    \node[] (minus) [left=of d1s]{$-$};

    \node[gauger, label=right:$D_1$] (D1t) [right=of d2rs]{};
    \node[gaugeb, label=right:$C_1$] (C1t) [below=of D1t]{};
    \node[gaugeb, label=below:$C_2$] (C2) [below=of C1t]{};
    \node[gauger, label=below:$D_1$] (D1l) [left=of C2]{};
    \node[gauger, label=below:$D_2$] (D2) [right=of C2]{};
    \node[gaugeb, label=below:$C_1$] (C1) [right=of D2]{};
    \node[gauger, label=below:$D_1$] (D1) [right=of C1]{};

    \draw[-] (D1t)--(C1t) (D1l)--(C2)--(D2)--(C1)--(D1);

    \draw[transform canvas={xshift=+1.5pt}] (C2)--(C1t);
    \draw[transform canvas={xshift=-1.5pt}] (C2)--(C1t);
    
    \node[] (topghost) [right=of d1r]{};
    \node[] (bottomghost) [right=of D1]{};

    \draw[->] (topghost)to [out=-45,in=45,looseness=1](bottomghost);

        \end{tikzpicture}
         \caption{}
        \label{fig:minE7SU31}
    \end{subfigure}
    \begin{subfigure}{\textwidth}
    \centering
    \begin{tikzpicture}
        \node[gauger, label=below:$D_1$,fill=red] (d1l) []{};
    \node[gaugeb, label=below:$C_1$,fill=blue] (c1l) [right=of d1l]{};
    \node[gauger, label=below:$D_2$,fill=red] (d2l) [right=of c1l]{};
    \node[gaugeb, label=below:$C_2$,fill=blue] (c2l) [right=of d2l]{};
    \node[gauger, label=below:$D_3$,fill=red] (d3l) [right=of c2l]{};
    \node[gaugeb, label=below:$C_2$,fill=blue] (c2r) [right=of d3l]{};
    \node[gauger, label=below:$D_2$,fill=red] (d2r) [right=of c2r]{};
    \node[gaugeb, label=below:$C_1$,fill=blue] (c1r) [right=of d2r]{};
    \node[gauger, label=below:$D_1$,fill=red] (d1r) [right=of c1r]{};
    
    \node[gaugeb, label=left:$C_1$,fill=blue] (c1t) [above=of d3l]{};
    \node[gauger, label=left:$D_1$,fill=red] (d1t) [above=of c1t]{};

    \draw[-] (d1l)--(c1l)--(d2l)--(c2l)--(d3l)--(c2r)--(d2r)--(c1r)--(d1r);
    \draw[-] (d1t)--(c1t)--(d3l);

    \node[gaugeb, label=below:$C_1$] (c1rs) [below=of c2r]{};
    \node[gauger, label=below:$D_2$] (d2rs) [left=of c1rs]{};
    \node[gaugeb, label=below:$C_2$] (c2s) [left=of d2rs]{};
    \node[gauger, label=below:$D_2$] (d2s) [left=of c2s]{};
    \node[gaugeb, label=below:$C_1$] (c1s) [left=of d2s]{};
    \node[gauger, label=below:$D_1$] (d1s) [left=of c1s]{};

    \draw[-] (c1rs)--(d2rs)--(c2s)--(d2s)--(c1s)--(d1s);

    \node[] (minus) [left=of d1s]{$-$};

    \node[gauger, label=left:$D_1$] (D1t) [below=of d2rs]{};
    \node[gaugeb, label=left:$C_1$] (C1tl) [below=of D1t]{};
    \node[gauger, label=below:$D_1$] (D1l) [below=of C1tl]{};
    \node[gaugeb, label=below:$C_1$] (C1l) [right=of D1l]{};
    \node[gauger, label=below:$D_2$] (D2) [right=of C1l]{};
    \node[gaugeb, label=below:$C_1$] (C1r) [right=of D2]{};
    \node[gauger, label=below:$D_1$] (D1r) [right=of C1r]{};
    \node[gaugeb, label=right:$C_1$] (C1tr) [above=of D2]{};

    \draw[-] (D1t)--(C1tl)--(D1l)--(C1l)--(D2)--(C1r)--(D1r) (D2)--(C1tr);

    \draw[transform canvas={yshift=+1.5pt}] (C1tl)--(C1tr);
    \draw[transform canvas={yshift=-1.5pt}] (C1tl)--(C1tr);
    
    \node[] (topghost) [right=of d1r]{};
    \node[] (bottomghost) [right=of D1r]{};

    \draw[->] (topghost)to [out=-45,in=45,looseness=1](bottomghost);
    
    \end{tikzpicture}
        \caption{}
        \label{fig:minE7SU32}
    \end{subfigure}
    \begin{subfigure}{\textwidth}
        \centering
        \begin{tikzpicture}
            \node[gauger, label=below:$D_1$] (D1l) []{};
            \node[gaugeb, label=below:$C_1$](C1l)[right=of D1l]{};
            \node[gauger, label=below:$D_2$] (D2) [right=of C1l]{};
            \node[gaugeb, label=below:$C_1$] (C1r) [right=of D2]{};
            \node[gauger, label=below:$D_1$] (D1r) [right=of C1r]{};
            \node[gaugeb, label=right:$C_1$] (C1t)[above=of C1l]{};
            \node[gauger, label=left:$D_1$] (D1t) [above=of C1t]{};

            \draw[-] (D1t)--(C1t)--(D1l)--(C1l)--(D2)--(C1r)--(D1r) (D2)--(C1t);
            \end{tikzpicture}
            \caption{}
            \label{fig:minE7SU3Int}
    \end{subfigure}
    \caption{Two alignments of the $\surm(3)$ quotient quiver against the magnetic orthosymplectic quiver for $\overline{min. E_7}$ producing \Quiver{fig:minE7SU31} and \Quiver{fig:minE7SU32}. Their intersection is \Quiver{fig:minE7SU3Int}.}
    \label{fig:minE7SU3}
\end{figure}

The Coulomb branch Hilbert series for \Quiver{fig:minE7SU31} is given in \eqref{HS:E7SU3a}, which identifies $\mathcal C(\text{\Quiver{fig:minE7SU31}})=\overline{\mathcal O}^{A_5}_{(2^3)}$.
\begin{equation}
\hs\left[\mathcal C\left(\text{\Quiver{fig:minE7SU31}}\right)\right]=\frac{(1 + t^2)^3 (1 + 14 t^2 + 72 t^4 + 133 t^6 + 72 t^8 + 14 t^{10} + 
   t^{12})}{(1 - t^2)^{18}}
\label{HS:E7SU3a}
\end{equation} 
For \Quiver{fig:minE7SU32}, the Coulomb branch Hilbert series in \eqref{HS:E7SU3b} similarly identifies $\mathcal C(\text{\Quiver{fig:minE7SU32}})=\overline{\mathcal O}^{A_5}_{(3,1^3)}$.
\begin{equation}
\hs\left[\mathcal C\left(\text{\Quiver{fig:minE7SU32}}\right)\right]=\frac{(1 + t^2) (1 + 16 t^2 + 136 t^4 + 416 t^6 + 626 t^8 + 416 t^{10} + 
   136 t^{12} + 16 t^{14} + t^{16})}{(1 - t^2)^{18}}
\label{HS:E7SU3b}
\end{equation}
Taking the intersection of $\mathcal C(\text{\Quiver{fig:minE7SU32}})$ and $\mathcal C(\text{\Quiver{fig:minE7SU31}})$ hence leads to \eqref{rln:E7SU3}, which is consistent with the result from the unitary quotient quiver subtraction \cite{Hanany:2023tvn}.
\begin{equation}
    \overline{min. E_7}///\surm(3)=\overline{\mathcal O}^{A_5}_{(2^3)}\cup \overline{\mathcal O}^{A_5}_{(3,1^3)}
    \label{rln:E7SU3}
\end{equation}

In particular, the unitary counterparts to \Quiver{fig:minE7SU31} and \Quiver{fig:minE7SU32} are \eqref{eq:A523Unitary} and \eqref{eq:A5313Unitary} respectively.

\begin{equation}
    \begin{tikzpicture}
        \node[gauge, label=below:$1$] (1l) at (0,0){};
        \node[gauge, label=below:$2$] (2l) at (1,0){};
        \node[gauge, label=below:$3$] (3) at (2,0){};
        \node[gauge, label=below:$2$] (2r) at (3,0){};
        \node[gauge, label=below:$1$] (1r) at (4,0){};
        \node[gauge, label=above:$1$] (1t) at (2,1){};

        \draw[-] (1l)--(2l)--(3)--(2r)--(1r);
        \draw[transform canvas={xshift= 1.5pt,yshift=0 pt}](3)--(1t);
        \draw[transform canvas={xshift= -1.5pt,yshift=0 pt}](3)--(1t);
    \end{tikzpicture}
    \label{eq:A523Unitary}
\end{equation}
\begin{equation}
    \begin{tikzpicture}
        \node[gauge, label=below:$1$] (1l) at (0,0){};
        \node[gauge, label=below:$2$] (2l) at (1,0){};
        \node[gauge, label=below:$2$] (2lm) at (2,0){};
        \node[gauge, label=below:$2$] (2rm) at (3,0){};
        \node[gauge, label=below:$2$] (2r) at (4,0){};
        \node[gauge, label=above:$1$] (1t) at (2.5,1){};

        \draw[-] (1l)--(2l)--(2lm)--(2rm)--(2r) (1t)--(2l);
        \draw[transform canvas={xshift=-0.55470019622 pt, yshift=-0.83205029433 pt}] (2r)--(1t);
        \draw[transform canvas={xshift=0.55470019622 pt, yshift=0.83205029433 pt}] (2r)--(1t);
    \end{tikzpicture}
    \label{eq:A5313Unitary}
\end{equation}

Computations of the Higgs branch Hilbert series give further evidence for this claim. The Higgs branch Hilbert series of \Quiver{fig:minE7SU31} can be computed with refinement of the $\sprm(1)$ rotating the two bifundamental hypermultiplets between the $C_1$ and $C_2$ gauge nodes. This gives \begin{equation}
    \hs\left[\mathcal H(\text{\Quiver{fig:minE7SU31}})\right]=\pe\left[[2]_{\sprm(1)}t^2+[2]_{\sprm(1)}t^4+[2]_{\sprm(1)}t^6-t^8-t^{10}-t^{12}\right]
\end{equation}where the Dynkin label is a shorthand for the character of the given representation. This is the same Hilbert series as for $\mathcal S^{A_5}_{\mathcal N,(3^2)}$ which is the Spaltenstein dual to $\overline{\mathcal O}^{A_5}_{(3^2)}$.

The Higgs branch Hilbert series of \Quiver{fig:minE7SU32} may also be computed with refinement of the $\sprm(1)$ symmetry rotating the two bifundamental hypermultiplets between the $C_1$ gauge nodes. This gives \begin{equation}
     \hs\left[\mathcal H(\text{\Quiver{fig:minE7SU32}})\right]=\pe\left[[2]_{\sprm(1)}t^2+2[1]_{\sprm(1)}t^5-t^{10}-t^{12}\right]
\end{equation}The Hilbert series of $\mathcal S^{A_5}_{\mathcal N,(4,1^2)}$, the Spaltenstein dual to $ \overline{\mathcal O}^{A_5}_{(3,1^3)}$, is given as \begin{equation}
    \hs\left[\mathcal S^{A_5}_{\mathcal N,(3,1^3)}\right]=\pe\left[[2]_{\sprm(1)}t^2+\left(q+1/q\right)[1]_{\sprm(1)}t^5-t^{10}-t^{12}\right],
\end{equation}where $q$ is a $\urm(1)$ fugacity. This Hilbert series in the limit $q\rightarrow 1$ matches that of the Higgs branch of \Quiver{fig:minE7SU32}. The Higgs branch dimension increases by $\textrm{rank}\;\surm(3)=2$ after $\surm(3)$ quotient quiver subtraction as expected.

The $\surm(3)$ quotient quiver subtraction in \Figref{fig:minE7SU3} hence provides new constructions for $\overline{\mathcal O}^{A_5}_{(2^3)}$ and $ \overline{\mathcal O}^{A_5}_{(3,1^3)}$ from the orthosymplectic magnetic quivers \Quiver{fig:minE7SU31} and \Quiver{fig:minE7SU32} respectively. It is interesting to note that neither of these quivers preserve the alternating $\sorm-\sprm$ pattern of gauge nodes typical of orthosymplectic theories.
\subsection{$\overline{min. E_8}///\surm(3)$ and $\overline{min. E_8}///\left(\surm(3)\times\surm(3)\right)$}
The subtraction of the $\surm(3)$ orthosymplectic quotient quiver against the orthosymplectic magnetic quiver for $\overline{min. E_8}$ to produce quiver $\mathcal Q_{\ref{fig:minE8A2Sub}}$ is shown in \Figref{fig:minE8A2Sub}. This is confirmed in \cite{Hanany:2022itc}. 

\begin{figure}[H]
    \centering
    \resizebox{\linewidth}{!}{\begin{tikzpicture}[main/.style={draw,circle}]

   \node[gauger, label=below:$D_1$] (d1l) []{};
   \node[gaugeb, label=below:$C_1$] (c1l) [right=of d1l]{};
   \node[gauger, label=below:$D_2$] (d2l) [right=of c1l]{};
   \node[gaugeb, label=below:$C_2$] (c2l) [right=of d2l]{};
   \node[gauger, label=below:$D_3$] (d3l) [right=of c2l]{};
   \node[gaugeb, label=below:$C_3$] (c3l) [right=of d3l]{};
   \node[gauger, label=below:$D_4$] (d4) [right=of c3l]{};
   \node[gaugeb, label=below:$C_3$] (c3r) [right=of d4]{};
   \node[gauger, label=below:$D_3$] (d3r) [right=of c3r]{};
   \node[gaugeb, label=below:$C_2$] (c2r) [right=of d3r]{};
   \node[gauger, label=below:$D_2$] (d2r) [right=of c2r]{};
   \node[gaugeb, label=below:$C_1$] (c1r) [right=of d2r]{};
   \node[gauger, label=below:$D_1$] (d1r) [right=of c1r]{};
    \node[gaugeb, label=left:$C_1$] (c1t) [above=of d4]{};

    \draw[-] (d1l)--(c1l)--(d2l)--(c2l)--(d3l)--(c3l)--(d4)--(c3r)--(d3r)--(c2r)--(d2r)--(c1r)--(d1r);
    \draw[-] (c1t)--(d4);

   \node[gauger, label=below:$D_1$] (d1ls) [below=of d1l]{};
   \node[gaugeb, label=below:$C_1$] (c1ls) [right=of d1ls]{};
   \node[gauger, label=below:$D_2$] (d2ls) [right=of c1ls]{};
   \node[gaugeb, label=below:$C_2$] (c2ls) [right=of d2ls]{};
   \node[gauger, label=below:$D_2$] (d2rs) [right=of c2ls]{};
   \node[gaugeb, label=below:$C_1$] (c1rs) [right=of d2rs]{};
    
    \draw[-] (d1ls)--(c1ls)--(d2ls)--(c2ls)--(d2rs)--(c1rs);

    \node[] (minus) [left=of d1ls]{$-$};

    \node[gaugeb, label=left:$C_1$] (C1t) [below=of c1rs]{};
   \node[gauger, label=below:$D_4$] (D4) [below=of C1t]{};
   \node[gaugeb, label=below:$C_2$] (C2l) [left=of D4]{};
   \node[gauger, label=below:$D_1$] (D1l) [left=of C2l]{};
   \node[gaugeb, label=below:$C_3$] (C3r) [right=of D4]{};
   \node[gauger, label=below:$D_3$] (D3r) [right=of C3r]{};
   \node[gaugeb, label=below:$C_2$] (C2r) [right=of D3r]{};
   \node[gauger, label=below:$D_2$] (D2r) [right=of C2r]{};
   \node[gaugeb, label=below:$C_1$] (C1r) [right=of D2r]{};
   \node[gauger, label=below:$D_1$] (D1r) [right=of C1r]{};
    \node[gaugeb, label=right:$C_1$] (C1tr) [above right=of D4]{};

    \draw[-] (C1tr)--(D4)--(C1t);
    \draw[-] (D1r)--(C1r)--(D2r)--(C2r)--(D3r)--(C3r)--(D4)--(C2l)--(D1l);

    \node[] (topghost) [below=of d1r]{};
    \node[] (bottomghost) [right=of D1r]{};

    \draw[->] (topghost)to [out=0,in=45,looseness=1](bottomghost);
        
    \end{tikzpicture}}
    \caption{Subtraction of the $\surm(3)$ orthosymplectic quotient quiver from the orthosymplectic magnetic quiver for $\overline{min. E_8}$.}
    \label{fig:minE8A2Sub}
\end{figure}

As mentioned in Section \ref{sec:E8///SU2SU2}, since the orthosymplectic magnetic quiver for $\overline{min. E_8}$ contains two long legs, a further $\surm(2)$ or $\surm(3)$ quotient quiver subtraction can be done.

For the case of doing a further $\surm(3)$ quotient quiver subtraction, the result is shown in \Figref{fig:minE8A2Sub2} producing \Quiver{fig:minE8A2Sub2}.

\begin{figure}[H]
    \centering
    \begin{tikzpicture}[main/.style={draw,circle}]
    \node[gaugeb, label=left:$C_1$] (C1t) []{};
   \node[gauger, label=below:$D_4$] (D4) [below=of C1t]{};
   \node[gaugeb, label=below:$C_2$] (C2l) [left=of D4]{};
   \node[gauger, label=below:$D_1$] (D1l) [left=of C2l]{};
   \node[gaugeb, label=below:$C_3$] (C3r) [right=of D4]{};
   \node[gauger, label=below:$D_3$] (D3r) [right=of C3r]{};
   \node[gaugeb, label=below:$C_2$] (C2r) [right=of D3r]{};
   \node[gauger, label=below:$D_2$] (D2r) [right=of C2r]{};
   \node[gaugeb, label=below:$C_1$] (C1r) [right=of D2r]{};
   \node[gauger, label=below:$D_1$] (D1r) [right=of C1r]{};
    \node[gaugeb, label=right:$C_1$] (C1tr) [above right=of D4]{};

    \draw[-] (C1tr)--(D4)--(C1t);
    \draw[-] (D1r)--(C1r)--(D2r)--(C2r)--(D3r)--(C3r)--(D4)--(C2l)--(D1l);

   \node[gauger, label=below:$D_1$] (d1ls) [below=of D1r]{};
   \node[gaugeb, label=below:$C_1$] (c1ls) [left=of d1ls]{};
   \node[gauger, label=below:$D_2$] (d2ls) [left=of c1ls]{};
   \node[gaugeb, label=below:$C_2$] (c2ls) [left=of d2ls]{};
   \node[gauger, label=below:$D_2$] (d2rs) [left=of c2ls]{};
   \node[gaugeb, label=below:$C_1$] (c1rs) [left=of d2rs]{};
    
    \draw[-] (d1ls)--(c1ls)--(d2ls)--(c2ls)--(d2rs)--(c1rs);

    \node[] (minus) [right=of d1ls]{$-$};

    \node[gaugeb, label=above:$C_1$] (c1tm) [below left=of c1rs]{};
   \node[gauger, label=below:$D_4$] (d4)[below=of c1tm]{}; 
   \node[gaugeb, label=below:$C_2$] (c2l) [left=of d4]{};
   \node[gauger, label=below:$D_1$] (d1l) [left=of c2l]{};
   \node[gaugeb, label=below:$C_2$] (c2r) [right=of d4]{};
   \node[gauger, label=below:$D_1$] (d1r) [right=of c2r]{};
    \node[gaugeb, label=left:$C_1$] (c1tl) [above left=of d4]{};
    \node[gaugeb, label=right:$C_1$] (c1tr) [above right=of d4]{};

    \draw[-] (d1l)--(c2l)--(d4)--(c2r)--(d1r);
    \draw[-] (c1tl)--(d4)--(c1tm);
    \draw[-] (c1tr)--(d4);
    
    \node[] (topghost) [left=of D1l]{};
    \node[] (bottomghost) [left=of d1l]{};

    \draw[->] (topghost)to [out=-135,in=+135,looseness=1](bottomghost);
        
    \end{tikzpicture}
    \caption{Subtraction of the $\surm(3)$ orthosymplectic quotient quiver from the orthosymplectic magnetic quiver for $\overline{min. E_8}///\surm(3)$. This gives $\overline{min. E_8}///\surm(3)\times \surm(3)$ as a moduli space of dressed monopole operators.}
    \label{fig:minE8A2Sub2}
\end{figure}

The Coulomb branch Hilbert series for both \Quiver{fig:minE8A2Sub} and \Quiver{fig:minE8A2Sub2} are given in \cite{Hanany:2022itc} with further discussion of the appearance of these quivers as magnetic quivers for six dimensional theories.

Here the Higgs branch Hilbert series of \Quiver{fig:minE8A2Sub} is presented. One subtlety to address in the magnetic quivers is the appearance of $\bigwedge^2\left([1]_{\sprm(1)}\right)$ hypermultiplets on the $C_1$ gauge nodes in the bouquet. In the Type IIA brane system these hypermultiplets come from D4 branes ending on the same NS5. These hypermultiplets are typically suppressed since they do not contribute to the monopole formula and the Coulomb branch Hilbert series. In the Higgs branch Hilbert series each $\bigwedge^2\left([1]_{\sprm(1)}\right)$ contributes a factor of $\mathbb H$. These will be ignored in the following computations.

The Higgs branch Hilbert series of \Quiver{fig:minE8A2Sub} is computed as \begin{equation}
    \hs\left[\mathcal H\left(\text{\Quiver{fig:minE8A2Sub}}\right)\right]=\left(1 + 3 t^{12} + 2 t^{14} + 2 t^{24} + 
 3 t^{26} + t^{38}\right)\pe\left[2 t^4 + t^6 + t^8 + t^{10} + t^{12}\right]\label{HS:minE8A2Sub}
\end{equation}This Hilbert series does not elucidate the moduli space. However, it is known that the Coulomb branch of \Quiver{fig:minE8A2Sub} is the double cover of $\overline{\mathcal O}^{E_6}_{[0,0,0,0,0,2]}$ (Bala-Carter label $A_2$) \cite{Hanany:2022itc}. One may expect that the Higgs branch is related to the dual slice to this orbit which is $\mathcal S^{E_6}_{\mathcal N,[2,0,2,0,2,0]}$ (Bala-Carter label $E_6(a_3)$) whose Hilbert series is \begin{equation}
    \hs\left[\mathcal S^{E_6}_{\mathcal N,[2,0,2,0,2,0]}\right]=\pe\left[2t^4+3t^6+2t^8+t^{10}+t^{12}-t^{16}-t^{18}-t^{24}\right]\label{HS:E6SliceE6a3}
\end{equation}
The volumes of the moduli spaces may be compared as \begin{equation}
    \frac{\mathrm{Vol}\left[ \mathcal H\left(\text{\Quiver{fig:minE8A2Sub}}\right) \right]}{\mathrm{Vol}\left[\mathcal S^{E_6}_{\mathcal N,[2,0,2,0,2,0]} \right]}=\lim_{t\rightarrow 1}\frac{ \hs\left[\mathcal H\left(\text{\Quiver{fig:minE8A2Sub}}\right)\right]}{ \hs\left[\mathcal S^{E_6}_{\mathcal N,[2,0,2,0,2,0]}\right]}=\frac{1}{2}
\end{equation}which suggests that $\mathcal H\left(\text{\Quiver{fig:minE8A2Sub}}\right)=\mathcal S^{E_6}_{\mathcal N,[2,0,2,0,2,0]}/\mathbb Z_2$. This suggests that the Hilbert series \eqref{HS:minE8A2Sub} can be written as the $\mathbb Z_2$ group average of \eqref{HS:E6SliceE6a3}. The even contribution to this average is \eqref{HS:E6SliceE6a3} itself whereas the odd contribution to the group average is \begin{equation}
     \hs\left[\mathcal S^{E_6}_{\mathcal N,[2,0,2,0,2,0]}\right]^-=\pe\left[2 t^4 - t^6 + t^{10} + 3 t^{12} + t^{18} - t^{24} - t^{36}\right]
\end{equation} The above Hilbert series has negative terms in the expansion which is allowed since it should not be thought of as a standalone object.

The two $C_1$ gauge nodes in the bouquet of \Quiver{fig:minE8A2Sub} may be discretely gauged and gives the following quiver \begin{equation}
    \begin{tikzpicture}
        \node[gaugeb, label=left:$C_2$] (C1t) []{};
   \node[gauger, label=below:$D_4$] (D4) [below=of C1t]{};
   \node[gaugeb, label=below:$C_2$] (C2l) [left=of D4]{};
   \node[gauger, label=below:$D_1$] (D1l) [left=of C2l]{};
   \node[gaugeb, label=below:$C_3$] (C3r) [right=of D4]{};
   \node[gauger, label=below:$D_3$] (D3r) [right=of C3r]{};
   \node[gaugeb, label=below:$C_2$] (C2r) [right=of D3r]{};
   \node[gauger, label=below:$D_2$] (D2r) [right=of C2r]{};
   \node[gaugeb, label=below:$C_1$] (C1r) [right=of D2r]{};
   \node[gauger, label=below:$D_1$] (D1r) [right=of C1r]{};

    \draw[-] (D4)--(C1t);
    \draw[-] (D1r)--(C1r)--(D2r)--(C2r)--(D3r)--(C3r)--(D4)--(C2l)--(D1l);
    \draw[-] (C1t) to[out=45,in=135,looseness=8] node[pos=0.5,above]{$\bigwedge^2$} (C1t);
    \end{tikzpicture}\label{eq:minE8A2SubZ2}
\end{equation}whose Coulomb branch is $\overline{\mathcal O}^{E_6}_{[0,0,0,0,0,2]}$. The Higgs branch is of dimension 4 so it is not the $\mathcal S^{E_6}_{\mathcal N,[2,0,2,0,2,0]}$ which is of dimension 3. Instead, the Higgs branch should be related to the $\mathcal S^{E_6}_{\mathcal N,[2,1,0,1,2,1]}$ (Bala-Carter label $A_5$) which is a slice from a non-special orbit whose Hasse diagram contains the special piece $\mathcal S^{E_6}_{\mathcal N,[2,0,2,0,2,0]}$ on top.

The Higgs branch Hilbert series is presented as \begin{align}
    \hs\left[\mathcal H\left(\eqref{eq:minE8A2SubZ2}\right)\right]&=\left(1+t^{12}+[1]_{\sprm(1)}t^{13}-[1]_{\sprm(1)}t^{27}-t^{28}-t^{40}\right)\nonumber\\&\times\pe\left[[2]_{\sprm(1)}t^2+[1]_{\sprm(1)}t^5+t^8+[1]_{\sprm(1)}t^{11}+[2]_{\sprm(1)}t^{14}-t^{16}-t^{24}\right]
\end{align}where the Dynkin labels are shorthand for the character of the $\sprm(1)$ representation. The Hilbert series for $\mathcal S^{E_6}_{\mathcal N,[2,0,2,0,2,0]}$ is computed as \begin{equation}
    \hs\left[\mathcal S^{E_6}_{\mathcal N,[2,0,2,0,2,0]}\right]=\pe\left[[2]_{\sprm(1)}t^2+[1]_{\sprm(1)}t^5+t^6+[1]_{\sprm(1)}t^7+t^8+[1]_{\sprm(1)}t^{11}-t^{16}-t^{18}-t^{24}\right]
\end{equation} Comparison of the volumes of the moduli space gives \begin{equation}
     \frac{\mathrm{Vol}\left[ \mathcal H\left(\eqref{eq:minE8A2SubZ2}\right) \right]}{\mathrm{Vol}\left[\mathcal S^{E_6}_{\mathcal N,[2,1,0,1,2,1]} \right]}=\lim_{t\rightarrow 1,\;x\rightarrow 1}\frac{ \hs\left[\mathcal H\left(\text{\Quiver{fig:minE8A2Sub}}\right)\right]}{ \hs\left[\mathcal S^{E_6}_{\mathcal N,[2,1,0,1,2,1]}\right]}=\frac{1}{2}
\end{equation}where $x$ is a fugacity for the $\sprm(1)$. This suggests that $ \mathcal H\left(\eqref{eq:minE8A2SubZ2}\right) =\mathcal S^{E_6}_{\mathcal N,[2,1,0,1,2,1]}/\mathbb Z_2$. Although the $\mathbb Z_2$ group average may in principle be performed through the $\hwg$ \cite{Bourget:2020bxh}, the particular form of the $\hwg$ is cumbersome to present.

Instead the odd contribution to the $\hs$ is presented as \begin{equation}
    \hs\left[\mathcal S^{E_6}_{\mathcal N,[2,0,2,0,2,0]}\right]^-=\pe\left[\begin{aligned}&[2]_{\sprm(1)} t^2 + [1]_{\sprm(1)} t^5-t^6-[1]_{\sprm(1)}t^7+t^8+[1]_{\sprm(1)}t^{11}+t^{12}\\&+\left([2]_{\sprm(1)}-1\right)t^{14}-t^{16}+t^{18}-t^{24}-t^{36}\end{aligned}\right]
\end{equation}




\subsection{$\left(\overline{min. F_4}\times\overline{min. F_4}\right)///\surm(3)$}
The next example of an $\surm(3)$ quotient quiver subtraction is on the $\overline{min. F_4}\times\overline{min. F_4}$ theory of \cite{Sperling:2021fcf}. This quiver admits an $\surm(3)$ quotient quiver subtraction shown in \Figref{fig:minF4SqSU3} resulting in quiver \Quiver{fig:minF4SqSU3}. The resulting unrefined Hilbert series of \Quiver{fig:minF4SqSU3} is \eqref{HS:F4F4SU3}. 
\begin{figure}[H]
    \centering
    \begin{tikzpicture}
        \node[gauger, label=below:$D_1$] (d1) []{};
        \node[gaugeb, label=below:$C_1$] (c1l) [right=of d1]{};
        \node[gauger, label=below:$D_2$] (d2l) [right=of c1l]{};
        \node[gaugeb, label=below:$C_2$] (c2) [right=of d2l]{};
        \node[gauger, label=below:$D_3$] (d3) [right=of c2]{};
        \node[gaugeb, label=below:$C_3$] (c3) [right=of d3]{};
        \node[gauger, label=right:$D_2$] (d2t) [above right=of c3]{};
        \node[gauger, label=right:$D_2$] (d2b) [below right=of c3]{};

        \draw[-] (d1)--(c1l)--(d2l)--(c2)--(d3)--(c3)--(d2t) (d2b)--(c3);

        \node[gauger, label=below:$D_1$] (d1ls) [below=of d1]{};
       \node[gaugeb, label=below:$C_1$] (c1ls) [right=of d1ls]{};
       \node[gauger, label=below:$D_2$] (d2ls) [right=of c1ls]{};
       \node[gaugeb, label=below:$C_2$] (c2ls) [right=of d2ls]{};
       \node[gauger, label=below:$D_2$] (d2rs) [right=of c2ls]{};
       \node[gaugeb, label=below:$C_1$] (c1rs) [right=of d2rs]{};
        
        \draw[-] (d1ls)--(c1ls)--(d2ls)--(c2ls)--(d2rs)--(c1rs);

        \node[gauger, label=above:$D_2$] (D2t) [below=of d2b]{};
        \node[gaugeb, label=right:$C_1$] (C1r) [below right=of D2t]{};
        \node[gauger, label=below:$D_2$] (D2b) [below left=of C1r]{};
        \node[gaugeb, label=below:$C_2$] (C2) [above left=of D2b]{};
        \node[gauger, label=below:$D_1$] (D1) [left=of C2]{};

        \draw[-] (D1)--(C2)--(D2b)--(C1r)--(D2t)--(C2);

         \node[](minus) [left=of d1ls]{$-$};

        \node[] (ghost) [right=of c3]{};
        \node[] (topghost) [right=of ghost]{};
        \node[] (bottomghost) [right=of C1r]{};

        \draw[->] (topghost)to [out=-45,in=45,looseness=1](bottomghost);

    \end{tikzpicture}
    \caption{Subtraction of the $\surm(3)$ quotient quiver from the orthosymplectic magnetic quiver for $\overline{min. F_4}\times\overline{min. F_4}$.}
    \label{fig:minF4SqSU3}
\end{figure}
\begin{align}
    \hs[\mathcal C(\text{\Quiver{fig:minF4SqSU3}})]&=\frac{\left(\begin{aligned}1 &+ 8 t^2 + 99 t^4 + 479 t^6 + 2095 t^8 + 5355 t^{10} \\&+ 11577 t^{12} + 
   16622 t^{14} + 20072 t^{16} +\cdots + 
   t^{32}\end{aligned}\right)}{(1 - t^2)^8 (1 - t^4)^8}\label{HS:F4F4SU3}\\
   \pl\left[\hs[\mathcal C(\text{\Quiver{fig:minF4SqSU3}})]\right]&=16 t^2 + 71 t^4 - 145 t^6 - 1359 t^8 + 6102 t^{10}+O(t^{12})\label{PL:F4F4SU3}
\end{align}
This leads to the conclusion\begin{equation}
    \left(\overline{min. F_4}\times\overline{min. F_4}\right)///\surm(3)=\mathcal C(\text{\Quiver{fig:minF4SqSU3}}),\label{eq:minF4SqSU3}
\end{equation}which can be checked using Weyl integration with the following embedding of $F_4\times F_4\hookleftarrow \surm(3)\times\surm(3)\times\surm(3)$ which decomposes the fundamental as \begin{equation}
    \left(\mu_1+\mu_1'\right)_{F_4\times F_4}\hookleftarrow \mu_1 \mu_2 + \nu_1\nu_2+ (\mu_2^2+\nu_2^2) \rho_1 + (\mu_1^2+\nu_1^2) \rho_2 + 2\rho_1 \rho_2
\end{equation}where the $\mu_i,\nu_i,$ and $\rho_i$ are all highest weight fugacities for $\surm(3)$. The $\surm(3)$ corresponding to the $\rho_i$ is the one by which the hyper-Kähler quotient is taken.

There is an additional way to realise the above result \eqref{eq:minF4SqSU3} using unitary quivers. Applying an $\surm(3)$ chain polymerisation \cite{Hanany:2024fqf} on two $F_4^{(1)}$ quivers gives \eqref{eq:minF4SqSU3Unitary} whose unrefined Coulomb branch Hilbert series is also found to be \eqref{HS:F4F4SU3}.

\begin{equation}
    \begin{tikzpicture}
        \node[gauge, label=below:$3$] (3R) []{};
        \node[gauge, label=below:$2$] (2lR) [left=of 3R]{};
        \node[gauge, label=below:$2$] (2rR) [right=of 3R]{};
        \node[gauge, label=below:$1$] (1lR) [left=of 2lR]{};
        \node[gauge, label=below:$1$] (1rR) [right=of 2rR]{};

        \draw[-] (1lR)--(2lR) (1rR)--(2rR);
        \draw [line width=1pt, double distance=3pt,
             arrows = {-Latex[length=0pt 2 0]}] (3R) -- (2lR);
        \draw [line width=1pt, double distance=3pt,
             arrows = {-Latex[length=0pt 2 0]}] (3R) -- (2rR);
    \end{tikzpicture}
    \label{eq:minF4SqSU3Unitary}
\end{equation}

The quiver \eqref{eq:minF4SqSU3Unitary} contains non-simply laced edges and so it is unknown how to compute the Higgs branch Hilbert series. However the quiver \Quiver{fig:minF4SqSU3} is simply laced and it is straight forward to compute the Higgs branch Hilbert series\begin{align}
    \hs\left[\mathcal H(\text{\Quiver{fig:minF4SqSU3}})\right]&=\frac{1 - t^2 + 2 t^6 - t^{10} + t^{12}}{(1 - t^2)  (1 - t^4)^2  (1 - t^6)}\\\pl\left[\hs\left[\mathcal H(\text{\Quiver{fig:minF4SqSU3}})\right]\right]&=2 t^4 + 3 t^6 + 2 t^8 + t^{10} - t^{12} - 2 t^{14} - 3 t^{16} - 2 t^{18} + t^{20}+O(t^{21})
\end{align}where the $t^2$ coefficient is zero indicating that the global symmetry of the moduli space is not continuous.

Although it is not known how to compute the Higgs branch Hilbert series of \eqref{eq:minF4SqSU3Unitary}, one may conjecture that it has the same Hilbert series above.

\subsection{$\left(\overline{min. E_6}\times\overline{min. E_6}\right)///\surm(3)$}
Another example of the $\surm(3)$ quotient quiver subtraction is provided by the $\overline{min. E_6}\times\overline{min. E_6}$ theory of \cite{Sperling:2021fcf}, as shown in \Figref{fig:minE6SqSU3Sub}. In this case, the resulting quiver \Quiver{fig:minE6SqSU3Sub} is star-shaped. The unrefined Coulomb branch Hilbert series is evaluated to be 
\begin{equation}
    \hs\left[\mathcal C(\text{\Quiver{fig:minE6SqSU3Sub}})\right]=\frac{\left(\begin{aligned}1 &+ 18 t^2 + 316 t^4 + 3261 t^6 + 26534 t^8 + 158590 t^{10} + 
  756809 t^{12} \\&+ 2858318 t^{14} + 8853594 t^{16} + 22509101 t^{18} + 
  47864400 t^{20} + 85223276 t^{22} \\&+ 128451799 t^{24} + 163786892 t^{26} + 
  177789766 t^{28} + \cdots + t^{56}\end{aligned}\right)}{(1-t^2)^{14}(1-t^4)^{14}}.
\end{equation} This moduli space is of dimension 14 and has $\surm(3)^4$ global symmetry.

\begin{figure}[H]
    \centering
    \begin{tikzpicture}
        \node[gauger, label=below:$D_1$] (d1l) []{};
        \node[gaugeb, label=below:$C_1$] (c1l) [right=of d1l]{};
        \node[gauger, label=below:$D_2$] (d2l) [right=of c1l]{};
        \node[gaugeb, label=below:$C_2$] (c2l) [right=of d2l]{};
        \node[gauger, label=below:$D_3$] (d3l) [right=of c2l]{};
        \node[gaugeb, label=below:$C_3$] (c3l) [right=of d3l]{};
        \node[gauger, label=below:$D_4$] (d4) [right=of c3l]{};
        \node[gaugeb, label=below:$C_2$] (c2r) [right=of d4]{};
        \node[gauger, label=below:$D_1$] (d1r) [right=of c2r]{};
        \node[gaugeb, label=right:$C_2$] (c2t) [above=of d4]{};
        \node[gauger, label=right:$D_1$] (d1t) [above=of c2t]{};

        \draw[-] (d1l)--(c1l)--(d2l)--(c2l)--(d3l)--(c3l)--(d4)--(c2r)--(d1r) (d4)--(c2t)--(d1t);

        \node[gaugeb, label=below:$C_1$] (c1rs) [below=of c3l]{};
        \node[gauger, label=below:$D_2$] (d2rs) [left=of c1rs]{};
        \node[gaugeb, label=below:$C_2$] (c2s) [left=of d2rs]{};
        \node[gauger, label=below:$D_2$] (d2s) [left=of c2s]{};
        \node[gaugeb, label=below:$C_1$] (c1s) [left=of d2s]{};
        \node[gauger, label=below:$D_1$] (d1s) [left=of c1s]{};

        \draw[-] (c1rs)--(d2rs)--(c2s)--(d2s)--(c1s)--(d1s);

        \node[] (minus) [left=of d1s]{$-$};

        \node[gauger, label=right:$D_1$] (D1t) [below=of d4]{};
        \node[gaugeb, label=right:$C_2$] (C2t) [below=of D1t]{};
        \node[gauger, label=below:$D_4$] (D4) [below=of C2t]{};
        \node[gaugeb, label=below:$C_2$] (C2l) [left=of D4]{};
        \node[gauger, label=below:$D_1$] (D1l) [left=of C2l]{};
        \node[gaugeb, label=below:$C_2$] (C2r) [right=of D4]{};
        \node[gauger, label=below:$D_1$] (D1r) [right=of C2r]{};
        \node[gaugeb, label=right:$C_1$] (D1tm) [above right=of D4]{};

        \draw[-] (D1t)--(C2t)--(D4)--(C2l)--(D1l) (D1tm)--(D4)--(C2r)--(D1r);

        \node[] (topghost) [right=of d1r]{};
        \node[] (bottomghost) [right=of D1r]{};

        \draw[->] (topghost)to [out=-45,in=45,looseness=1](bottomghost);
    \end{tikzpicture}
    \caption{Subtraction of the $\surm(3)$ quotient quiver from the orthosymplectic quiver for $\overline{min. E_6}\times\overline{min. E_6}$.}
    \label{fig:minE6SqSU3Sub}
\end{figure}

The conclusion is that \begin{equation}
    \left(\overline{min. E_6}\times \overline{min. E_6}\right)///\surm(3)=\mathcal C(\text{\Quiver{fig:minE6SqSU3Sub}}).\label{eq:minE6SqSU3Sub}
\end{equation}

This can be checked with the embedding of $E_6\times E_6\hookleftarrow \surm(3)^5$ which decomposes the fundamental as \begin{equation}
    \left(\mu_1+\mu_5'\right)_{E_6}\hookleftarrow \mu_1\nu_2+\mu_2\rho_2+\nu_1\rho_1+\sigma_1\rho_1+\gamma_2\rho_2+\sigma_2\gamma_1
\end{equation}where the $\mu_i, \nu_i,\rho_i, \sigma_i,$ and $\gamma_i$ are highest weight fugacities for $\surm(3)$. The hyper-Kähler quotient is computed w.r.t. the $\surm(3)$ corresponding to the $\rho_i$.

There is an additional way the result \eqref{eq:minE6SqSU3Sub} can be obtained from unitary magnetic quivers. This is through the $\surm(3)$ chain polymerisation of two $E_6^{(1)}$ \cite{Benini:2010uu} producing \eqref{eq:minE6SqSU3Unitary}, which is also known as the $T_4[\surm(3)]$ theory.

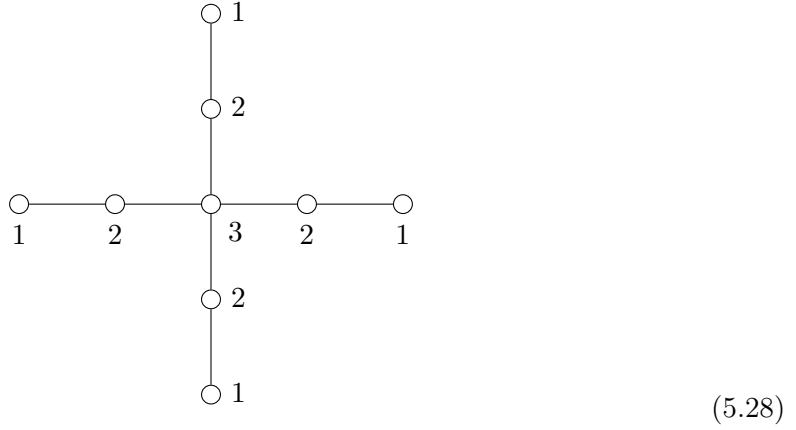
\begin{equation}
    \begin{tikzpicture}
        \node[gauge, label=right:$1$] (1rt) []{};
        \node[gauge, label=right:$2$] (2rt) [below=of 1rt]{};
        \node[gauge, label=below right:$3$] (3r)[below=of 2rt]{};
        \node[gauge, label=below:$2$] (2rr) [right=of 3r]{};
        \node[gauge, label=below:$1$] (1rr) [right=of 2rr]{};
        \node[gauge, label=below:$2$] (2rl) [left=of 3r]{};
        \node[gauge, label=below:$1$] (1rl) [left=of 2rl]{};
        \node[gauge, label=right:$2$] (2rb) [below=of 3r]{};
        \node[gauge, label=right:$1$] (1rb) [below=of 2rb]{};

        \draw[-] (1rt)--(2rt)--(3r)--(2rr)--(1rr) (1rl)--(2rl)--(3r)--(2rb)--(1rb);
    \end{tikzpicture}\label{eq:minE6SqSU3Unitary}
\end{equation}

The quiver \eqref{eq:minE6SqSU3Unitary} is the unitary counterpart to \Quiver{fig:minE6SqSU3Sub}. Interestingly, the unitary magnetic quiver \eqref{eq:minE6SqSU3Unitary} has an $S_4$ permutation symmetry but the orthosymplectic magnetic quiver \Quiver{fig:minE6SqSU3Sub} only manifests $S_3\subset S_4$.

\subsection{$\overline{min. E_6}\times\overline{min. F_4}///\surm(3)$}
\begin{figure}[h!]
    \centering
    \begin{tikzpicture}
        \node[gauger, label=below:$D_1$] (d1l) []{};
        \node[gaugeb, label=below:$C_1$] (c1l) [right=of d1l]{};
        \node[gauger, label=below:$D_2$] (d2l) [right=of c1l]{};
        \node[gaugeb, label=below:$C_2$] (c2l) [right=of d2l]{};
        \node[gauger, label=below:$D_3$] (d3l) [right=of c2l]{};
        \node[gaugeb, label=below:$C_3$] (c3l) [right=of d3l]{};
        \node[gauger, label=below:$D_4$] (d4) [right=of c3l]{};
        \node[gaugeb, label=below:$C_2$] (c2r) [right=of d4]{};
        \node[gauger, label=below:$D_1$] (d1r) [right=of c2r]{};

        \draw[-] (d1l)--(c1l)--(d2l)--(c2l)--(d3l)--(c3l)--(d4) (c2r)--(d1r);

         \draw [line width=1pt, double distance=3pt,
             arrows = {-Latex[length=0pt 2 0]}] (d4) --  (c2r);

        \node[gaugeb, label=below:$C_1$] (c1rs) [below=of c3l]{};
        \node[gauger, label=below:$D_2$] (d2rs) [left=of c1rs]{};
        \node[gaugeb, label=below:$C_2$] (c2s) [left=of d2rs]{};
        \node[gauger, label=below:$D_2$] (d2s) [left=of c2s]{};
        \node[gaugeb, label=below:$C_1$] (c1s) [left=of d2s]{};
        \node[gauger, label=below:$D_1$] (d1s) [left=of c1s]{};

        \draw[-] (c1rs)--(d2rs)--(c2s)--(d2s)--(c1s)--(d1s);

        \node[] (minus) [left=of d1s]{$-$};

        \node[gaugeb, label=below:$C_2$] (C2l) [below=of c1rs]{};
        \node[gauger, label=below:$D_4$] (D4) [right=of C2l]{};
        \node[gauger, label=below:$D_1$] (D1l) [left=of C2l]{};
        \node[gaugeb, label=below:$C_2$] (C2r) [right=of D4]{};
        \node[gauger, label=below:$D_1$] (D1r) [right=of C2r]{};
        \node[gaugeb, label=right:$C_1$] (D1tm) [above =of D4]{};

        \draw[-] (D4)--(C2l)--(D1l) (D1tm)--(D4) (C2r)--(D1r);

          \draw [line width=1pt, double distance=3pt,
             arrows = {-Latex[length=0pt 2 0]}] (D4) --  (C2r);

        \node[] (topghost) [right=of d1r]{};
        \node[] (bottomghost) [right=of D1r]{};

        \draw[->] (topghost)to [out=-45,in=45,looseness=1](bottomghost);
    \end{tikzpicture}
    \caption{Subtraction of the $\surm(3)$ quotient quiver from the orthosymplectic quiver for $\overline{min. E_6}\times\overline{min. F_4}$.}
    \label{fig:minE6F4SU3Sub}
\end{figure}
The $\surm(3)$ orthosymplectic quiver may be subtracted from the magnetic orthosymplectic quiver for $\overline{min. E_6}\times\overline{min. F_4}$ \cite{Sperling:2021fcf}, as shown in \Figref{fig:minE6F4SU3Sub} to produce \Quiver{fig:minE6F4SU3Sub}. The presence of the non-simply laced edge does not affect the method of $\surm(3)$ quotient quiver subtraction.

This quiver \Quiver{fig:minE6F4SU3Sub} can also be derived from folding \Quiver{fig:minE6SqSU3Sub}. It is important to note that orthosymplectic quotient quiver subtraction commutes with the action of folding, since the legs that are folded do not participate in subtraction.

The Coulomb branch Hilbert series is evaluated as\begin{equation}
    \hs\left[\mathcal C(\text{\Quiver{fig:minE6F4SU3Sub}})\right]= 
    \frac{\left(\begin{aligned}1 &+ 13 t^2 + 186 t^4 + 1418 t^6 + 
   8627 t^8 + 36761 t^{10} + 124228 t^{12}+321308 t^{14} \\ &+ 672441 t^{16} + 
   1118860 t^{18} + 1526957 t^{20} + 1682240 t^{22} + \cdots + t^{44}\end{aligned}\right)}{(1-t^2)^{11}(1-t^4)^{11}}.
\end{equation} This moduli space is of dimension 11 and has $\surm(3)^3$ global symmetry.

The conclusion is that \begin{equation}
    \left(\overline{min. E_6}\times \overline{min. F_4}\right)///\surm(3)=\mathcal C(\text{\Quiver{fig:minE6F4SU3Sub}}).
\end{equation}This can be checked using the following embedding of $E_6\times F_4\hookleftarrow\surm(3)^4$ which decomposes the fundamental as \begin{equation}
    ({\mu_1})_{E_6}+({\mu_1'})_{F_4}\rightarrow \mu_2\nu_2+\mu_1\rho_1+\nu_1\rho_2+\rho_1\rho_2+\sigma_1\rho_1+\sigma_2\rho_2
\end{equation}where the $\mu_i, \nu_i,\rho_i,$ and $\sigma_i$ are highest weight fugacities for $\surm(3)$. The hyper-Kähler quotient is computed w.r.t. the $\surm(3)$ corresponding to the $\rho_i$.

The same result can be computed from unitary magnetic quivers. This is seen diagrammatically from the $\surm(3)$ chain polymerisation of the $E_6^{(1)}$ and $F_4^{(1)}$ quivers producing \eqref{eq:minE6F4SU3Unitary}.
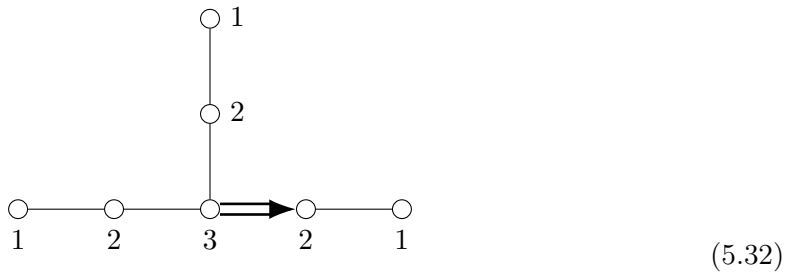
\begin{equation}
    \begin{tikzpicture}
        \node[gauge, label=right:$1$] (1rt) []{};
        \node[gauge, label=right:$2$] (2rt) [below=of 1rt]{};
        \node[gauge, label=below:$3$] (3r)[below=of 2rt]{};
        \node[gauge, label=below:$2$] (2rr) [right=of 3r]{};
        \node[gauge, label=below:$1$] (1rr) [right=of 2rr]{};
        \node[gauge, label=below:$2$] (2rl) [left=of 3r]{};
        \node[gauge, label=below:$1$] (1rl) [left=of 2rl]{};
    
        \draw[-] (1rt)--(2rt)--(3r) (2rr)--(1rr) (1rl)--(2rl)--(3r);

         \draw [line width=1pt, double distance=3pt,
             arrows = {-Latex[length=0pt 2 0]}] (3r) --  (2rr);
    \end{tikzpicture}\label{eq:minE6F4SU3Unitary}
\end{equation}

In fact the quiver \eqref{eq:minE6F4SU3Unitary} is the unitary counterpart to \Quiver{fig:minE6F4SU3Sub}. Although these quivers are counterparts in the sense that their Coulomb branches are the same, it is not known how to compute the Higgs branch Hilbert series of either quiver. Interestingly, the $S_2$ outer automorphism acting on \eqref{eq:minE6F4SU3Unitary} to permute its two legs has no obvious analogue in \Quiver{fig:minE6F4SU3Sub}.

\subsection{$\overline{min.E_8}///\left(\surm(2)\times\surm(3)\right)$}
\label{sec:minE8A1A2Quot}
As discussed in Section \ref{sec:E8///SU2SU2}, the orthosymplectic magnetic quiver for $\overline{min. E_8}$ can admit either an $\surm(2)$ or $\surm(3)$ quotient quiver subtraction on each long leg. The case where the $\surm(2)$ quotient quiver was subtracted from both legs was studied in Section \ref{sec:E8///SU2SU2}. The case where the $\surm(3)$ quotient quiver was subtracted from both legs was studied in \cite{Hanany:2022itc}. Now the remaining option is studied which is subtracting the $\surm(2)$ quotient quiver from one leg and the $\surm(3)$ quotient quiver on the other.

This is shown in \Figref{fig:minE8SU2SU3} resulting in $\mathcal{Q}_{\ref{fig:minE8SU2SU3}}$.  Much like the $\overline{min.E_8}///\left(\surm(2)\times\surm(2)\right)$ case in Section \ref{sec:E8///SU2SU2}, which does not admit an exact Hilbert series calculation owing to its high rank; the perturbative Hilbert series will shed light on the generators and relations of the moduli space.
\begin{figure}[H]
    \centering
    \resizebox{\textwidth}{!}{\begin{tikzpicture}[main/.style={draw,circle}]

        \node[gauger, label=below:$D_1$] (d1l) []{};
        \node[gaugeb, label=below:$C_1$] (c1l) [right=of d1l]{};
        \node[gauger, label=below:$D_2$] (d2l) [right=of c1l]{};
        \node[gaugeb, label=below:$C_2$] (c2l) [right=of d2l]{};
        \node[gauger, label=below:$D_3$] (d3l) [right=of c2l]{};
        \node[gaugeb, label=below:$C_3$] (c3l) [right=of d3l]{};
        \node[gauger, label=below:$D_4$] (d4) [right=of c3l]{};
        \node[gaugeb, label=below:$C_3$] (c3r) [right=of d4]{};
        \node[gauger, label=below:$D_3$] (d3r) [right=of c3r]{};
        \node[gaugeb, label=below:$C_2$] (c2r) [right=of d3r]{};
        \node[gauger, label=below:$D_2$] (d2r) [right=of c2r]{};
        \node[gaugeb, label=below:$C_1$] (c1r) [right=of d2r]{};
        \node[gauger, label=below:$D_1$] (d1r) [right=of c1r]{};
        \node[gaugeb, label=right:$C_1$] (c1t) [above=of d4]{};

        \draw[-] (d1l)--(c1l)--(d2l)--(c2l)--(d3l)--(c3l)--(d4)--(c3r)--(d3r)--(c2r)--(d2r)--(c1r)--(d1r) (d4)--(c1t);
        
    \node[gauger, label=below:$D_1$] (d1rj) [below=of d1r]{};
    \node[gaugeb, label=below:$C_1$] (c1rj) [left=of d1rj]{};
    \node[gauger, label=below:$D_1$] (d1lj) [left=of c1rj]{};
    \node[gaugeb, label=below:$C_1$] (c1lj) [left=of d1lj]{};

    \draw[-] (d1rj)--(c1rj)--(d1lj)--(c1lj);
    
    \node[gauger, label=below:$D_1$] (d1lk) [below=of d1l]{};
   \node[gaugeb, label=below:$C_1$] (c1lk) [right=of d1lk]{};
   \node[gauger, label=below:$D_2$] (d2lk) [right=of c1lk]{};
   \node[gaugeb, label=below:$C_2$] (c2lk) [right=of d2lk]{};
   \node[gauger, label=below:$D_2$] (d2rk) [right=of c2lk]{};
   \node[gaugeb, label=below:$C_1$] (c1rk) [right=of d2rk]{};
    
    \draw[-] (d1lk)--(c1lk)--(d2lk)--(c2lk)--(d2rk)--(c1rk);
    
    \node[gaugeb, label=left:$C_1$] (C1ts) [below right ={3} of d1rk]{};
    \node[gauger, label=below:$D_4$] (D4s) [below right=of C1ts]{};
    \node[gaugeb, label=below:$C_2$] (C2ls) [left=of D4s]{};
    \node[gauger, label=below:$D_1$] (D1ls) [left=of C2ls]{};
    \node[gaugeb, label=below:$C_3$] (C3rs) [right=of D4s]{};
    \node[gauger, label=below:$D_3$] (D3rs) [right=of C3rs]{};
    \node[gaugeb, label=below:$C_1$] (C1rs) [right=of D3rs]{};
    \node[gauger, label=below:$D_1$] (D1rs) [right=of C1rs]{};
    \node[gaugeb, label=right:$C_1$] (C1trrs) [above=of D3rs]{};
    \node[gaugeb, label=right:$C_1$] (C1trs) [above right=of D4s]{};

    \draw[-] (C1trs)--(D4s)--(C1ts);
    \draw[-] (D1rs)--(C1rs)--(D3rs)--(C3rs)--(D4s)--(C2ls)--(D1ls);
    \draw[-] (C1trrs)--(D3rs);
    \node[] (minus) [right=of d1rj]{$-$};
    \node[] (minus) [left=of d1lk]{$-$};
    \node[] (topghost) [below=of d1lk]{};
    \node[] (bottomghost) [left=of D1ls]{};
    \draw[->] (topghost)to [out=-90,in=+135,looseness=1](bottomghost);
    \end{tikzpicture}}
    \caption{Subtraction of the $\surm(3)$ and $\surm(2)$ quotient quivers from each leg of the orthosymplectic quiver for $\overline{min. E_8}\times\overline{min. E_8}$.}
    \label{fig:minE8SU2SU3}
\end{figure} 
\begin{align}
\hs\left[\mathcal C(\text{\Quiver{fig:minE8SU2SU3}})\right]&=1 + 35 t^2 + 854 t^4 + 15524 t^6 + 226316 t^8 + 2734454 t^{10}+28109718 t^{12}+O\left(t^{13}\right)\\
\pl\left[\hs\left[\mathcal C(\text{\Quiver{fig:minE8SU2SU3}})\right]\right]&=35 t^2 + 224 t^4 - 86 t^6 - 10809 t^8 - 12024 t^{10} + 952909 t^{12}+O\left(t^{13}\right)
\end{align}
The unrefined Hilbert series for $\overline{min. E_8}///\left(\surm(2)\times\surm(3)\right)$ may be computed as \begin{align}
    &\hs\left[\overline{min. E_8}///\left(\surm(2)\times\surm(3)\right)\right]\\&=\frac{\left(\begin{aligned}1 &+ 17 t^2 + 359 t^4 + 4385 t^6 + 45413 t^8 + 355769 t^{10} + 
 2300271 t^{12} + 12090539 t^{14} + 53342987 t^{16} \\&+ 197855300 t^{18} + 
 626993788 t^{20} + 1704044024 t^{22} + 4010026707 t^{24} + 
 8196956823 t^{26} \\&+ 14642063141 t^{28} + 22901954086 t^{30} + 
 31482549792 t^{32} + 38064403005 t^{34} + 40558624938 t^{36}\\&+\cdots+t^{72}\end{aligned}\right)}{(1-t^2)^{18}(1-t^4)^{18}}\\&=1+35 t^2+854 t^4+15524 t^6+226316 t^8+2734454 t^{10}+28109718 t^{12}+O(t^{13})
\end{align}using the following embedding of $E_8\hookleftarrow \surm(6)\times\surm(3)\times\surm(2)$ which decomposes the adjoint of $E_8$ as \begin{equation}
    \left(\mu_7\right)_{E_8}\rightarrow \mu_1\mu_5 + \nu_1\nu_2+\rho^2+\mu_4\nu_1 + \mu_2\nu_2  + \mu_3\rho + \mu_1\nu_1\rho + \mu_5\nu_2\rho 
\end{equation}where the $\left(\mu_7\right)_{E_8}$ is a highest weight fugacity for the adjoint of $E_8$ and the $\mu_i,\;\nu_i,$ and $\rho$ on the right-hand side are highest weight fugacities for $\surm(6),\;\surm(3),$ and $\surm(2)$ respectively. The hyper-Kähler quotient is taken w.r.t the $\surm(3)$ and $\surm(2)$.  
\newpage
\section{Examples of $G_2$ Orthosymplectic Quotient Quiver Subtraction}
\label{sec:G2Examples}





\subsection{$\overline{min. E_8}///G_2$}
Hyper-K\"ahler quotients of larger groups naturally find application to magnetic quivers describing moduli spaces of higher dimension -- for $G_2$ quotient quiver subtraction, the most straightforward variety to begin with is $\overline{min. E_8}$.

Shown in \Figref{fig:minE8G2}, the $G_2$ quotient quiver has two possible alignments against the orthosymplectic quiver for $\overline{min. E_8}$, giving two quivers \Quiver{fig:minE8G21} and \Quiver{fig:minE8G22}.

\begin{figure}[h]
    \centering
    \begin{subfigure}{\textwidth}
    \centering
        \resizebox{\linewidth}{!}{\begin{tikzpicture}
        \node[gauger, label=below:$D_1$] (d1l) []{};
        \node[gaugeb, label=below:$C_1$] (c1l) [right=of d1l]{};
        \node[gauger, label=below:$D_2$] (d2l) [right=of c1l]{};
        \node[gaugeb, label=below:$C_2$] (c2l) [right=of d2l]{};
        \node[gauger, label=below:$D_3$] (d3l) [right=of c2l]{};
        \node[gaugeb, label=below:$C_3$] (c3l) [right=of d3l]{};
        \node[gauger, label=below:$D_4$] (d4) [right=of c3l]{};
        \node[gaugeb, label=below:$C_3$] (c3r) [right=of d4]{};
        \node[gauger, label=below:$D_3$] (d3r) [right=of c3r]{};
        \node[gaugeb, label=below:$C_2$] (c2r) [right=of d3r]{};
        \node[gauger, label=below:$D_2$] (d2r) [right=of c2r]{};
        \node[gaugeb, label=below:$C_1$] (c1r) [right=of d2r]{};
        \node[gauger, label=below:$D_1$] (d1r) [right=of c1r]{};
        \node[gaugeb, label=right:$C_1$] (c1t) [above=of d4]{};

        \draw[-] (d1l)--(c1l)--(d2l)--(c2l)--(d3l)--(c3l)--(d4)--(c3r)--(d3r)--(c2r)--(d2r)--(c1r)--(d1r) (d4)--(c1t);

        \node[gaugeb, label=right:$C_1$] (c1ts) [below=of d4]{};
        \node[gauger, label=below:$D_2$] (d2rs) [below=of c1ts]{};
        \node[gaugeb, label=below:$C_3$] (c3s) [left=of d2rs]{};
        \node[gauger, label=below:$D_3$] (d3s) [left=of c3s]{};
        \node[gaugeb, label=below:$C_2$] (c2s) [left=of d3s]{};
        \node[gauger, label=below:$D_2$] (d2ls) [left=of c2s]{};
        \node[gaugeb, label=below:$C_1$] (c1ls) [left=of d2ls]{};
        \node[gauger, label=below:$D_1$] (d1ls) [left=of c1ls]{};

        \draw[-] (c1ts)--(d2rs)--(c3s)--(d3s)--(c2s)--(d2ls)--(c1ls)--(d1ls);

        \node[] (ghost) [below=of d2rs]{};

        \node[gauger, label=below:$D_2$] (D2l) [below=of ghost]{};
        \node[gaugeb, label=below:$C_3$] (C3) [right=of D2l]{};
        \node[gauger, label=below:$D_3$] (D3) [right=of C3]{};
        \node[gaugeb, label=below:$C_2$] (C2) [right=of D3]{};
        \node[gauger, label=below:$D_2$] (D2) [right=of C2]{};
        \node[gaugeb, label=below:$C_1$] (C1) [right=of D2]{};
        \node[gauger, label=below:$D_1$] (D1) [right=of C1]{};
        \node[gaugeb, label=right:$C_1$] (C1t) [above=of C3]{};

        \draw[-] (D2l)--(C3)--(D3)--(C2)--(D2)--(C1)--(D1);
        \draw[transform canvas={xshift=+1.5pt}] (C3)--(C1t);
        \draw[transform canvas={xshift=-1.5pt}] (C3)--(C1t);

        \node[](minus) [left=of d1ls]{$-$};

        \node[] (topghost) [right=of d1r]{};
        \node[] (bottomghost) [right=of D1]{};

        \draw[->] (topghost)to [out=-45,in=45,looseness=1](bottomghost);

        \end{tikzpicture}}
        \caption{}
        \label{fig:minE8G21}
    \end{subfigure}
    \begin{subfigure}{\textwidth}
    \centering
        \resizebox{\linewidth}{!}{\begin{tikzpicture}
        \node[gauger, label=below:$D_1$] (d1l) []{};
        \node[gaugeb, label=below:$C_1$] (c1l) [right=of d1l]{};
        \node[gauger, label=below:$D_2$] (d2l) [right=of c1l]{};
        \node[gaugeb, label=below:$C_2$] (c2l) [right=of d2l]{};
        \node[gauger, label=below:$D_3$] (d3l) [right=of c2l]{};
        \node[gaugeb, label=below:$C_3$] (c3l) [right=of d3l]{};
        \node[gauger, label=below:$D_4$] (d4) [right=of c3l]{};
        \node[gaugeb, label=below:$C_3$] (c3r) [right=of d4]{};
        \node[gauger, label=below:$D_3$] (d3r) [right=of c3r]{};
        \node[gaugeb, label=below:$C_2$] (c2r) [right=of d3r]{};
        \node[gauger, label=below:$D_2$] (d2r) [right=of c2r]{};
        \node[gaugeb, label=below:$C_1$] (c1r) [right=of d2r]{};
        \node[gauger, label=below:$D_1$] (d1r) [right=of c1r]{};
        \node[gaugeb, label=right:$C_1$] (c1t) [above=of d4]{};

        \draw[-] (d1l)--(c1l)--(d2l)--(c2l)--(d3l)--(c3l)--(d4)--(c3r)--(d3r)--(c2r)--(d2r)--(c1r)--(d1r) (d4)--(c1t);

        \node[gaugeb, label=below:$C_1$] (c1rs) [below=of c3r]{};
        \node[gauger, label=below:$D_2$] (d2rs) [left=of c1rs]{};
        \node[gaugeb, label=below:$C_3$] (c3s) [left=of d2rs]{};
        \node[gauger, label=below:$D_3$] (d3s) [left=of c3s]{};
        \node[gaugeb, label=below:$C_2$] (c2s) [left=of d3s]{};
        \node[gauger, label=below:$D_2$] (d2ls) [left=of c2s]{};
        \node[gaugeb, label=below:$C_1$] (c1ls) [left=of d2ls]{};
        \node[gauger, label=below:$D_1$] (d1ls) [left=of c1ls]{};

        \draw[-] (c1rs)--(d2rs)--(c3s)--(d3s)--(c2s)--(d2ls)--(c1ls)--(d1ls);

        \node[gaugeb, label=left:$C_1$] (C1tl)[below=of d2rs]{};
        \node[gauger, label=below:$D_2$] (D2l) [below=of C1tl]{};
        \node[gaugeb, label=below:$C_2$] (C2l) [right=of D2l]{};
        \node[gauger, label=below:$D_3$] (D3) [right=of C2l]{};
        \node[gaugeb, label=below:$C_2$] (C2r) [right=of D3]{};
        \node[gauger, label=below:$D_2$] (D2r) [right=of C2r]{};
        \node[gaugeb, label=below:$C_1$] (C1) [right=of D2r]{};
        \node[gauger, label=below:$D_1$] (D1) [right=of C1]{};
        \node[gaugeb, label=right:$C_1$] (C1tr) [above=of D3]{};

        \draw[-] (C1tl)--(D2l)--(C2l)--(D3)--(C2r)--(D2r)--(C1)--(D1) (D3)--(C1tr);

         \draw[transform canvas={yshift=+1.5pt}] (C1tl)--(C1tr);
        \draw[transform canvas={yshift=-1.5pt}] (C1tl)--(C1tr);

        \node[](minus) [left=of d1ls]{$-$};

        \node[] (topghost) [right=of d1r]{};
        \node[] (bottomghost) [right=of D1]{};

        \draw[->] (topghost)to [out=-45,in=45,looseness=1](bottomghost);
        \end{tikzpicture}}
        \caption{}
        \label{fig:minE8G22}
    \end{subfigure}
    \begin{subfigure}{\textwidth}
    \centering
        \begin{tikzpicture}
        \node[gauger, label=below:$D_2$] (D2l) []{};
        \node[gaugeb, label=below:$C_2$] (C2l) [right=of D2l]{};
        \node[gauger, label=below:$D_3$] (D3) [right=of C2l]{};
        \node[gaugeb, label=below:$C_2$] (C2r) [right=of D3]{};
        \node[gauger, label=below:$D_2$] (D2r) [right=of C2r]{};
        \node[gaugeb, label=below:$C_1$] (C1) [right=of D2r]{};
        \node[gauger, label=below:$D_1$] (D1) [right=of C1]{};
        \node[gaugeb, label=above:$C_1$] (C1t) [above=of C2l]{};

        \draw[-] (D2l)--(C2l)--(D3)--(C2r)--(D2r)--(C1)--(D1) (D2l)--(C1t)--(D3);
        \end{tikzpicture}
        \caption{}
        \label{fig:minE8G2Int}
    \end{subfigure}
    \caption{Two alignments of the $G_2$ quotient quiver against the orthosymplectic quiver for $\overline{min. E_8}$ producing \Quiver{fig:minE8G21} and \Quiver{fig:minE8G22}. Their intersection is \Quiver{fig:minE8G2Int}.}
    \label{fig:minE8G2}
\end{figure}

The Hilbert series for the Coulomb branch of \Quiver{fig:minE8G21} is \begin{equation}
    \hs\left[\text{\Quiver{fig:minE8G21}}\right]=\frac{(1 + t^2) \left(\begin{aligned}1 &+ 21 t^2 + 231 t^4 + 1498 t^6 + 6219 t^8 + 16834 t^{10} \\&+ 
   30420 t^{12} + 36972 t^{14} + \cdots + t^{28}\end{aligned}\right)}{(1 - t^2)^{30}},
\end{equation}from which the Coulomb branch can be identified as $\mathcal C(\text{\Quiver{fig:minE8G21}})=\overline{\mathcal O}^{F_4}_{[2,0,0,0]}$ (Bala-Carter label $A_2$). The Hilbert series for the Coulomb branch of \Quiver{fig:minE8G22} is computed as \begin{equation}
    \hs\left[\text{\Quiver{fig:minE8G22}}\right]=\frac{(1 + t^2) \left(\begin{aligned}1 &+ 21 t^2 + 257 t^4 + 2018 t^6 + 
    9573 t^8 + 28261 t^{10} \\&+ 53781 t^{12} + 66651 t^{14} + \cdots + t^{28}\end{aligned}\right)}{(1 - t^2)^{30}},
\end{equation}from which the Coulomb branch can be identified as $\mathcal C(\text{\Quiver{fig:minE8G22}})=\textrm{Normalisation}\left[\overline{\mathcal O}^{F_4}_{[0,0,0,2]}\right]$ (Bala-Carter label $\tilde{A}_2$), where the normalisation of the nilpotent orbit closure is computed by the localisation formula \cite{Hanany:2017ooe}.

The proposed intersection of the quivers \Quiver{fig:minE8G21} and \Quiver{fig:minE8G22} is \Quiver{fig:minE8G2Int}, with Coulomb branch Hilbert series \begin{equation}
     \hs\left[\text{\Quiver{fig:minE8G2Int}}\right]=\frac{(1 + t^2) \left(\begin{aligned}1 &+ 22 t^2 + 254 t^4 + 1773 t^6 + 
    7171 t^8 \\&+ 16619 t^{10} + 22030 t^{12} + \cdots + t^{24}\end{aligned}\right)}{(1 - t^2)^{28}},
\end{equation}from which there is the identification $\mathcal C(\text{\Quiver{fig:minE8G2Int}})=\overline{\mathcal O}^{F_4}_{[0,1,0,0]}$ (Bala-Carter label $A_1+\tilde A_1$).

The conclusion is hence that \begin{equation}
    \overline{min. E_8}///G_2=\overline{\mathcal O}^{F_4}_{[2,0,0,0]}\cup \textrm{Normalisation}\left[\overline{\mathcal O}^{F_4}_{[0,0,0,2]}\right],
\end{equation}with the following Hilbert series \begin{align}
    \hs\left[\overline{min. E_8}///G_2\right]&=\frac{\left(\begin{aligned}1 &+ 22 t^2 + 278 t^4 + 2275 t^6 + 12644 t^8 + 
   47792 t^{10} + 121354 t^{12} \\&+ 202670 t^{14} + 217542 t^{16} + 
   144142 t^{18} + 52945 t^{20} + 6213 t^{22} \\&- 2660 t^{24} - 1199 t^{26} - 
   208 t^{28} - 20 t^{30} - t^{32}\end{aligned}\right)}{(1-t^2)^{30}}\\&=\hs\left[\text{\Quiver{fig:minE8G21}}\right]+\hs\left[\text{\Quiver{fig:minE8G22}}\right]-\hs\left[\text{\Quiver{fig:minE8G2Int}}\right]
\end{align}
This is verified using Weyl integration and the embedding $E_8\hookleftarrow F_4\times G_2$, in which the fundamental of $E_8$ decomposes as\begin{equation}
    (\mu_7)_{E_8}\hookleftarrow \mu_1 + \nu_1 + 
\mu_4\nu_2,
\end{equation} where $\mu_i$ and $\nu_i$ are highest weight fugacities for $F_4$ and $G_2$ respectively.

It is challenging to find constructions of nilpotent orbit closures of height three or more either as a Higgs branch or a Coulomb branch from unitary or orthosymplectic quivers. The nilpotent orbits of $F_4$ with characteristics $[2,0,0,0]$ and $[0,0,0,2]$ are of height 4 and the orbit with characteristic $[0,1,0,0]$ is of height three. It is non-trivial to have found the quivers \Quiver{fig:minE8G21}, \Quiver{fig:minE8G22}, and \Quiver{fig:minE8G2Int} whose Coulomb branches are (normalisations of) these nilpotent orbits and are the first of their kind.

It is also straightforward to compute the Higgs branch Hilbert series of these quivers with refinement of the global symmetry. The Higgs branch Hilbert series of \Quiver{fig:minE8G21} is computed, with refinement of the $\sprm(1)$ rotating the two bifundamental hypermultiplets between the $C_1$ and $C_3$ gauge node, as \begin{equation}
    \hs\left[\mathcal H(\text{\Quiver{fig:minE8G21}})\right]=\pe\left[[2]_{\sprm(1)}t^2+[4]_{\sprm(1)}t^8-t^{16}-t^{24}\right]
\end{equation}where the Dynkin label is short-hand for the character of the given $\sprm(1)$ representation. This is the same Hilbert series as for $\mathcal S^{F_4}_{\mathcal N,[2,2,0,0]}$ (Bala-Carter label $B_3$) which is the Spaltenstein dual to $\overline{\mathcal O}^{F_4}_{[2,0,0,0]}$.

The Higgs branch Hilbert series of \Quiver{fig:minE8G22} is computed with refinement of the two bifundamental hypermultiplets between the two $C_1$ gauge nodes. 

\begin{equation}
    \hs\left[\mathcal H(\text{\Quiver{fig:minE8G22}})\right]=\pe\left[[2]_{\sprm(1)}t^2+[1]_{\sprm(1)}t^5+t^8+[1]_{\sprm(1)}t^{11}-t^{16}-t^{24}\right]
\end{equation} This Hilbert series is the same as for $\mathcal S^{F_4}_{\mathcal N,[1,0,1,2]}$ (Bala-Carter label $C_3$) which is the Spaltenstein dual to $\overline{\mathcal O}^{F_4}_{[0,0,0,2]}$.

The Higgs branch Hilbert series of \Quiver{fig:minE8G2Int} is computed without refinement.

\begin{equation}
     \hs\left[\mathcal H(\text{\Quiver{fig:minE8G2Int}})\right]=\pe\left[2 t^4 + t^6 + t^8 + t^{10} + t^{12} - t^{16} - t^{24}\right]
\end{equation}
This Hilbert series is the same as for $\mathcal S^{F_4}_{\mathcal N,[0,2,0,2]}$ (Bala-Carter label $F_4(a_2)$) which is the Spaltenstein dual to $\overline{\mathcal O}^{F_4}_{[0,1,0,0]}$.

There are no known unitary counterparts to these three quivers and remains a challenge to find them.

\subsection{$(\overline{min. E_6}\times\overline{min. E_6})///G_2$}

The next natural example of $G_2$ quotient quiver subtraction is on the product space $\overline{min. E_6}\times \overline{min. E_6}$ of \cite{Chacaltana:2011ze,Sperling:2021fcf}.
Similar to the previous case, the $G_2$ quotient quiver has two alignments against the product theory, as shown in \Figref{fig:E6E6G21}.

\begin{figure}
    \centering
    \begin{subfigure}{\textwidth}
    \centering
    \begin{tikzpicture}
         \node[gauger, label=below:$D_1$] (d1l) []{};
        \node[gaugeb, label=below:$C_1$] (c1l) [right=of d1l]{};
        \node[gauger, label=below:$D_2$] (d2l) [right=of c1l]{};
        \node[gaugeb, label=below:$C_2$] (c2l) [right=of d2l]{};
        \node[gauger, label=below:$D_3$] (d3l) [right=of c2l]{};
        \node[gaugeb, label=below:$C_3$] (c3l) [right=of d3l]{};
        \node[gauger, label=below:$D_4$] (d4) [right=of c3l]{};
        \node[gaugeb, label=above:$C_2$] (c2t) [above right=of d4]{};
        \node[gauger, label=above:$D_1$] (d1t) [right=of c2t]{};
        \node[gaugeb, label=above:$C_2$] (c2b) [below right=of d4]{};
        \node[gauger, label=above:$D_1$] (d1b) [right=of c2b]{};
        \draw[-] (d1l)--(c1l)--(d2l)--(c2l)--(d3l)--(c3l)--(d4) (d1t)--(c2t)--(d4)--(c2b)--(d1b);

        \node[gaugeb, label=right:$C_1$] (c1ts) [below=of c2b]{};
        \node[gauger, label=below:$D_2$] (d2rs) [above left=of c1ts]{};
        \node[gaugeb, label=below:$C_3$] (c3s) [left=of d2rs]{};
        \node[gauger, label=below:$D_3$] (d3s) [left=of c3s]{};
        \node[gaugeb, label=below:$C_2$] (c2s) [left=of d3s]{};
        \node[gauger, label=below:$D_2$] (d2ls) [left=of c2s]{};
        \node[gaugeb, label=below:$C_1$] (c1ls) [left=of d2ls]{};
        \node[gauger, label=below:$D_1$] (d1ls) [left=of c1ls]{};

        \draw[-] (c1ts)--(d2rs)--(c3s)--(d3s)--(c2s)--(d2ls)--(c1ls)--(d1ls);

        \node[gaugeb, label=above:$C_2$] (C2t) [below=of c1ts]{};
        \node[gauger, label=above:$D_1$] (D1t) [right=of C2t]{};
        \node[gauger, label=left:$D_2$] (D2) [below left=of C2t]{};
        \node[gaugeb, label=below:$C_1$] (C1) [below right=of D2]{};
        \node[gauger, label=below:$D_1$] (D1) [right=of C1]{};
        \node[gaugeb, label=right:$C_1$] (C1r) [below=of D1t]{};

        \draw[-] (D1t)--(C2t)--(D2)--(C1)--(D1)--(C1r);

        \draw[transform canvas={xshift=-1.3pt,yshift=-1.3pt}] (C2t)--(C1r);
    \draw[transform canvas={xshift=+1.3pt,yshift=+1.3pt}] (C2t)--(C1r);

    \node[](minus) [left=of d1ls]{$-$};

    \node[] (topghost) at (d4-|d1t) {};
    \node[] (bottomghost) [right=of C1r]{};

    \draw[->] (topghost)to [out=-45,in=45,looseness=1](bottomghost);

    \end{tikzpicture}
    \caption{}
    \label{fig:E6E6G21}
    \end{subfigure}
    \begin{subfigure}{\textwidth}
    \centering
        \begin{tikzpicture}
         \node[gaugeb, label=above:$C_2$] (C2t) []{};
        \node[gauger, label=above:$D_1$] (D1t) [right=of C2t]{};
        \node[gauger, label=left:$D_2$] (D2) [below left=of C2t]{};
        \node[gaugeb, label=below:$C_1$] (C1) [below right=of D2]{};
        \node[gauger, label=below:$D_1$] (D1) [right=of C1]{};
        \node[gaugeb, label=right:$C_1$] (C1r) [right=of D2]{};

        \draw[-] (D1t)--(C2t)--(D2)--(C1)--(D1)--(C1r)--(D1t) (D2)--(C1r);
            
        \end{tikzpicture}
        \caption{}
        \label{fig:E6E6G2Int}
    \end{subfigure}
    \caption{One alignment of the $G_2$ quotient quiver against the orthosymplectic quiver for $\overline{min. E_6}\times\overline{min. E_6}$. This produces \Quiver{fig:E6E6G21}. The other alignment gives the same quiver. The intersection of two \Quiver{fig:E6E6G21} is \Quiver{fig:E6E6G2Int}.}
    \label{fig:E6E6GG2}
\end{figure}

In this case, both of these alignments give the same resulting quiver \Quiver{fig:E6E6G21}, with Coulomb branch Hilbert series \begin{align}
    \hs[\mathcal C(\text{\Quiver{fig:E6E6G21}})]&=\frac{\left(\begin{aligned}1 &+ 8 t^2 + 91 t^4 + 415 t^6 + 
   1684 t^8 + 4131 t^{10} + 8599 t^{12} + 12168 t^{14} + 14548 t^{16} \\&+ 
   12168 t^{18} + 8599 t^{20} + 4131 t^{22} + 1684 t^{24} + 415 t^{26} + 
   91 t^{28} + 8 t^{30} + t^{32}\end{aligned}\right)}{(1 - t^2)^{8} (1 - t^4)^8}\\\pl\left[\hs[\mathcal C(\text{\Quiver{fig:E6E6G21}})]\right]&=16 t^2 + 63 t^4 - 145 t^6 - 1006 t^8 + 5662 t^{10}+O(t^{11})
\end{align}

The intersection of two \Quiver{fig:E6E6G21} is \Quiver{fig:E6E6G2Int}, with Coulomb branch Hilbert series
\begin{align}
     \hs[\mathcal C(\text{\Quiver{fig:E6E6G2Int}})]&=\frac{\left(\begin{aligned}1 &+ 9 t^2 + 101 t^4 + 459 t^6 + 
   1659 t^8 + 3591 t^{10} + 6087 t^{12} + 6858 t^{14} \\&+ 6087 t^{16} + 
   3591 t^{18} + 1659 t^{20} + 459 t^{22} + 101 t^{24} + 9 t^{26} + t^{28}\end{aligned}\right)}{(1 - t^2)^7(1 - t^4)^7}\\\pl\left[\hs[\mathcal C(\text{\Quiver{fig:E6E6G2Int}})]\right]&=16 t^2 + 63 t^4 - 210 t^6 - 1062 t^8 + 9468 t^{10}+O(t^{11})
\end{align}
The Hilbert series of the hyper-Kähler quotient is computed as \begin{align}
    &\hs\left[(\overline{min. E_6}\times\overline{min. E_6})///G_2\right]\nonumber\\&=  \frac{\left(\begin{aligned}1 &+ 8 t^2 + 91 t^4 + 480 t^6 + 
   2260 t^8 + 6688 t^{10} + 15902 t^{12} + 25497 t^{14} + 32363 t^{16} \\&+ 
   27603 t^{18} + 18359 t^{20} + 6966 t^{22} + 1794 t^{24} - 278 t^{26} - 
   168 t^{28} - 75 t^{30} - 6 t^{32} - t^{34}\end{aligned}\right)}{(1-t^2)^8(1-t^4)^8}\\&=2\hs[\mathcal C(\text{\Quiver{fig:E6E6G21}})]- \hs[\mathcal C(\text{\Quiver{fig:E6E6G2Int}})]\nonumber\\&\pl\left[\hs\left[(\overline{min. E_6}\times\overline{min. E_6})///G_2\right]\right]\nonumber\\&=16 t^2 + 63 t^4 - 80 t^6 - 950 t^8 + 1856 t^{10}+O(t^{11})
\end{align}
using the embedding of $E_6\times E_6\hookleftarrow G_2\times \surm(3)\times\surm(3)$ which decomposes the fundamental as \begin{equation}
    (\mu_6)_{E_6}+(\mu'_6)_{E_6}\rightarrow 2\nu_1 + (\mu_1 \mu_2+\mu'_1\mu'_2) (1 + \nu_2)
\end{equation}where the $\mu_i$ and $\mu'_i$ are highest weight fugacities for the two $\surm(3)$ and $\nu_i$ are highest weight fugacities for $G_2$.

This result may also be interpreted as a magnetic quiver realisation for the Higgs branch of the diagonal $G_2$ flavour symmetry gauging of two $4d\;\mathcal N=2$ rank-1 $E_6$ SCFTs.
\subsection{$(\overline{min. E_7}\times\overline{min. E_7})///G_2$}
The magnetic quiver for $\overline{min. E_7}\times\overline{min. E_7}$ admits subtraction of the $G_2$ quotient quiver from the maximal leg. This is shown in \Figref{fig:minE7SqG2} to produce \Quiver{fig:minE7SqG2}.
\begin{figure}[H]
    \centering
    \resizebox{\linewidth}{!}{\begin{tikzpicture}[main/.style={draw,circle}]

   \node[gauger, label=below:$D_1$] (d1l) []{};
   \node[gaugeb, label=below:$C_1$] (c1l) [right=of d1l]{};
   \node[gauger, label=below:$D_2$] (d2l) [right=of c1l]{};
   \node[gaugeb, label=below:$C_2$] (c2l) [right=of d2l]{};
   \node[gauger, label=below:$D_3$] (d3l) [right=of c2l]{};
   \node[gaugeb, label=below:$C_3$] (c3l) [right=of d3l]{};
   \node[gauger, label=below:$D_4$] (d4) [right=of c3l]{};
   \node[gaugeb, label=below:$C_4$] (c4) [right=of d4]{};
   \node[gauger, label=below:$D_5$] (d5) [right=of c3r]{};
   \node[gaugeb, label=below:$C_3$] (c3r) [right=of d3r]{};
   \node[gauger, label=below:$D_2$] (d2r) [right=of c2r]{};
   \node[gaugeb, label=below:$C_1$] (c1r) [right=of d2r]{};
   \node[gauger, label=below:$D_1$] (d1r) [right=of c1r]{};
    \node[gaugeb, label=left:$C_2$] (c2t) [above=of d5]{};

    \draw[-] (d1l)--(c1l)--(d2l)--(c2l)--(d3l)--(c3l)--(d4)--(c4)--(d5)--(c3r)--(d3r)--(c2r)--(d2r)--(c1r)--(d1r);
    \draw[-] (c2t)--(d5);

    \node[gaugeb, label=below:$C_1$] (c1ts) [below=of c4]{};
    \node[gauger, label=below:$D_2$] (d2rs) [below=of d4]{};
    \node[gaugeb, label=below:$C_3$] (c3s) [below=of c3l]{};
    \node[gauger, label=below:$D_3$] (d3s) [below=of d3l]{};
    \node[gaugeb, label=below:$C_2$] (c2s) [below=of c2l]{};
    \node[gauger, label=below:$D_2$] (d2ls) [below=of d2l]{};
    \node[gaugeb, label=below:$C_1$] (c1ls) [below=of c1l]{};
    \node[gauger, label=below:$D_1$] (d1ls) [below=of d1l]{};

    \draw[-] (c1ts)--(d2rs)--(c3s)--(d3s)--(c2s)--(d2ls)--(c1ls)--(d1ls);

    \node[] (minus) [left=of d1ls]{$-$};

   \node[gaugeb, label=left:$C_2$] (fc1u) [below=of c1ts]{};
   \node[gauger, label=below:$D_5$] (fd2l) [below=of fc1u]{};
   \node[gaugeb, label=below:$C_3$] (fc2l) [right=of fd2l]{};
   \node[gaugeb, label=right:$C_1$] (fc2u) [above=of fc2l]{};
   \node[gaugeb, label=below:$C_3$] (fc1l) [left=of fd2l]{};
   \node[gauger, label=below:$D_2$] (fd1l) [left=of fc1l]{};
   \node[gauger, label=below:$D_2$] (fd3l) [right=of fc2l]{};
   \node[gaugeb, label=below:$C_1$] (fc3l) [right=of fd3l]{};
   \node[gauger, label=below:$D_1$] (fd4) [right=of fc3l]{};

    \draw[-] (fd1l)--(fc1l)--(fd2l)--(fc2l)--(fd3l)--(fd4);
    \draw[-] (fc1u) -- (fd2l) -- (fc2u);
    
    \node[] (topghost) [below left=of d1r]{};
    \node[] (bottomghost) [right=of fd4]{};

    \draw[->] (topghost)to [out=-45,in=45,looseness=1](bottomghost);
        
    \end{tikzpicture}}
    \caption{Subtraction of the $G_2$ orthosymplectic quotient quiver from the orthosymplectic magnetic quiver for $\overline{min. E_7} \times \overline{min. E_7}$.}
    \label{fig:minE7SqG2}
\end{figure}
The Coulomb branch Hilbert series is computed perturbatively up to order $t^{10}$ together with its $\pl$.
\begin{align}
     \hs[\mathcal C(\text{\Quiver{fig:minE7SqG2}})]&= 1+42 t^2+1098 t^4+21238 t^6+329230 t^8+4255866 t^{10}+47097064 t^{12}+O\left(t^{13}\right)\\
     \pl\left[\hs[\mathcal C(\text{\Quiver{fig:minE7SqG2}})]\right]&=42 t^2 + 195 t^4 - 196 t^6 - 6728 t^8 - 2304 t^{10}+460312 t^{12}+O\left(t^{13}\right)
\end{align}The $t^2$ coefficient indicates a $\sorm(7)\times\sorm(7)$ global symmetry. This is verified with Weyl integration using the following embedding of $E_7\times E_7\hookleftarrow \sorm(7)\times\sorm(7)\times G_2$ which decomposes the fundamental as \begin{equation}
    {\mu_1}_{E_7}+{\mu_1'}_{E_7}\rightarrow \mu_1^2 + \mu_1'^2 + 2 \nu_1 + (\mu_2 + \mu'_2) \nu_2
\end{equation}where the $\mu_i$ and $\mu_i'$ are highest weight fugacities for the two $\sorm(7)$ and the $\nu_i$ are highest weight fugacities for the $G_2$ being gauged.

This quiver \Quiver{fig:minE7SqG2} may also be interpreted as a magnetic quiver for the class $\mathcal S$ theory specified by algebra $\sorm(10)$ on a four punctured sphere with puncture data $\{(3^3,1),(2^2,1^6),(2^4,1^2),$ $(5,2^2,1)\}$. This quiver may also be a magnetic quiver for the Higgs branch of the diagonal $G_2$ flavour symmetry gauging of two $4d\;\mathcal N=2$ rank-1 $E_7$ SCFTs \cite{Chacaltana:2011ze}.
\section{Example of $\sorm(7)$ Orthosymplectic Quotient Quiver Subtraction}
\label{sec:SO7Examples}
\subsection{$(\overline{min. E_7}\times\overline{min. E_7})///\sorm(7)$}
The orthosymplectic magnetic quiver for $\overline{min. E_7}\times\overline{min. E_7}$ admits two possible alignments of the $\sorm(7)$ orthosymplectic quotient quiver. These are shown in \Figref{fig:minE7SqSO71} and \Figref{fig:minE7SqSO72} to give \Quiver{fig:minE7SqSO71} and \Quiver{fig:minE7SqSO72} respectively.

\begin{figure}
    \centering
    \begin{subfigure}{\textwidth}
    \centering
    \resizebox{\linewidth}{!}{\begin{tikzpicture}
    \node[gauger, label=below:$D_1$] (d1l) []{};
        \node[gaugeb, label=below:$C_1$] (c1l) [right=of d1l]{};
        \node[gauger, label=below:$D_2$] (d2l) [right=of c1l]{};
        \node[gaugeb, label=below:$C_2$] (c2l) [right=of d2l]{};
        \node[gauger, label=below:$D_3$] (d3l) [right=of c2l]{};
        \node[gaugeb, label=below:$C_3$] (c3l) [right=of d3l]{};
        \node[gauger, label=below:$D_4$] (d4l) [right=of c3l]{};
        \node[gaugeb, label=below:$C_4$] (c4l) [right=of d4l]{};
        \node[gauger, label=below:$D_5$] (d5) [right=of c4l]{};
        \node[gaugeb, label=below:$C_3$] (c3r) [right=of d5]{};
        \node[gauger, label=below:$D_2$] (d2r) [right=of c3r]{};
        \node[gaugeb, label=below:$C_1$] (c1r) [right=of d2r]{};
        \node[gauger, label=below:$D_1$] (d1r) [right=of c1r]{};
        \node[gaugeb, label=right:$C_2$] (c2t) [above=of d5]{};

        \draw[-] (d1l)--(c1l)--(d2l)--(c2l)--(d3l)--(c3l)--(d4l)--(c4l)--(d5)--(c3r)--(d2r)--(c1r)--(d1r) (d5)--(c2t);

        \node[gaugeb, label=right:$C_1$] (c1ts) [below=of d5]{};
        \node[gauger, label=below:$D_2$] (d2rs) [below=of c1ts]{};
        \node[gaugeb, label=below:$C_3$] (c3rs) [left=of d2rs]{};
        \node[gauger, label=below:$D_4$] (d4s) [left=of c3rs]{};
        \node[gaugeb, label=below:$C_3$] (c3ls) [left=of d4s]{};
        \node[gauger, label=below:$D_3$] (d3ls) [left=of c3ls]{};
        \node[gaugeb, label=below:$C_2$] (c2ls) [left=of d3ls]{};
        \node[gauger, label=below:$D_2$] (d2ls) [left=of c2ls]{};
        \node[gaugeb, label=below:$C_1$] (c1ls) [left=of d2ls]{};
        \node[gauger, label=below:$D_1$] (d1ls) [left=of c1ls]{};

        \draw[-] (c1ts)--(d2rs)--(c3rs)--(d4s)--(c3ls)--(d3ls)--(c2ls)--(d2ls)--(c1ls)--(d1ls);

        \node[gaugeb, label=right:$C_1$] (C1t) [below=of d2rs]{};
        \node[gauger, label=below:$D_3$] (D3) [below=of C1t]{};
        \node[gaugeb, label=below:$C_1$] (C1l) [left=of D3]{};
        \node[gaugeb, label=below:$C_3$] (C3r) [right=of D3]{};
        \node[gauger, label=below:$D_2$] (D2r) [right=of C3r]{};
        \node[gaugeb, label=below:$C_1$] (C1r) [right=of D2r]{};
        \node[gauger, label=below:$D_1$] (D1r) [right=of C1r]{};
        \node[gaugeb, label=right:$C_1$] (C1t2) [above=of C3r]{};

        \draw[-] (C1t)--(D3)--(C1l) (D3)--(C3r)--(D2r)--(C1r)--(D1r);
        \draw[transform canvas={xshift=+1.5pt}] (C3r)--(C1t2);
        \draw[transform canvas={xshift=-1.5pt}] (C3r)--(C1t2);

        \node[](minus) [left=of d1ls]{$-$};

        \node[] (topghost) [right=of d1r]{};
        \node[] (bottomghost) [right=of D1r]{};

        \draw[->] (topghost)to [out=-45,in=45,looseness=1](bottomghost);
        
    \end{tikzpicture}}
     \caption{}
     \label{fig:minE7SqSO71}
    \end{subfigure}
    \begin{subfigure}{\textwidth}
        \resizebox{\linewidth}{!}{\begin{tikzpicture}
    \node[gauger, label=below:$D_1$] (d1l) []{};
        \node[gaugeb, label=below:$C_1$] (c1l) [right=of d1l]{};
        \node[gauger, label=below:$D_2$] (d2l) [right=of c1l]{};
        \node[gaugeb, label=below:$C_2$] (c2l) [right=of d2l]{};
        \node[gauger, label=below:$D_3$] (d3l) [right=of c2l]{};
        \node[gaugeb, label=below:$C_3$] (c3l) [right=of d3l]{};
        \node[gauger, label=below:$D_4$] (d4l) [right=of c3l]{};
        \node[gaugeb, label=below:$C_4$] (c4l) [right=of d4l]{};
        \node[gauger, label=below:$D_5$] (d5) [right=of c4l]{};
        \node[gaugeb, label=below:$C_3$] (c3r) [right=of d5]{};
        \node[gauger, label=below:$D_2$] (d2r) [right=of c3r]{};
        \node[gaugeb, label=below:$C_1$] (c1r) [right=of d2r]{};
        \node[gauger, label=below:$D_1$] (d1r) [right=of c1r]{};
        \node[gaugeb, label=right:$C_2$] (c2t) [above=of d5]{};

        \draw[-] (d1l)--(c1l)--(d2l)--(c2l)--(d3l)--(c3l)--(d4l)--(c4l)--(d5)--(c3r)--(d2r)--(c1r)--(d1r) (d5)--(c2t);

        \node[gaugeb, label=right:$C_1$] (c1ts) [below=of c3r]{};
        \node[gauger, label=below:$D_2$] (d2rs) [left=of c1ts]{};
        \node[gaugeb, label=below:$C_3$] (c3rs) [left=of d2rs]{};
        \node[gauger, label=below:$D_4$] (d4s) [left=of c3rs]{};
        \node[gaugeb, label=below:$C_3$] (c3ls) [left=of d4s]{};
        \node[gauger, label=below:$D_3$] (d3ls) [left=of c3ls]{};
        \node[gaugeb, label=below:$C_2$] (c2ls) [left=of d3ls]{};
        \node[gauger, label=below:$D_2$] (d2ls) [left=of c2ls]{};
        \node[gaugeb, label=below:$C_1$] (c1ls) [left=of d2ls]{};
        \node[gauger, label=below:$D_1$] (d1ls) [left=of c1ls]{};

        \draw[-] (c1ts)--(d2rs)--(c3rs)--(d4s)--(c3ls)--(d3ls)--(c2ls)--(d2ls)--(c1ls)--(d1ls);

        \node[gaugeb, label=left:$C_2$] (C2t) [below=of d2rs]{};
        \node[gauger, label=below:$D_3$] (D3) [below=of C2t]{};
        \node[gaugeb, label=below:$C_1$] (C1l) [left=of D3]{};
        \node[gaugeb, label=below:$C_2$] (C2r) [right=of D3]{};
        \node[gauger, label=below:$D_2$] (D2r) [right=of C2r]{};
        \node[gaugeb, label=below:$C_1$] (C1r) [right=of D2r]{};
        \node[gauger, label=below:$D_1$] (D1r) [right=of C1r]{};
        \node[gaugeb, label=right:$C_1$] (C1t) [above=of D2r]{};

        \draw[-] (C2t)--(D3)--(C1l) (D3)--(C2r)--(D2r)--(C1r)--(D1r) (D2r)--(C1t);
        \draw[transform canvas={yshift=+1.5pt}] (C2t)--(C1t);
        \draw[transform canvas={yshift=-1.5pt}] (C2t)--(C1t);

        \node[](minus) [left=of d1ls]{$-$};

        \node[] (topghost) [right=of d1r]{};
        \node[] (bottomghost) [right=of D1r]{};

        \draw[->] (topghost)to [out=-45,in=45,looseness=1](bottomghost);
        
    \end{tikzpicture}}
    \caption{}
    \label{fig:minE7SqSO72}
    \end{subfigure}
    \begin{subfigure}{\textwidth}
    \centering
    \begin{tikzpicture}
        \node[gaugeb, label=right:$C_1$] (C1t) []{};
        \node[gauger, label=below:$D_3$] (D3) [below=of C1t]{};
        \node[gaugeb, label=below:$C_1$] (C1l) [left=of D3]{};
        \node[gaugeb, label=below:$C_2$] (C2r) [right=of D3]{};
        \node[gauger, label=below:$D_2$] (D2r) [right=of C2r]{};
        \node[gaugeb, label=below:$C_1$] (C1r) [right=of D2r]{};
        \node[gauger, label=below:$D_1$] (D1r) [right=of C1r]{};
        \node[gaugeb, label=right:$C_1$] (C1t2) [above=of C2r]{};

        \draw[-] (C1t)--(D3)--(C1l) (D3)--(C2r)--(D2r)--(C1r)--(D1r) (D3)--(C1t2)--(D2r);
        \end{tikzpicture}
        \caption{}
        \label{fig:minE7SqSO7Int}
    \end{subfigure}
    
    \caption{Two alignments of the $\sorm(7)$ quotient quiver against the orthosymplectic magnetic quiver for $\overline{min. E_7}\times\overline{min. E_7}$ producing \Quiver{fig:minE7SqSO71} and \Quiver{fig:minE7SqSO71}. Their intersection is \Quiver{fig:minE7SqSO7Int}.}
    \label{fig:minE7SqSO7}
\end{figure}

The Coulomb branch Hilbert series of \Quiver{fig:minE7SqSO71} is computed perturbatively up to order $t^{10}$ as \begin{align}
    \hs\left[\mathcal C\left(\text{\Quiver{fig:minE7SqSO71}}\right)\right]&=1 + 26 t^2 + 438 t^4 + 5385 t^6 + 52704 t^8 + 427872 t^{10} + O(t^{11})
 \\\pl\left[\hs\left[\mathcal C\left(\text{\Quiver{fig:minE7SqSO71}}\right)\right]\right]&=26 t^2 + 87 t^4 - 153 t^6 - 1434 t^8 + 5124 t^{10}+O(t^{11})
\end{align}

The Coulomb branch Hilbert series of \Quiver{fig:minE7SqSO72} is also computed perturbatively up to order $t^{20}$ as 
\begin{align}
    \hs\left[\mathcal C\left(\text{\Quiver{fig:minE7SqSO72}}\right)\right]&=1 + 26 t^2 + 439 t^4 + 5505 t^6 + 55520 t^8 + 466241 t^{10} + 
 3347718 t^{12} + O(t^{13})
 \\\pl\left[\hs\left[\mathcal C\left(\text{\Quiver{fig:minE7SqSO72}}\right)\right]\right]&=26 t^2 + 88 t^4 - 59 t^6 - 1501 t^8 - 1442 t^{10} + 52349 t^{12}+O(t^{13})
\end{align}

The Coulomb branch Hilbert series of \Quiver{fig:minE7SqSO7Int}, which is the intersection of \Quiver{fig:minE7SqSO71} and \Quiver{fig:minE7SqSO72}, is also computed perturbatively up to order $t^{20}$ as 
\begin{align}
    \hs\left[\mathcal C\left(\text{\Quiver{fig:minE7SqSO7Int}}\right)\right]&=1 + 26 t^2 + 438 t^4 + 5385 t^6 + 52505 t^8 + 421837 t^{10} + O(t^{11})
 \\\pl\left[\hs\left[\mathcal C\left(\text{\Quiver{fig:minE7SqSO7Int}}\right)\right]\right]&=26 t^2 + 87 t^4 - 153 t^6 - 1633 t^8 + 4263 t^{10}+O(t^{11})
\end{align}

The Hilbert series of the union of Coulomb branches is computed as \begin{align}
    \hs\left[\mathcal C\left(\text{\Quiver{fig:minE7SqSO71}}\right)\cup C\left(\text{\Quiver{fig:minE7SqSO72}}\right)\right]&= \hs\left[\mathcal C\left(\text{\Quiver{fig:minE7SqSO71}}\right)\right]+ \hs\left[\mathcal C\left(\text{\Quiver{fig:minE7SqSO72}}\right)\right]- \hs\left[\mathcal C\left(\text{\Quiver{fig:minE7SqSO7Int}}\right)\right]\nonumber\\&=1 + 26 t^2 + 439 t^4 + 5505 t^6 + 55719 t^8 + 472276 t^{10} + 
 3441602 t^{12} \nonumber\\&+O(t^{13})
 \\\pl\left[ \hs\left[\mathcal C\left(\text{\Quiver{fig:minE7SqSO71}}\right)\cup C\left(\text{\Quiver{fig:minE7SqSO72}}\right)\right]\right]&=26 t^2 + 88 t^4 - 59 t^6 - 1302 t^8 - 581 t^{10} + 36486 t^{12}+O(t^{13})
\end{align}

The result from the unions of the Coulomb branches has an $\sorm(5)^2\times\surm(2)^2$ global symmetry as indicated from the $t^2$ coefficient of the Hilbert series. This is confirmed with Weyl integration which uses the following embedding of $E_7\times E_7\hookleftarrow \sorm(5)^2\times\surm(2)^2\times\sorm(7)$ which decomposes the bifundamental as \begin{equation}
    {\mu_1}_{E_7}+{\mu_1'}_{E_7}\rightarrow \nu^2 + {\nu'}^2 + \mu_2^2 + {\mu_2^2}' + \mu_1 \rho_1 + \mu'_1 \rho_1 + 2 \rho_2 + \nu \mu_2 \rho_3 + \nu' \mu'_2 \rho_3
\end{equation}where the $\mu_i$ and $\mu_i'$ are highest weight fugacities for the two $\sorm(5)$, $\nu$ and $\nu'$ are highest weight fugacities for each $\surm(2)$, and the $\rho_i$ are highest weight fugacities for the $\sorm(7)$ which is gauged.

This result may be viewed as the diagonal flavour symmetry gauging of $\sorm(7)$ from two $4d\;\mathcal N=1$ rank-1 $E_7$ SCFTs realised through magnetic quivers \cite{Chacaltana:2011ze}.
\section{Magnetic Quivers for one M5-brane on $E_6$ Klein Singularity}
\label{sec:E6M5}
When an M5-brane probes an $E_6$ Klein singularity there is fractionation to at most four $\frac{1}{4}$M5 branes along the direction the singularity extends \cite{DelZotto:2014hpa}, as shown in \Figref{fig:4QuarterM5}. The fraction refers to the three-form flux each fractional M5 brane carries.

\begin{figure}[h!]
    \centering
    \begin{subfigure}{0.45\textwidth}
    \centering
        \begin{tikzpicture}
        \node[label=above:$\frac{1}{4}$ M5] (2) at (1.5,0){$\times$};
        \node  at (2.5,0){$\times$};
        \node  at (3.5,0){$\times$};
        \node at (4.5,0){$\times$};
    
        \node[] (E6) at (0,0){$E_6$};

        \draw[-] (0.5,0)--(5,0);

        \draw[-] (5.5,0)--(6.5,0) (6.4,0.1)--(6.5,0)--(6.4,-0.1);
        \draw[-] (5.5,0)--(5.5,1) (5.4,0.9)--(5.5,1)--(5.6,0.9);

        \node[] (x6) at (7,0){$x^6$};
        \node[] (x78910) at (5.5,1.5){$x^{7,8,9,10}$};        
        \end{tikzpicture}
        \caption{}
        \label{fig:4QuarterM5Branes}
    \end{subfigure}
    \centering
    \begin{subfigure}{0.45\textwidth}
    \centering
    \begin{tikzpicture}
        \node[gaugeb] (spol) {};
        \node[align=center,anchor=north] (lab) at (spol.south) {$\sprm(0)$\\$-1$};
        \node[gauge] (surm3) [right=of spol]{};
        \node[align=center,anchor=north] (lab2) at (surm3.south) {$\surm(3)$\\$-3$};
        \node[gaugeb] (spor) [right=of surm3]{};
        \node[align=center,anchor=north] (lab3) at (spor.south) {$\sprm(0)$\\$-1$};
        \node[flavour, label=left:$E_6$] (l) [above=of spol]{};
        \node[flavour, label=right:$E_6$] (r) [above=of spor]{};

        \draw[-] (l)--(spol)--(surm3)--(spor)--(r);
        
    \end{tikzpicture}
        \caption{}
        \label{fig:4QuarterM5Electric}
    \end{subfigure}
    \caption{Four $\frac{1}{4}$M5-branes on an $E_6$ Klein singularity. The directions $x^{0,1,2,3,4,5}$ are suppressed. The corresponding electric quiver \Quiver{fig:4QuarterM5Electric} comes from the F-theory dual description.}
    \label{fig:4QuarterM5}
\end{figure}

In general, the light spectrum of the low energy $6d\;\mathcal N=(1,0)$ field theory changes with the separation between the fractional M5-branes; as in the case for unitary 6d quivers \cite{Hanany:2018vph,Bourget:2022tmw}. In particular, the Higgs branch of the field theory changes dramatically between finite and infinite gauge coupling. This is a reflection of the presence of tensionless strings in the spectrum which changes the Higgs branch. This effect is difficult to see in general. However one can study this physics through a Type IIA brane system and magnetic quiver. Both of which are perturbative descriptions and moreover the magnetic quiver is a Lagrangian description.

Unlike the case of M5-branes on $A/D$ Klein singularities, there is no perturbative Type IIA description for the M-theory background on Klein singularities of $E$-type. Instead, F-theory translates the notion of bringing together fractional M5-branes to the collapse of curves of negative self-intersection \cite{Aspinwall:1998xj,DelZotto:2014hpa}. The following section will propose a pair of orthosymplectic magnetic quivers -- and the corresponding Type IIA brane system -- whose Coulomb branches are conjectured to be the Higgs branches of the worldvolume $6d\;\mathcal N=(1,0)$ theories of two $\frac{1}{2}$M5-branes and one M5-brane on a Klein $E_6$ singularity respectively. The latter is referred to in the Literature as $6d\;\mathcal N=(1,0)\;(E_6,E_6)$ conformal matter and is believed to be a strongly interacting SCFT \cite{DelZotto:2014hpa}.

\subsection{Two $\frac{1}{2}$M5-branes on $E_6$ Klein singularity}
First consider the phase in which the four $\frac{1}{4}$M5-branes on the $E_6$ Klein singularity coincide pairwise to become two $\frac{1}{2}$M5-branes, represented by the brane system and electric quiver of \Figref{fig:2HalfM5}.

\begin{figure}[h!]
    \centering
    \begin{subfigure}{0.45\textwidth}
    \centering
        \begin{tikzpicture}
        \node[label=above:$\frac{1}{2}$ M5] (2) at (2,0){$\times$};
        \node[] (3) at (3.5,0){$\times$};

        \node[] (E6) at (0,0){$E_6$};

        \draw[-] (0.5,0)--(5,0);

        \draw[-] (5.5,0)--(6.5,0) (6.4,0.1)--(6.5,0)--(6.4,-0.1);
        \draw[-] (5.5,0)--(5.5,1) (5.4,0.9)--(5.5,1)--(5.6,0.9);

        \node[] (x6) at (7,0){$x^6$};
        \node[] (x78910) at (5.5,1.5){$x^{7,8,9,10}$};
        
        \end{tikzpicture}
        \caption{}
        \label{fig:2HalfM5Branes}
    \end{subfigure}
    \centering
    \begin{subfigure}{0.45\textwidth}
    \centering
    \begin{tikzpicture}
        \node[gaugeb] (spol) {};
        \node[align=center,anchor=north] (lab) at (spol.south) {$\sprm(0)$};
        \node[gauge] (surm3) [right=of spol]{};
        \node[align=center,anchor=north] (lab2) at (surm3.south) {$\surm(3)$\\$-1$};
        \node[gaugeb] (spor) [right=of surm3]{};
        \node[align=center,anchor=north] (lab3) at (spor.south) {$\sprm(0)$};
        \node[flavour, label=left:$E_6$] (l) [above=of spol]{};
        \node[flavour, label=right:$E_6$] (r) [above=of spor]{};

        \draw[-] (l)--(spol)--(surm3)--(spor)--(r);
        
    \end{tikzpicture}
        \caption{}
        \label{fig:2HalfM5Electric}
    \end{subfigure}
    \caption{Two $\frac{1}{2}$ M5 branes on an $E_6$ Klein singularity. The directions $x^{0,1,2,3,4,5}$ are suppressed. The corresponding electric quiver \Quiver{fig:2HalfM5Electric} comes from the F-theory dual description.}
    \label{fig:2HalfM5}
\end{figure}

This transition removes two of the tensor multiplets and tunes the coupling of both $\sprm(0)$ to infinite coupling, increasing the dimension of the Higgs branch due to tensionless strings arising. The gauge coupling of the $\surm(3)$ in $\mathcal{Q}_{\ref{fig:2HalfM5Electric}}$ remains finite. The corresponding operation in F-theory collapses both $(-1)$-curves and simultaneously collapses the $(-3)$-curve supporting the $\surm(3)$ gauge symmetry to a $(-1)$-curve. Moreover, the F-theory picture suggests that the theory is described by two rank-1 E-strings (each with Higgs branch $\overline{min. E_8}$) coupled to a diagonal $\surm(3)$ gauge symmetry which remains at finite coupling. From this argument, the Higgs branch of \Quiver{fig:2HalfM5Electric} is thus predicted to be $\left(\overline{min. E_8}\times\overline{min. E_8}\right)///\surm(3)$.

The construction of a suitable magnetic quiver for the Higgs branch of \Quiver{fig:2HalfM5Electric} begins with the $\overline{min. E_8}\times \overline{min. E_8}$ theory of \cite{Sperling:2021fcf}, which matches the $E_8 \times E_8$ global symmetry supported by the two $(-1)$-curves, and then performing $\surm(3)$ quotient quiver subtraction which is illustrated in \Figref{fig:2HalfM5Magnetic} to give \Quiver{fig:2HalfM5Magnetic}.

\begin{figure}[h!]
    \centering
    \resizebox{\linewidth}{!}{\begin{tikzpicture}
        \node[gauger, label=below:$D_1$] (d1l) []{};
        \node[gaugeb, label=below:$C_1$] (c1l)[right=of d1l]{};
        \node[gauger, label=below:$D_2$] (d2l) [right=of c1l]{};
        \node[gaugeb, label=below:$C_2$] (c2l)[right=of d2l]{};
        \node[gauger, label=below:$D_3$] (d3l) [right=of c2l]{};
        \node[gaugeb, label=below:$C_3$] (c3l)[right=of d3l]{};
        \node[gauger, label=below:$D_4$] (d4l) [right=of c3l]{};
        \node[gaugeb, label=below:$C_4$] (c4l)[right=of d4l]{};
        \node[gauger, label=below:$D_5$] (d5l) [right=of c4l]{};
        \node[gaugeb, label=below:$C_5$] (c5l)[right=of d5l]{};
        \node[gauger, label=below:$D_6$] (d6l) [right=of c5l]{};
        \node[gaugeb, label=below:$C_6$] (c6l)[right=of d6l]{};
        \node[gauger, label=below:$D_7$] (d7) [right=of c6l]{};
        \node[gaugeb, label=below:$C_4$] (c4r) [right=of d7]{};
        \node[gauger, label=below:$D_2$] (d2r) [right=of c4r]{};
        \node[gaugeb, label=right:$C_3$] (c3t) [above=of d7]{};

        \draw[-] (d1l)--(c1l)--(d2l)--(c2l)--(d3l)--(c3l)--(d4l)--(c4l)--(d5l)--(c5l)--(d6l)--(c6l)--(d7)--(c4r)--(d2r) (d7)--(c3t);

        \node[gaugeb, label=below:$C_1$] (c1rs) [below=of c3l]{};
        \node[gauger, label=below:$D_2$] (d2rs) [left=of c1rs]{};
        \node[gaugeb, label=below:$C_2$] (c2s) [left=of d2rs]{};
        \node[gauger, label=below:$D_2$] (d2s) [left=of c2s]{};
        \node[gaugeb, label=below:$C_1$] (c1s) [left=of d2s]{};
        \node[gauger, label=below:$D_1$] (d1s) [left=of c1s]{};

        \draw[-] (c1rs)--(d2rs)--(c2s)--(d2s)--(c1s)--(d1s);

        \node[] (minus) [left=of d1s]{$-$};

        \node (ghost) [below=of d4l]{};

        \node[gaugeb, label=left:$C_1$] (C1t) [below=of ghost]{};
        \node[gauger, label=below:$D_4$] (D4l) [below=of C1t]{};
        \node[gaugeb, label=below:$C_2$] (C2l) [left=of D4l]{};
        \node[gauger, label=below:$D_1$] (D1l) [left=of C2l]{};
        \node[gaugeb, label=below:$C_4$] (C4l)[right=of D4l]{};
        \node[gauger, label=below:$D_5$] (D5l) [right=of C4l]{};
        \node[gaugeb, label=below:$C_5$] (C5l)[right=of D5l]{};
        \node[gauger, label=below:$D_6$] (D6l) [right=of C5l]{};
        \node[gaugeb, label=below:$C_6$] (C6l)[right=of D6l]{};
        \node[gauger, label=below:$D_7$] (D7) [right=of C6l]{};
        \node[gaugeb, label=below:$C_4$] (C4r) [right=of D7]{};
        \node[gauger, label=below:$D_2$] (D2r) [right=of C4r]{};
        \node[gaugeb, label=right:$C_3$] (C3t) [above=of D7]{};

        \draw[-] (D1l)--(C2l)--(D4l)--(C4l)--(D5l)--(C5l)--(D6l)--(C6l)--(D7)--(C4r)--(D2r) (D7)--(C3t) (D4l)--(C1t);
    \end{tikzpicture}}
    \caption{The subtraction of the $\surm(3)$ quotient quiver on the orthosymplectic quiver for $\overline{min. E_8}\times\overline{min. E_8}$ to give \Quiver{fig:2HalfM5Magnetic}. This is a magnetic quiver for the $6d\;\mathcal N=(1,0)$ worldvolume theory of two separated $\frac{1}{2}$M5 branes on $E_6$ Klein singularity.}
    \label{fig:2HalfM5Magnetic}
\end{figure}

Although the high rank of the resulting theory $\mathcal{Q}_{\ref{fig:2HalfM5Magnetic}}$ makes computation of an exact Hilbert series challenging, a calculation to order $t^{10}$ can be made, shown in \eqref{HS:2HalfM5Mag}, and its $\pl$ \eqref{HS:2HalfM5MagPL}. From this, a conjecture of which representations the generators and relations transform in are given in \eqref{HS:2HalfM5MagPLRefined} up to order $t^6$ as well as the $\pl$ of the $\hwg$ \eqref{HS:2HalfM5MagneticHWG}.
\begin{align}
    \hs\left[\mathcal C(\text{\Quiver{fig:2HalfM5Magnetic}})\right]&=1 + 156 t^2 + 13859 t^4 + 893669 t^6 + 45609733 t^8+1923636761t^{10} + O(t^{11})\label{HS:2HalfM5Mag}\\
    \pl\left[\hs\left[\mathcal C(\text{\Quiver{fig:2HalfM5Magnetic}})\right]\right]&=156 t^2 + 1613 t^4 - 2915 t^6 - 627017 t^8-1911458t^{10}+O(t^{11})
    \label{HS:2HalfM5MagPL}\\\pl\left[\hs\left[\mathcal C(\text{\Quiver{fig:2HalfM5Magnetic}})\right]\right]&=
    \left(\nu_6+\nu_6'\right)t^2+\left(\nu_1\nu_5'+\nu_5\nu_1'+\nu_6+\nu_6'-1\right)t^4\nonumber\\&-\left(\nu_1\nu_5'+\nu_5\nu_1'+\nu_6+\nu_6'+\nu_1\nu_5+\nu_1'\nu_5'+1\right)t^6+O(t^8)\label{HS:2HalfM5MagPLRefined}\\\pl\left[\hwg\left[\mathcal C(\text{\Quiver{fig:2HalfM5Magnetic}})\right]\right]&=\left(\mu_6 + \mu_6'\right) t^2 + (1 + \mu_1 \mu_5 + \mu_1' \mu_5 + \mu_1 \mu_5' + \mu_1' \mu_5' + \mu_6 + 
    \mu_6') t^4 \nonumber\\&+ \big(1 + \mu_1' \mu_2 + \mu_1 \mu_2' + 2 \mu_3 + 2 \mu_3' + \mu_1 \mu_5 + \mu_1' \mu_5 + 
    \mu_4' \mu_5 + \mu_1 \mu_5' + \mu_1' \mu_5' \nonumber\\&+ \mu_4 \mu_5' + \mu_6 + \mu_6'\big)t^6+O(t^8)\label{HS:2HalfM5MagneticHWG}
\end{align}
The $\nu_i$ and $\nu_i'$ in \eqref{HS:2HalfM5MagPLRefined} are highest weight fugacities for each factor of $E_6$ and are shorthand for the character of the corresponding representation in the $\pl$ of the $\hs$. The $\mu_i$ and $\mu_i'$ are also highest weight fugacities for each factor of $E_6$ and appear in the $\pl$ of the $\hwg$ \eqref{HS:2HalfM5MagneticHWG}.

The $t^2$ term in \eqref{HS:2HalfM5MagPL} shows that there is a generator transforming in the adjoint of $E_6\times E_6$ which correctly identifies an $E_6\times E_6$ global symmetry. The $t^4$ term suggests that there is a generator in the bifundamental as well as another generator in the adjoint. The relation at $t^4$ sets the second Casimir of each $E_6$ the same. At order $t^6$ relations appear which transform in the bifundamental, adjoint, singlet, and $\nu_1\nu_5+\nu_1'\nu_5'$.

This matches the Coulomb branch Hilbert series of the unitary counterpart which is constructed through the $\surm(3)$ chain polymerisation of two unitary affine $E_8^{(1)}$ quivers \cite{Hanany:2024fqf}. The result is \eqref{eq:2HalfM5MagneticUnitary}.

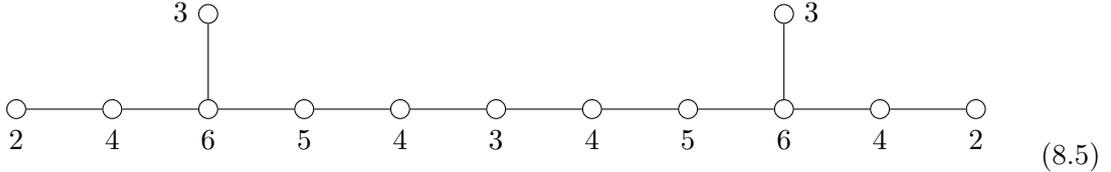
\begin{equation}
    \begin{tikzpicture}
        \node[gauge,label=right:$3$] (3btr) []{};
            \node[gauge, label=below:$6$] (6br) [below=of 3btr]{};
            \node[gauge, label=below:$4$] (4br) [right=of 6br]{};
            \node[gauge, label=below:$2$] (2br) [right=of 4br]{};
            \node[gauge, label=below:$5$] (5br) [left=of 6br]{};
            \node[gauge, label=below:$4$] (4bmr) [left=of 5br]{};
            \node[gauge, label=below:$3$] (3b) [left=of 4bmr]{};
            \node[gauge, label=below:$4$] (4bml) [left=of 3b]{};
            \node[gauge, label=below:$5$] (5bl) [left=of 4bml]{};
            \node[gauge, label=below:$6$] (6bl) [left=of 5bl]{};
            \node[gauge, label=below:$4$] (4bl) [left=of 6bl]{};
            \node[gauge, label=below:$2$] (2bl) [left=of 4bl]{};
            \node[gauge, label=left:$3$] (3btl) [above=of 6bl]{};

            \draw[-] (2bl)--(4bl)--(6bl)--(5bl)--(4bml)--(3b)--(4bmr)--(5br)--(6br)--(4br)--(2br) (3btr)--(6br) (3btl)--(6bl);
    \end{tikzpicture}\label{eq:2HalfM5MagneticUnitary}
\end{equation}

It is particularly interesting to note that although the orthosymplectic quiver \Quiver{fig:2HalfM5Magnetic} and its unitary counterpart take the same shape, the unitary quiver has a clear $S_2$ outer automorphism symmetry whereas the orthosymplectic quiver does not.

One must emphasise that a significant challenge with studying the physics of strongly interacting M5 brane on Klein $E_{6,7,8}$ singularities is the lack of a perturbative description as a magnetic quiver or in Type IIA. The problem of finding a magnetic quiver has been addressed with the derivation of \Quiver{fig:2HalfM5Magnetic} and \eqref{eq:2HalfM5MagneticUnitary}. It is challenging to find a brane construction of \eqref{eq:2HalfM5MagneticUnitary}. However, it is straightforward to give a Type IIA brane system from \Quiver{fig:2HalfM5Magnetic} as shown in \Figref{fig:2HalfM5MagneticBrane}. Importantly, the $C_1$ gauge node in \Quiver{fig:2HalfM5Magnetic} is interpreted as arising from a D4 brane ending between the NS5 and D6.
\begin{figure}[H]
\centering
\begin{tikzpicture}
     \def\x{1.5cm};
         \draw[-] (0,-\x)--(0,\x);
         \draw[-] (1,-\x)--(1,\x);
         \draw[-] (2,-\x)--(2,\x);
         \draw[-] (3,-\x)--(3,\x);
         \draw[-] (4,-\x)--(4,\x);
         \draw[-] (5,-\x)--(5,\x);
         \draw[-] (5,-\x)--(5,\x);
         \draw[-] (6,-\x)--(6,\x);
         \draw[-] (7,-\x)--(7,\x);
         \draw[-] (8,-\x)--(8,\x);
         \draw[-] (9,-\x)--(9,\x);
         \draw[-] (10,-\x)--(10,\x);
         \draw[-] (11,-\x)--(11,\x);
         
         \ns{2.5,1.2};
         \ns{2.5,-1.2};
         \ns{11,1};
         \ns{11,-1};
         \draw[-] (0,0.25)--(9,0.25);
         \draw[-] (0,-0.25)--(9,-0.25);

         \draw[-] (9,0.5)--(10,0.5);
         \draw[-] (9,-0.5)--(10,-0.5);

         \draw[-] (9,1)--(10.8,1);
         \draw[-] (10.8,1)--(10,0.9);
         \draw[-] (9,-1)--(10.8,-1);
         \draw[-] (10.8,-1)--(10,-0.9);

         \draw[-] (1,0)--(2,0) (3,0)--(4,0) (5,0)--(6,0) (7,0)--(8,0) (9,0)--(10,0);

	   \node[circle, draw, fill=white, label=right:$\mathrm{ON}^{-}$] at (11,0) {};

        \node[label=right:$\times 3$] at (11,1){};
        \node[label=right:$\times 3$] at (11,-1){};
        \node[label=above:$1$] at (0.5,0.2){};
        \node[label=above:$2$] at (1.5,0.2){};
         \node[label=above:$4$] at (2.5,0.2){};
         \node[label=above:$4$] at (3.5,0.2){};
         \node[label=above:$5$] at (4.5,0.2){};
         \node[label=above:$5$] at (5.5,0.2){};
         \node[label=above:$6$] at (6.5,0.2){};
         \node[label=above:$6$] at (7.5,0.2){};
         \node[label=above:$7$] at (8.5,0.2){};
         \node[label=above:$3$] at (9.5,0.3){};
         \node[label=above:$4$] at (10.5,0.9){};

\end{tikzpicture}
\caption{The Type IIA brane system for the magnetic quiver $\mathcal{Q}_{\ref{fig:2HalfM5Magnetic}}$. This is a magnetic quiver for the $6d\;\mathcal N=(1,0)$ worldvolume theory of two $\frac{1}{2}$M5 branes on a Klein $E_6$ singularity.}
\label{fig:2HalfM5MagneticBrane}
\end{figure}

\subsection{One M5 brane on $E_6$ Klein Singularity -- $6d\;\mathcal N=(1,0)\;(E_6,E_6)$ Conformal Matter}
When all four $\frac{1}{4}$M5 brane fractions are taken to be coincident, as shown in \Figref{fig:1M5Branes}, the resulting gauge theory, termed $6d\;\mathcal N=(1,0)\;(E_6,E_6)$ \emph{conformal matter}, is believed to be a strongly coupled $6d\;\mathcal N=(1,0)$ SCFT described by the electric quiver \Quiver{fig:1M5Electric} \cite{DelZotto:2014hpa}.

\begin{figure}[h!]
    \centering
    \begin{subfigure}{0.45\textwidth}
    \centering
        \begin{tikzpicture}
        \node[label=above:M5] (2) at (3,0){$\times$};
        \draw[-] (0.5,0)--(5,0);

        \draw[-] (5.5,0)--(6.5,0) (6.4,0.1)--(6.5,0)--(6.4,-0.1);
        \draw[-] (5.5,0)--(5.5,1) (5.4,0.9)--(5.5,1)--(5.6,0.9);

        \node[] (x6) at (7,0){$x^6$};
        \node[] (x78910) at (5.5,1.5){$x^{7,8,9,10}$};
        
        \end{tikzpicture}
        \caption{}
        \label{fig:1M5Branes}
    \end{subfigure}
    \centering
    \begin{subfigure}{0.45\textwidth}
    \centering
    \begin{tikzpicture}
        \node[gaugeb] (spol) {};
        \node[align=center,anchor=north] (lab) at (spol.south) {$\sprm(0)$};
        \node[gauge] (surm3) [right=of spol]{};
        \node[align=center,anchor=north] (lab2) at (surm3.south) {$\surm(3)$};
        \node[gaugeb] (spor) [right=of surm3]{};
        \node[align=center,anchor=north] (lab3) at (spor.south) {$\sprm(0)$};
        \node[flavour, label=left:$E_6$] (l) [above=of spol]{};
        \node[flavour, label=right:$E_6$] (r) [above=of spor]{};

        \draw[-] (l)--(spol)--(surm3)--(spor)--(r);
        
    \end{tikzpicture}
        \caption{}
        \label{fig:1M5Electric}
    \end{subfigure}
    \caption{One M5 brane on an $E_6$ Klein singularity. The directions $x^{0,1,2,3,4,5}$ are suppressed. The corresponding electric quiver \Quiver{fig:1M5Electric} comes from the F-theory dual description.}
    \label{fig:1M5}
\end{figure}

Although the two electric quivers \Quiver{fig:2HalfM5Electric} and \Quiver{fig:1M5Electric} may look similar, the latter results from shrinking the $(-1)$-curve in the former to zero size, which in field theory corresponds to tuning the gauge coupling associated to the $\surm(3)$ to infinity. This is known to give a small $E_8$ instanton transition \cite{Ganor:1996mu}. In the Type IIA brane system, this is seen by taking an NS5-brane onto the $O8^{-}$-plane \cite{Hanany:2018uhm} accounting for any brane creation \cite{Hanany:1996ie}. The small $E_8$ instanton transition can also be seen through the magnetic quiver by ``sliding off'' a $C_1$ gauge node and adding an affine $E_8^{(1)}$ shaped orthosymplectic quiver with node ranks given by the dual Coxeter numbers of $E_8^{(1)}$ \cite{Hanany:2018uhm}. The resulting brane system is shown in \Figref{fig:1M5MagneticBrane} and the corresponding magnetic quiver $\mathcal{Q}_{\ref{fig:E6E6ConfMatMagnetic}}$ is shown in \Figref{fig:E6E6ConfMatMagnetic}. In particular, \Quiver{fig:E6E6ConfMatMagnetic} is conjectured to be the magnetic quiver for $6d\;\mathcal N=(1,0)\;(E_6,E_6)$ conformal matter.

\begin{figure}[h!]
\centering
\begin{tikzpicture}
     \def\x{1.5cm};
         \draw[-] (0,-\x)--(0,\x);
         \draw[-] (1,-\x)--(1,\x);
         \draw[-] (2,-\x)--(2,\x);
         \draw[-] (3,-\x)--(3,\x);
         \draw[-] (4,-\x)--(4,\x);
         \draw[-] (5,-\x)--(5,\x);
         \draw[-] (5,-\x)--(5,\x);
         \draw[-] (6,-\x)--(6,\x);
         \draw[-] (7,-\x)--(7,\x);
         \draw[-] (8,-\x)--(8,\x);
         \draw[-] (9,-\x)--(9,\x);
         \draw[-] (10,-\x)--(10,\x);
         \draw[-] (11,-\x)--(11,\x);
         
         \ns{11,1};
         \ns{11,-1};
         \draw[-] (0,0.25)--(9,0.25);
         \draw[-] (0,-0.25)--(9,-0.25);

         \draw[-] (9,0.5)--(10,0.5);
         \draw[-] (9,-0.5)--(10,-0.5);

         \draw[-] (9,1)--(10.8,1);
         \draw[-] (10.8,1)--(10,0.9);
         \draw[-] (9,-1)--(10.8,-1);
         \draw[-] (10.8,-1)--(10,-0.9);

         \draw[-] (1,0)--(2,0) (3,0)--(4,0) (5,0)--(6,0) (7,0)--(8,0) (9,0)--(10,0);

	   \node[circle, draw, fill=white, label=right:$\mathrm{ON}^{-}$] at (11,0) {};

        \node[label=right:$\times 4$] at (11,1){};
        \node[label=right:$\times 4$] at (11,-1){};
        \node[label=above:$1$] at (0.5,0.2){};
        \node[label=above:$2$] at (1.5,0.2){};
         \node[label=above:$4$] at (2.5,0.2){};
         \node[label=above:$5$] at (3.5,0.2){};
         \node[label=above:$7$] at (4.5,0.2){};
         \node[label=above:$8$] at (5.5,0.2){};
         \node[label=above:$10$] at (6.5,0.2){};
         \node[label=above:$11$] at (7.5,0.2){};
         \node[label=above:$13$] at (8.5,0.2){};
         \node[label=above:$6$] at (9.5,0.2){};
         \node[label=above:$8$] at (10.5,1){};

\end{tikzpicture}
\caption{The Type IIA brane system from which the magnetic quiver \Quiver{fig:E6E6ConfMatMagnetic} is read. This is a magnetic quiver for $6d\;\mathcal N=(1,0)\;(E_6,E_6)$ conformal matter. This configuration can be obtained from that of \Figref{fig:2HalfM5MagneticBrane} by moving the NS5-brane onto the $O8^{-}$-plane.}
\label{fig:1M5MagneticBrane}
\end{figure}

\begin{figure}[h!]
    \centering
    \begin{tikzpicture}
        \node[gaugeb, label=left:$C_1$] (C1t) []{};
        \node[gauger, label=below:$D_4$] (D4l) [below=of C1t]{};
        \node[gaugeb, label=below:$C_2$] (C2l) [left=of D4l]{};
        \node[gauger, label=below:$D_1$] (D1l) [left=of C2l]{};
        \node[gaugeb, label=below:$C_4$] (C4l)[right=of D4l]{};
        \node[gauger, label=below:$D_5$] (D5l) [right=of C4l]{};
        \node[gaugeb, label=below:$C_5$] (C5l)[right=of D5l]{};
        \node[gauger, label=below:$D_6$] (D6l) [right=of C5l]{};
        \node[gaugeb, label=below:$C_6$] (C6l)[right=of D6l]{};
        \node[gauger, label=below:$D_7$] (D7) [right=of C6l]{};
        \node[gaugeb, label=below:$C_4$] (C4r) [right=of D7]{};
        \node[gauger, label=below:$D_2$] (D2r) [right=of C4r]{};
        \node[gaugeb, label=right:$C_3$] (C3t) [above=of D7]{};

        \draw[-] (D1l)--(C2l)--(D4l)--(C4l)--(D5l)--(C5l)--(D6l)--(C6l)--(D7)--(C4r)--(D2r) (D7)--(C3t) (D4l)--(C1t);

        \node[] (ghost) [below=of D7]{};

        \node[gaugeb, label=right:$C_6$] (c6t) [below=of ghost]{};
        \node[gauger, label=below:$D_{13}$] (d13) [below=of c6t]{};
        \node[gaugeb, label=below:$C_{11}$] (c11) [left=of d13]{};
        \node[gauger, label=below:$D_{10}$] (d10) [left=of c11]{};
        \node[gaugeb, label=below:$C_8$] (c8) [left=of d10]{};
        \node[gauger, label=below:$D_7$] (d7) [left=of c8]{};
        \node[gaugeb, label=below:$C_5$] (c5) [left=of d7]{};
        \node[gauger, label=below:$D_4$] (d4) [left=of c5]{};
        \node[gaugeb, label=below:$C_2$] (c2) [left=of d4]{};
        \node[gauger, label=below:$D_1$] (d1) [left=of c2]{};
        \node[gaugeb, label=below:$C_8$] (c8r) [right=of d13]{};
        \node[gauger, label=below:$D_4$] (d4r) [right=of c8r]{};

        \draw[-] (d1)--(c2)--(d4)--(c5)--(d7)--(c8)--(d10)--(c11)--(d13)--(c8r)--(d4r) (d13)--(c6t);

        \node[] (topghost) [below=of C5l]{};
        \node[] (bottomghost) [above=of c8]{};

        \draw[->] (topghost)to node[pos=0.5,left]{Small $E_8$ instanton transition}(bottomghost);
    \end{tikzpicture}
    \caption{The effect of a small $E_8$-instanton transition on the magnetic quiver \Quiver{fig:2HalfM5Magnetic} associated to the brane system in \Figref{fig:2HalfM5MagneticBrane} as the NS5-brane is taken onto the $O8^{-}$-plane. The resulting quiver \Quiver{fig:E6E6ConfMatMagnetic} is the magnetic quiver for $6d\;\mathcal N=(1,0)\;(E_6,E_6)$ conformal matter.}
    \label{fig:E6E6ConfMatMagnetic}
\end{figure}

Several pieces of evidence support this conjecture. The most basic is the dimension of the moduli space, which is easily read from $\mathcal{Q}_{\ref{fig:E6E6ConfMatMagnetic}}$ to be $\text{dim}\;E_6+1=79$, as expected. The high dimension makes it difficult for exact calculation of the Hilbert series under current techniques. However the first three terms of the $\hs$, can be computed as \eqref{HS:E6E6ConfMatter} and the $\pl$ \eqref{HS:E6E6ConfMatterPL}. It is then simple to conjecture representations the generators and relations transform in up to order $t^6$, which is given in \eqref{HS:E6E6ConfMatterPLRefined} and also conjecture a $\hwg$ whose $\pl$ is given in \eqref{HS:E6E6ConfMatterHWG}. Due to the balance of the nodes the next term in the Hilbert series is expected to show up at order $t^8$.

\begin{align}
     \hs\left[\mathcal C(\text{\Quiver{fig:E6E6ConfMatMagnetic}})\right]&=1+156 t^2+13703 t^4+876875 t^6+ O(t^{8}) \label{HS:E6E6ConfMatter}\\
    \pl\left[\hs\left[\mathcal C(\text{\Quiver{fig:E6E6ConfMatMagnetic}})\right]\right]&=156 t^2 + 1457 t^4 + 4627 t^6+O(t^{8}).
\label{HS:E6E6ConfMatterPL}\\
\pl\left[\hs\left[\mathcal C(\text{\Quiver{fig:E6E6ConfMatMagnetic}})\right]\right]&=
\left(\nu_6+\nu_6'\right)t^2+\left(\nu_1\nu_5'+\nu_5\nu_1'-1\right)t^4+\left(\nu_6\nu_6'+1-\nu_1\nu_5'-\nu_5\nu_1'\right)+O(t^8)\label{HS:E6E6ConfMatterPLRefined}\\\pl\left[\hwg\left[\mathcal C(\text{\Quiver{fig:E6E6ConfMatMagnetic}})\right]\right]&=\left(\mu_6+\mu_6'\right)t^2+\left(1 + \mu_1 \mu_5 + \mu_1' \mu_5 + \mu_1 \mu_5' +\mu_1'\mu_5'\right)t^4\nonumber\\&+\big(1 + \mu_1 \mu_2'+ \mu_2\mu_1' +\mu_3 + \mu_3' + \mu_1\mu_5 + \mu_1'\mu_5'+ \mu_1\mu_5'+ \mu_5\mu_1' \nonumber\\&+
   \mu_4\mu_5'+ \mu_5\mu_4'    + \mu_6\mu_6'\big)t^6+O(t^8)\label{HS:E6E6ConfMatterHWG}
\end{align}

The $\nu_i$ and $\nu_i'$ in \eqref{HS:E6E6ConfMatterPLRefined} are highest weight fugacities for each $E_6$ and are shorthand for the character of the given representation. The $\mu_i$ and $\mu_i'$ are also highest weight fugacities for each $E_6$ and appear in the $\pl$ of the $\hwg$ in \eqref{HS:E6E6ConfMatterHWG}.

The specific representations that appear in \eqref{HS:E6E6ConfMatterPLRefined} have simple physical interpretations. The $t^{2}$ term identifies a generator transforming in the adjoint of $E_6\times E_6$ indicating an $E_6\times E_6$ global symmetry. The $t^{4}$ term is interpreted as a generator transforming in the bifundamental representation of $E_6\times E_6$ with a relation. This relation expresses the equality of the second Casimir of each $E_6$. At order $t^6$ there are generators transforming in the biadjoint and singlet. The relations at this order transform in the bifundamental and come from the fact that the tensor product representation of the adjoint and bifundamental of $E_6\times E_6$ produce two identical bifundamentals (among other representations).

It is interesting to note that the magnetic quiver \Quiver{fig:E6E6ConfMatMagnetic} for $6d\;\mathcal N=(1,0)\;(E_6,E_6)$ conformal matter is star-shaped. Therefore this quiver \Quiver{fig:E6E6ConfMatMagnetic} is also a magnetic quiver for the Higgs branch of the class $\mathcal S$ theory specified by algebra $\sorm(26)$ on a three-punctured sphere with puncture data given by the following list of partitions of $26$, $\{(3^8,1^2),(9^2,7,1),(13^2)\}$. This class $\mathcal S$ theory is apparently different to the $T^2$ compactification of $6d\;\mathcal N=(1,0)\;(E_6,E_6)$ conformal matter \cite{Ohmori:2015pua} which is also specified by a three-punctured sphere but with algebra $E_6$ and two maximal and one minimal puncture.
\section{Discussion}
In this work the method of orthosymplectic quotient quiver subtraction is introduced. This is an operation on orthosymplectic magnetic quivers which has the effect of gauging either a $\surm(2),\;\surm(3),\;G_2,$ or $\sorm(7)$ subgroup of the Coulomb branch global symmetry. The effect on the Coulomb branch is a hyper-Kähler quotient by the corresponding group. The technique involves the subtraction of certain orthosymplectic quivers with negatively balanced nodes called the orthosymplectic quotient quivers. The full explanation of the algorithm is in Section \ref{sec:Method}.

The method of orthosymplectic quotient quiver subtraction shares much in common with its unitary counterpart \cite{Hanany:2023tvn}. They both involve subtraction of quotient quivers which contain gauge nodes of negative balance. Similarly, if the quotient quiver goes past a junction then the result is a union of the Coulomb branches coming from the different alignments. In the orthosymplectic case rebalancing of gauge nodes is done with a $C_1$ gauge node, this breaks the alternating $\sorm-\sprm$ pattern of gauge nodes typically seen in orthosymplectic quivers. In all cases studied here, the resulting intersection is found through an $A_1$ Kraft-Procesi (KP) transition \cite{Kraft1980MinimalGLn,Kraft1982OnGroups}. Although there is no established quiver subtraction for KP transitions on orthosymplectic quivers, subtraction of \eqref{eq:A1Slice} is used here.

In contrast to the unitary quotient quiver subtraction \cite{Hanany:2023tvn} where quotient quivers are known for the entire $\surm(n)$ family, identification of orthosymplectic quotient quivers has thus-far resisted attempts at generalisation beyond the four cases considered in this work. This is because they were derived from studying the various flavour symmetry gaugings of the non-anomalous $6d\;\mathcal N=(1,0)\;\sprm(k)$ electric gauge theory at infinite coupling. The consideration of anomaly cancellation and requirement of linearity of the electric $6d$ theory narrows down the possible flavour symmetry subgroups that can be gauged. In Section \ref{sec:Derivation}, the orthosymplectic quotient quivers for $\surm(3),\;G_2,$ and $\sorm(7)$ were derived from comparing magnetic quivers for these electric theories before and after gauging those flavour symmetry subgroups. The $\surm(2)$ orthosymplectic quotient quiver was conjectured through the natural Higgsing pattern of $\sorm(7)\rightarrow G_2\rightarrow \surm(3)\rightarrow\surm(2)$ observed in $6d$.


It hence remains to find a generalisation of the quotient quivers in this work to the classical groups, alongside an explanation from the 3d $\mathcal{N}=4$ side for the emergence of the series $\surm(2)$, $\surm(3)$, $G_2$, $\sorm(7)$. It is interesting to note that three of the four groups $\surm(2),\;\surm(3),$ and $G_2$ have a non-trivial sixth homotopy group, $\pi_6$, and therefore may suffer a global gauge anomaly \cite{Bershadsky:1997sb}. Whether there is a relationship between this anomaly and the linearity of the electric quiver remains unclear.

\paragraph{Nilpotent Orbits} The method of quotient quiver subtraction was applied to magnetic quivers for nilpotent orbit closures in Sections \ref{sec:SU2Examples}, \ref{sec:SU3Examples}, \ref{sec:G2Examples}, and \ref{sec:SO7Examples}. In many cases the result from orthosymplectic quotient quiver subtraction either matched unitary quotient quiver subtraction or from quiver polymerisation \cite{Hanany:2024fqf}. This gave orthosymplectic counterparts to some known unitary quivers, with both the Coulomb branch and Higgs branch matching. The counterparts which are explicitly checked are summarised in Tables \ref{tab:UnitaryOrthoCounter} and \ref{tab:UnitaryOrthoCounter2}. Many other unitary counterparts were proposed but were not completely checked either due to the large dimension of the moduli space or because one or both quivers contained non-simply laced edges.

The $G_2$ and $\sorm(7)$ orthosymplectic quotient quiver subtraction is the first step in understanding systematically the gauging of these Coulomb branch global symmetry subgroups. These do not have realisations from unitary quivers. For example, using $G_2$ quotient quiver subtraction on the orthosymplectic magnetic quiver for $\overline{min. E_8}$ identifies the moduli space as a union of two Coulomb branches. This gave two orthosymplectic magnetic quivers \Quiver{fig:minE8G21} and \Quiver{fig:minE8G22} whose Coulomb branches are the the height four nilpotent orbit closures $\overline{\mathcal O}^{F_4}_{[2,0,0,0]}$ and (normalisation of) $\overline{\mathcal O}^{F_4}_{[0,0,0,2]}$ respectively. The intersection \Quiver{fig:minE8G2Int} has a Coulomb branch which is the height three nilpotent closure $\overline{\mathcal O}^{F_4}_{[0,1,0,0]}$. All three quivers are non-trivial findings as there is no prescription to find magnetic quivers for nilpotent orbit closures of height greater than three. These quivers are a welcome addition to the catalogue.

\begin{landscape}
    \begin{table}[]
        \centering
        \begin{tabular}{cccc}
        \toprule
             Unitary Quiver& Orthosymplectic Quiver& Coulomb Branch& Higgs Branch \\
             \midrule
             \raisebox{-0.5\height}{\begin{tikzpicture}
              \node[gauge, label=right:$2$] (2) at (0,1){};
    \node[gauge, label=below:$4$] (4) at (0,0){};
    \node[gauge,label=below:$2$] (2L) at (-1,0){};
    \node[gauge,label=below:$3$] (3) at (1,0){};
    \node[gauge, label=below:$2$] (2R) at (2,0){};
    \node[gauge, label=below:$1$] (1) at (3,0){};
    \node[gauge,label=left:$1$] (1F) at ({-cos(45)},{sin(45)}){};
    
    \draw[-] (2L)--(4)--(3)--(2R)--(1);
    \draw[-] (1F)--(4)--(2);
        \end{tikzpicture}}& \raisebox{-0.5\height}{\begin{tikzpicture}\node[gauger, label=right:$D_1$,fill=red] (D1t) at (0,2){};
    \node[gaugeb, label=right:$C_1$,fill=blue] (C1t) at(0,1){};
    \node[gauger, label=below:$D_3$,fill=red] (D3) at (0,0){};
    \node[gaugeb, label=below:$C_1$,fill=blue] (C1l) at (-1,0){};
    \node[gauger, label=below:$D_1$,fill=red] (D1l) at (-2,0){};
    \node[gaugeb, label=left:$C_1$,fill=blue] (C1diag) at ({-sin(45)},{cos(45)}){};

    \node[gaugeb, label=below:$C_2$,fill=blue] (C2r) at (1,0){};
    \node[gauger, label=below:$D_2$,fill=red] (D2r) at (2,0){};
    \node[gaugeb, label=below:$C_1$,fill=blue] (C1r) at (3,0){};
    \node[gauger, label=below:$D_1$,fill=red] (D1r) at (4,0){};

    \draw[-] (D1t)--(C1t)--(D3)--(C1l)--(D1l);
    \draw[-] (C1diag)--(D3)--(C2r)--(D2r)--(C1r)--(D1r); \end{tikzpicture}}& $\overline{n. n. min. D_6}$ & $\mathcal S^{D_6}_{\mathcal N,(7,5)}$\\
    \raisebox{-0.5\height}{\begin{tikzpicture} \node[gauge, label=below:$1$] (1l) at (0,0){};
        \node[gauge, label=below:$2$] (2l) at (1,0){};
        \node[gauge, label=below:$2$] (2m) at (2,0){};
        \node[gauge, label=below:$2$] (2r) at (3,0){};
        \node[gauge, label=below:$1$] (1r) at (4,0){};
        \node[gauge, label=above:$1$] (1t) at (2,1){};

        \draw[-] (1l)--(2l)--(2m)--(2r)--(1r) (2l)--(1t)--(2r);\end{tikzpicture}}& \raisebox{-0.5\height}{\begin{tikzpicture} \node[gauger, label=left:$D_1$] (d1t) at (1,1){};
   \node[gaugeb, label=below:$C_1$] (c2) at (1,0){};
   \node[gauger, label=below:$D_1$] (d1l) at (0,0){};
   \node[gauger, label=below:$D_2$] (d2r) at (2,0){};
   \node[gaugeb, label=below:$C_1$] (c1r) at (3,0){};
   \node[gauger, label=below:$D_1$] (d1r) at (4,0){};
    \node[gaugeb, label=right:$C_1$] (c1t) at (2,1){};

    \draw[-] (d1r)--(c1r)--(d2r)--(c1t)--(d1t)--(c2)--(d1l);
    \draw[-] (c2)--(d2r); \end{tikzpicture}}&$\overline{n. min. A_5}$ & $\mathcal S^{A_5}_{\mathcal N,(4,2)}$\\
     \raisebox{-0.5\height}{\begin{tikzpicture} \node[gauge, label=below:$1$] (1l) at (0,0) {};
        \node[gauge, label=below:$2$] (2l) at (1,0) {};
        \node[gauge, label=below:$2$] (2r) at (2,0){};
        \node[gauge, label=above:$1$] (1T) at ({1+cos(60)},{sin(60)}){};
        \draw[-] (1l)--(2l)--(2r) (2l)--(1T);
        \draw[transform canvas={xshift=-1.29903810568 pt, yshift=-0.75 pt}] (2r)--(1T);
        \draw[transform canvas={xshift=+1.29903810568 pt, yshift=+0.75 pt}] (2r)--(1T);\end{tikzpicture}}&  \raisebox{-0.5\height}{\begin{tikzpicture} \node[gauger, label=above:$D_1$] (D1t) at ({1+cos(60)},{sin(60)}){};
    \node[gaugeb, label=below:$C_1$] (C1l) at (1,0){};
    \node[gaugeb, label=below:$C_1$] (C1r) at (2,0){};
    \node[gauger, label=below:$D_1$] (D1l) at (0,0){};
    \node[gauger, label=below:$D_1$] (D1r) at (3,0){};

    \draw[-] (D1l)--(C1l)--(D1t)--(C1r)--(D1r);

    \draw[transform canvas={yshift=1.5pt}] (C1l)--(C1r);
    \draw[transform canvas={yshift=-1.5pt}] (C1l)--(C1r); \end{tikzpicture}} & $\overline{\mathcal O}^{A_3}_{(3,1)}$ &$\mathcal S^{A_3}_{\mathcal N,(2,1^2)}$\\
        \raisebox{-0.5\height}{\begin{tikzpicture} \node[gauge, label=below:$1$] (1l) at (0,0) {};
        \node[gauge, label=below:$2$] (2l) at (1,0) {};
        \node[gauge, label=below:$1$] (1r) at (2,0){};
        \node[gauge, label=right:$1$] (1tr) at ({1+cos(60)},{sin(60)}){};
        \node[gauge, label=left:$1$] (1tl) at ({1-cos(60)},{sin(60)}){};
        \draw[-] (1l)--(2l)--(2r) (1tl)--(2l)--(1tr);
        \draw[transform canvas={xshift=0 pt, yshift=1.5 pt}] (1tl)--(1tr);
        \draw[transform canvas={xshift=0 pt, yshift=-1.5 pt}] (1tl)--(1tr);\end{tikzpicture}}& \raisebox{-0.5\height}{\begin{tikzpicture} \node[gauger, label=below:$D_1$] (D1r) at (2,0){};
    \node[gaugeb, label=below:$C_1$] (C1l) at (1,0){};
    \node[gauger, label=below:$D_1$] (D1l) at (0,0){};
    \node[gauger, label=left:$D_1$] (D1t) at ({cos(60)},{sin(60)}){};
    \node[gaugeb, label=right:$C_1$] (C1t) at ({1+cos(60)},{sin(60)}){};

    \draw[-] (D1l)--(C1l)--(D1r) (D1t)--(C1l) (D1t)--(C1t);
     \draw[transform canvas={xshift=+0.75pt,yshift=-1.29903810568pt}] (C1l)--(C1t);
    \draw[transform canvas={xshift=-0.75pt,yshift=+1.29903810568pt}] (C1l)--(C1t);\end{tikzpicture}}&$\overline{[\mathcal W_{D_4}]}^{[0,1,0,2]}_{[0,0,0,2]}$& No particular name\\
    \raisebox{-0.5\height}{\begin{tikzpicture}\node[gauge, label=below:$1$] (1l) at (0,0){};
        \node[gauge, label=below:$2$] (2) at (1,0){};
        \node[gauge, label=below:$1$] (1r) at (2,0){};
        \node[gauge, label=above:$1$] (1t) at (1,1){};
        \draw[-] (1l)--(2)--(1r);
        \draw[transform canvas={xshift=1.5 pt, yshift=0 pt}] (2)--(1t);
        \draw[transform canvas={xshift=-1.5 pt, yshift=0 pt}] (2)--(1t);\end{tikzpicture}}&\raisebox{-0.5\height}{\begin{tikzpicture}
            \node[gaugeb, label=below:$C_1$] (C1)at (0,0){};
            \node[gauger, label=below:$D_1$] (D1l) at (-1,0){};
            \node[gauger, label=below:$D_1$] (D1r) at (1,0){};
            \node[gauger, label=above:$D_1$] (D1t) at (0,1){};
            \draw[-] (D1l)--(C1)--(D1r);
            \draw[transform canvas={xshift=+1.5pt}] (C1)--(D1t);
            \draw[transform canvas={xshift=-1.5pt}] (C1)--(D1t);
        \end{tikzpicture}}&$\overline{\mathcal O}^{A_2}_{(2^2)}$&$\mathcal S^{A_2}_{\mathcal N,(2^2)}$\\
         \raisebox{-0.5\height}{\begin{tikzpicture}
             \node[gauge,label=right:$1$] (oneF) at (1,1){};
    \node[gauge, label=below:$4$] (four) at (1,0){};
    \node[gauge, label=below:$2$] (twoL) at (0,0){};
    \node[gauge, label=below:$5$] (five) at (2,0){};
    \node[gauge, label=below:$6$] (six) at (3,0){};
    \node[gauge, label=below:$4$] (fourR) at (4,0){};
    \node[gauge,label=below:$2$] (twoR) at (5,0){};
    \node[gauge,label=right:$3$] (three) at (3,1){};
    
    \draw[-] (twoL)--(four)--(five)--(six)--(fourR)--(twoR);
    \draw[-] (three)--(six);
    \draw[-] (oneF)--(four);
         \end{tikzpicture}}& \raisebox{-0.5\height}{\begin{tikzpicture}
             \node[gauger, label=below:$D_1$,fill=red] (d1ls) at (0,0) {};
    \node[gaugeb, label=below:$C_1$,fill=blue] (c1ls) at (1,0){};
    \node[gauger, label=below:$D_3$,fill=red] (d3ls) at (2,0){};
    \node[gaugeb, label=below:$C_3$,fill=blue] (c3ls) at (3,0){};
    \node[gauger, label=below:$D_4$,fill=red] (d4s) at (4,0){};
    \node[gaugeb, label=below:$C_3$,fill=blue] (c3rs) at (5,0){};
    \node[gauger, label=below:$D_3$,fill=red] (d3rs) at (6,0){};
    \node[gaugeb, label=below:$C_2$] (c2rs) at (7,0){};
    \node[gauger, label=below:$D_2$] (d2rs) at (8,0){};
    \node[gaugeb, label=below:$C_1$,fill=blue] (c1rs) at (9,0){};
    \node[gauger, label=below:$D_1$,fill=red] (d1rs) at (10,0){};
    
    \node[gaugeb, label=left:$C_1$,fill=blue] (c1t1s) at (2,1){};
    \node[gaugeb, label=right:$C_1$,fill=blue] (c1t2s) at (4,1){};

    \draw[-] (d1ls)--(c1ls)--(d3ls)--(c3ls)--(d4s)--(c3rs)--(d3rs)--(c2rs)--(d2rs)--(c1rs)--(d1rs);
    \draw[-] (c1t1s)--(d3ls);
    \draw[-] (c1t2s)--(d4s);
    
         \end{tikzpicture}}&$\overline{n. min. E_7}$ &$\mathcal S^{E_7}_{\mathcal N,[2, 2, 0, 2, 0, 2, 2]}$\\\bottomrule
        \end{tabular}
        \caption{Some unitary quivers and their orthosymplectic counterparts.}
        \label{tab:UnitaryOrthoCounter}
    \end{table}
\end{landscape}
\begin{landscape}
    \begin{table}[]
        \centering
        \begin{tabular}{cccc}
        \toprule
             Unitary Quiver& Orthosymplectic Quiver& Coulomb Branch& Higgs Branch \\
             \midrule
             \raisebox{-0.5\height}{\begin{tikzpicture}
                 \node[gauge, label=below:$1$] (1l) at (0,0){};
        \node[gauge, label=below:$2$] (2l) at (1,0){};
        \node[gauge, label=below:$3$] (3) at (2,0){};
        \node[gauge, label=below:$2$] (2r) at (3,0){};
        \node[gauge, label=below:$1$] (1r) at (4,0){};
        \node[gauge, label=above:$1$] (1t) at (2,1){};

        \draw[-] (1l)--(2l)--(3)--(2r)--(1r);
        \draw[transform canvas={xshift= 1.5pt,yshift=0 pt}](3)--(1t);
        \draw[transform canvas={xshift= -1.5pt,yshift=0 pt}](3)--(1t);
             \end{tikzpicture}}&\raisebox{-0.5\height}{\begin{tikzpicture}
                 \node[gauger, label=right:$D_1$] (D1t) at (0,2){};
    \node[gaugeb, label=right:$C_1$] (C1t) at (0,1){};
    \node[gaugeb, label=below:$C_2$] (C2) at (0,0){};
    \node[gauger, label=below:$D_1$] (D1l) at (-1,0){};
    \node[gauger, label=below:$D_2$] (D2) at (1,0){};
    \node[gaugeb, label=below:$C_1$] (C1) at (2,0){};
    \node[gauger, label=below:$D_1$] (D1) at (3,0){};

    \draw[-] (D1t)--(C1t) (D1l)--(C2)--(D2)--(C1)--(D1);

    \draw[transform canvas={xshift=+1.5pt}] (C2)--(C1t);
    \draw[transform canvas={xshift=-1.5pt}] (C2)--(C1t);
             \end{tikzpicture}}& $\overline{\mathcal O}^{A_5}_{(2^3)}$ &$\mathcal S^{A_5}_{\mathcal N,(3^2)}$\\
             \raisebox{-0.5\height}{\begin{tikzpicture}
                 \node[gauge, label=below:$1$] (1l) at (0,0){};
        \node[gauge, label=below:$2$] (2l) at (1,0){};
        \node[gauge, label=below:$2$] (2lm) at (2,0){};
        \node[gauge, label=below:$2$] (2rm) at (3,0){};
        \node[gauge, label=below:$2$] (2r) at (4,0){};
        \node[gauge, label=above:$1$] (1t) at (2.5,1){};

        \draw[-] (1l)--(2l)--(2lm)--(2rm)--(2r) (1t)--(2l);
        \draw[transform canvas={xshift=-0.55470019622 pt, yshift=-0.83205029433 pt}] (2r)--(1t);
        \draw[transform canvas={xshift=0.55470019622 pt, yshift=0.83205029433 pt}] (2r)--(1t);
             \end{tikzpicture}}&\raisebox{-0.5\height}{\begin{tikzpicture}
                 \node[gauger, label=left:$D_1$] (D1t) at (0,2){};
    \node[gaugeb, label=left:$C_1$] (C1tl) at (0,1){};
    \node[gauger, label=below:$D_1$] (D1l) at (0,0){};
    \node[gaugeb, label=below:$C_1$] (C1l) at (1,0){};
    \node[gauger, label=below:$D_2$] (D2) at (2,0){};
    \node[gaugeb, label=below:$C_1$] (C1r) at (3,0){};
    \node[gauger, label=below:$D_1$] (D1r) at (4,0){};
    \node[gaugeb, label=right:$C_1$] (C1tr) at (2,1){};

    \draw[-] (D1t)--(C1tl)--(D1l)--(C1l)--(D2)--(C1r)--(D1r) (D2)--(C1tr);

    \draw[transform canvas={yshift=+1.5pt}] (C1tl)--(C1tr);
    \draw[transform canvas={yshift=-1.5pt}] (C1tl)--(C1tr);
             \end{tikzpicture}}& $\overline{\mathcal O}^{A_5}_{(3,1^3)}$&$\mathcal S^{A_5}_{(4,1^2)}$\\
             \raisebox{-0.5\height}{\begin{tikzpicture}
                 \node[gauge, label=below:$2$] (2l) at (0,0){};
                 \node[gauge, label=below:$4$] (4l) at (1,0){};
                 \node[gauge,label=below left:$6$] (6) at (2,0){};
                 \node[gauge, label=below:$4$] (4r) at (3,0){};
                 \node[gauge, label=below:$2$] (2r) at (4,0) {};
                 \node[gauge, label=above:$3$] (3t) at (2,1){};
                 \node[gauge, label=below:$1$] (1) at (2,-1) {};

                \draw[-] (2l)--(4l)--(6)--(4r)--(2r) (3t)--(6)--(1);
             \end{tikzpicture}}&\raisebox{-0.5\height}{\begin{tikzpicture}
                  \node[gaugeb, label=left:$C_1$] (C1t) at ({-cos(60)},{sin(60)}){};
   \node[gauger, label=below:$D_4$] (D4) at (0,0){};
   \node[gaugeb, label=below:$C_2$] (C2l) at (-1,0){};
   \node[gauger, label=below:$D_1$] (D1l) at (-2,0){};
   \node[gaugeb, label=below:$C_3$] (C3r) at (1,0){};
   \node[gauger, label=below:$D_3$] (D3r) at (2,0){};
   \node[gaugeb, label=below:$C_2$] (C2r) at (3,0){};
   \node[gauger, label=below:$D_2$] (D2r) at (4,0){};
   \node[gaugeb, label=below:$C_1$] (C1r) at (5,0){};
   \node[gauger, label=below:$D_1$] (D1r) at (6,0){};
    \node[gaugeb, label=right:$C_1$] (C1tr) at ({cos(60)},{sin(60)}){};

    \draw[-] (C1tr)--(D4)--(C1t);
    \draw[-] (D1r)--(C1r)--(D2r)--(C2r)--(D3r)--(C3r)--(D4)--(C2l)--(D1l);
             \end{tikzpicture}}& $\overline{\mathcal O}^{E_6}_{[0,0,0,0,0,2]}$ Double Cover& $\mathcal S^{E_6}_{\mathcal N,[2,0,2,0,2,0]}/\mathbb Z_2$ \\
             \raisebox{-0.5\height}{\begin{tikzpicture}
                 \node[gauge, label=right:$1$] (1rt) at (0,2){};
        \node[gauge, label=right:$2$] (2rt) at (0,1){};
        \node[gauge, label=below left:$3$] (3r)at (0,0){};
        \node[gauge, label=below:$2$] (2rr) at (1,0){};
        \node[gauge, label=below:$1$] (1rr) at (2,0){};
        \node[gauge, label=below:$2$] (2rl) at (-1,0){};
        \node[gauge, label=below:$1$] (1rl) at (-2,0){};
        \node[gauge, label=right:$2$] (2rb) at (0,-1){};
        \node[gauge, label=right:$1$] (1rb) at (0,-2){};

        \draw[-] (1rt)--(2rt)--(3r)--(2rr)--(1rr) (1rl)--(2rl)--(3r)--(2rb)--(1rb);
             \end{tikzpicture}}&\raisebox{-0.5\height}{\begin{tikzpicture}
                 \node[gauger, label=left:$D_1$] (D1t) at (0,2) {};
        \node[gaugeb, label=left:$C_2$] (C2t) at (0,1){};
        \node[gauger, label=below:$D_4$] (D4) at (0,0){};
        \node[gaugeb, label=below:$C_2$] (C2l) at (-1,0){};
        \node[gauger, label=below:$D_1$] (D1l) at (-2,0){};
        \node[gaugeb, label=below:$C_2$] (C2r) at (1,0){};
        \node[gauger, label=below:$D_1$] (D1r) at (2,0){};
        \node[gaugeb, label=right:$C_1$] (D1tm) at ({cos(45)},{sin(45)}){};

        \draw[-] (D1t)--(C2t)--(D4)--(C2l)--(D1l) (D1tm)--(D4)--(C2r)--(D1r);
             \end{tikzpicture}}&No particular name&No particular name\\
             \raisebox{-0.5\height}{\begin{tikzpicture}
                 
             \end{tikzpicture}}&\raisebox{-0.5\height}{\begin{tikzpicture}
                 
             \end{tikzpicture}}& &\\
             \raisebox{-0.5\height}{\begin{tikzpicture}
                 
             \end{tikzpicture}}&\raisebox{-0.5\height}{\begin{tikzpicture}
                 
             \end{tikzpicture}}& &\\
             \bottomrule
        \end{tabular}
        \caption{Table \ref{tab:UnitaryOrthoCounter} continued.}
        \label{tab:UnitaryOrthoCounter2}
        \end{table}
\end{landscape}

\paragraph{Magnetic Quivers for an M5 brane on an $E_6$ Klein singularity}When an M5 brane probes an $E_6$ Klein singularity it may fractionate into at most four $\frac{1}{4}$M5 branes. The various separations between the fractions correspond to a different light spectra of the $6d\;\mathcal N=(1,0)$ theory. It has been a longstanding challenge to find magnetic descriptions for the finite and infinite coupling limit of this theory and additionally a description in Type IIA. 

In Section \ref{sec:E6M5} a magnetic quiver for the worldvolume gauge theory of two $\frac{1}{2}$M5 branes on $E_6$ Klein singularity is found. The F-theory description of this gauge theory is as the diagonal $\surm(3)$ gauging of two rank-1 E-strings. Therefore the derivation of the magnetic quiver \Quiver{fig:2HalfM5Magnetic} comes from application of the $\surm(3)$ orthosymplectic quotient quiver subtraction on the magnetic quiver for $\overline{min. E_8}\times\overline{min. E_8}$. The Coulomb branch Hilbert series of \Quiver{fig:2HalfM5Magnetic} matches that of the unitary magnetic quiver \eqref{eq:2HalfM5MagneticUnitary}. A suitable Type IIA brane description is drawn in \Figref{fig:2HalfM5MagneticBrane} based off this magnetic quiver \Quiver{fig:2HalfM5Magnetic}.

The magnetic quiver and brane system for the phase where all four M5 brane fractions are coincident is also derived. This theory is referred to as $6d\;\mathcal N=(1,0)\;(E_6,E_6)$ conformal matter and is believed to be a strongly coupled SCFT. The F-theory description suggests that a collapse of a $(-1)$-curve makes the two $\frac{1}{2}$M5 branes coincide. In field theory this causes a small $E_8$ instanton transition and is realised in the Type IIA brane system by bringing an NS5 brane onto the $O8^{-}$ accounting for brane creation. This can also be realised on the magnetic quiver. The Type IIA brane system is drawn in \Figref{fig:1M5MagneticBrane} and the resulting magnetic quiver is \Quiver{fig:E6E6ConfMatMagnetic}. 

The unrefined Hilbert series for both theories is also computed, \eqref{HS:2HalfM5Mag} for \Quiver{fig:2HalfM5Magnetic} and \eqref{HS:E6E6ConfMatter} for \Quiver{fig:E6E6ConfMatMagnetic}, together with the $\pl$ \eqref{HS:2HalfM5MagPL} and \eqref{HS:E6E6ConfMatterPL}. From this, a conjecture of which representations the generators transform in was made. The generators up to order $t^6$ are shown in Table \ref{tab:E6E6Gens}. Importantly, there is a clear difference in the generators of the Higgs branch between these two theories. This reflects the presence of tensionless strings as the gauge coupling of the $\surm(3)$ goes from finite to infinite.

\begin{table}[h!]
    \centering
    \begin{tabular}{ccc}
    \toprule
         R-charge & Higgs branch generators (Finite Coupling) & Higgs branch generators (Infinite Coupling) \\\midrule
         1 & $(\mathbf{78},\mathbf{1}),\;(\mathbf{1},\mathbf{78})$ & $(\mathbf{78},\mathbf{1}),\;(\mathbf{1},\mathbf{78})$\\
         2 & $(\mathbf{27},\overline{\mathbf{27}}),\;(\overline{\mathbf{27}},\mathbf{27}),\;(\mathbf{78},\mathbf{1}),\;(\mathbf{1},\mathbf{78})$&$(\mathbf{27},\overline{\mathbf{27}}),\;(\overline{\mathbf{27}},\mathbf{27})$\\
         3& None & $(\mathbf{78},\mathbf{78}),\;(\mathbf{1},\mathbf{1})$\\\bottomrule
    \end{tabular}
    \caption{The $E_6\times E_6$ representations Higgs branch generators transform in, of the $6d\;\mathcal N=(1,0)$ worldvolume theory of two $\frac{1}{2}$M5 branes (finite coupling) and one M5 brane (infinite coupling) on Klein $E_6$ singularity.}
    \label{tab:E6E6Gens}
\end{table}

There are further curiosities to note about \Quiver{fig:E6E6ConfMatMagnetic}. Firstly, it is a star-shaped quiver and therefore is a magnetic quiver for the class $\mathcal S$ theory \cite{Gaiotto:2009we} specified by an $\sorm(26)$ algebra and a three punctured sphere specified by partitions $\{(3^8,1^2),(9^2,7,1),(13^2)\}$. This is apparently different to a result in \cite{Ohmori:2015pua} which showed that the compactification of $6d\;\mathcal N=(1,0)\;(E_6,E_6)$ conformal matter on $T^2$ results in the $4d\;\mathcal N=2$ class $\mathcal S$ theory defined by an $E_6$ algebra and a three punctured sphere with two maximal and one minimal puncture. There is no known magnetic quiver for class $\mathcal S$ theories specified by exceptional algebras. Could \Quiver{fig:E6E6ConfMatMagnetic} be the magnetic for this class $\mathcal S$ theory with $E_6$ algebra?

The appearance of the $\sorm(26)$ algebra does not appear to be a coincidence as it is related to $E_6$ in the following way. There is an embedding of $E_6\hookleftarrow F_4$ through the folding action and also and embedding of $\sorm(26)\hookleftarrow F_4$ through the fundamental of $F_4$ since this representation (like all representations of $F_4$) is real. It remains a challenge to see if this is a general behaviour and whether magnetic quivers for class $\mathcal S$ theories specified by exceptional algebras can be found systematically.

It is worth re-emphasising that the magnetic quiver and Type IIA brane system for $6d\;\mathcal N=(1,0)\;(E_6,E_6)$ conformal matter conjectured in this paper are the first of their kind, and that a similar set of magnetic quivers may also exist for the case of $(E_7,E_7)$ and $(E_8,E_8)$ but it is not yet clear how to find these.

\paragraph{Future Directions}
Orthosymplectic quotient quiver subtraction gauges four possible subgroups of the Coulomb branch global symmetry; $\surm(2),\;\surm(3),\;G_2,$ and $\sorm(7)$. The derivation from $6d\;\mathcal N=(1,0)$ theories does not extend easily since it is not known how to couple additional matter in the magnetic quiver. If this was known then $\sorm(n)$ orthosymplectic quotient quivers for $n\geq 8$ may easily be found.

It is more challenging to find $\surm(n)$ quotient quivers for $n\geq 4$ due to a lack of magnetic quivers. In principle, if the magnetic quivers were known then the method applied here extends simply. 

Physics in other dimensions may also provide further inspiration for quotient quivers. In particular from $5d\;\mathcal N=1$ theories which have descriptions from brane webs \cite{Aharony:1997bh}. Since there are no gauge anomalies in $5d$ there may be families of quotient quivers that may be found.
\paragraph{Acknowledgements.}
We thank Guillermo Arias-Tamargo, Rudolph Kalveks, Lorenzo Mansi, Shlomo Razamat, Marcus Sperling, and Gabi Zafrir for helpful discussions. We also thank Marcus Sperling for comments on the draft. We are grateful to Rudolph Kalveks for Mathematica help. The work of SB, AH, and GK is partially supported by STFC Consolidated Grants ST/T000791/1 and ST/X000575/1. The work of SB is supported by the STFC DTP research studentship grant ST/Y509231/1. The work of GK is supported by STFC DTP research studentship grant ST/X508433/1.
\bibliographystyle{JHEP}
\bibliography{references}
\end{document}